%% file: main.tex
\documentclass[acmsmall]{acmart}

\usepackage{color}
\usepackage{enumitem}
\usepackage{algorithm}
\usepackage{algpseudocode}
\usepackage{graphicx} %use graph format
\usepackage{epstopdf}
\usepackage{amsmath,bm}
\usepackage{amsfonts}
\usepackage{multirow}
\usepackage{array}
\usepackage{hyperref}
\hypersetup{hidelinks}
\AtBeginDocument{%
  \providecommand\BibTeX{{%
    \normalfont B\kern-0.5em{\scshape i\kern-0.25em b}\kern-0.8em\TeX}}}

\newcommand{\extension}{\color{black}}%blue
\newcommand{\ChengComment}{}
\newcommand{\ICDEComment}{}
\newcommand{\nnComment}{}
\newcommand{\revision}{}

\begin{document}

\title{Billiards Sports Analytics: Datasets and Tasks}

\author{Qianru Zhang}
\authornote{Both authors contributed equally to this research.}
\affiliation{%
  \institution{The University of Hong Kong}
  \country{Hong Kong SAR}
}
\email{qrzhang@cs.hku.hk}

\author{Zheng Wang}
\authornotemark[1]
\affiliation{%
  \institution{Nanyang Technological University}
  \country{Singapore}
}
\email{zheng011@e.ntu.edu.sg}

\author{Cheng~Long}
\authornote{Corresponding author.}
\affiliation{%
  \institution{Nanyang Technological University}
  \country{Singapore}}
\email{c.long@ntu.edu.sg}

\author{Siu-Ming~Yiu}
\authornotemark[2]
\affiliation{%
  \institution{The University of Hong Kong}
  \country{Hong Kong SAR}
}
\email{smyiu@cs.hku.hk}


\input{abstract}






\maketitle

\input{introduction2}
\input{related}
\input{dataset}
\input{tasks}

\input{method}
\input{experiments}

\input{conclusion}

\bibliographystyle{ACM-Reference-Format}
\bibliography{ref}

\end{document}

%% file: abstract.tex
\begin{abstract}
%{\color{red}
Nowadays, it becomes a common practice to capture some data of sports games with devices such as GPS sensors and cameras and then use the data to perform various analyses on sports games, including tactics discovery, similar game retrieval, performance study, etc. While this practice has been conducted to many sports such as basketball and soccer, it remains largely unexplored on the billiards sports, which is mainly due to the lack of publicly available datasets. 
Motivated by this, we collect a dataset of billiards sports, which includes the layouts (i.e., locations) of billiards balls after performing break shots, called break shot layouts, the traces of the balls as a result of strikes (in the form of trajectories), and detailed statistics and performance indicators.
We then study and develop techniques for three tasks on the collected dataset, including (1) prediction and (2) generation on the layouts data, 
% (3) similarity search on the strikes data, 
and (3) similar billiards layout retrieval on the layouts data, which can serve different users such as coaches, players and fans.
%the \emph{similar billiards layout retrieval} problem, which is to find from a database of many billiards layouts those that are similar to a user-specified one called \emph{query layout}. This problem can be used for many different purposes including game recommendation, prediction and clustering analysis, etc. This problem relies on a similarity measurement on billiards layouts, which is non-trivial. In this paper, we propose a novel solution equipped with deep metric learning, called BL2Vec, to learn the billiards' representations, which can effectively capture the unique characteristics in a billiards layout and takes linear time wrt the number of balls. 
We conduct extensive experiments on the collected dataset and the results show that our methods perform effectively and efficiently. 
% compared with baseline methods.
\end{abstract}

%% file: introduction2.tex
\section{INTRODUCTION}
\label{sec:introduction}

\begin{figure}
\vspace*{-2mm}
\centering
\begin{tabular}{c}
  \begin{minipage}{0.6\linewidth}
	\includegraphics[width=\linewidth]{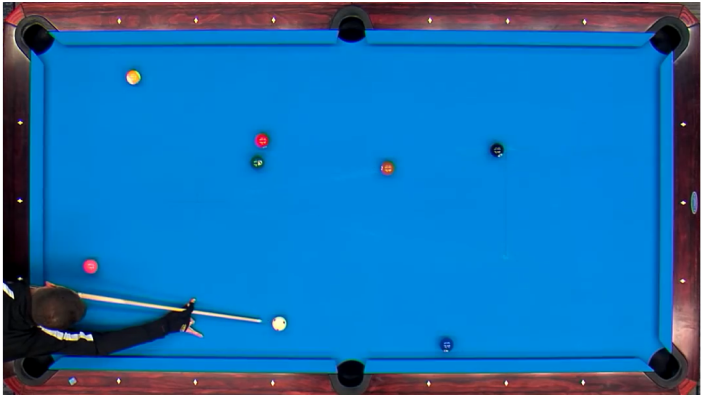}
  \end{minipage}
\end{tabular}
\vspace*{-2mm}
\caption{A real 9-ball billiards layout.}
\vspace*{-3mm}
\label{fig:example_intro}
\end{figure}

Nowadays, it becomes a common practice to collect some data from sports games using tracking devices such as cameras and/or GPS sensors. The collected data facilitates various analytic tasks such as similar game retrieval~\cite{wang2019effective,sha2016chalkboarding}, tactics detection~\cite{decroos2018automatic} and score prediction~\cite{aoki2017luck,yue2014learning}, which cover sports including basketball, soccer, volleyball and handball. Yet these analytic tasks have not been explored on billiards sports, and this is mainly due to the lack of publicly available datasets of billiards sports.
The billiards sport is a two-player game. In each game, there are a certain number of rounds (called \emph{frames}). In each frame, two players take turns to strike a billiards ball (called the cue ball) so as to pot other balls (called object balls) to the pockets of a pool table (See Figure~\ref{fig:example_intro} for example). The object balls are usually associated with numbers and/or colors. %
Billiards sport, which can be played in halls, hotels, or as often in the club houses of other organisations, becomes a popular recreational sport \cite{smith2007pickpocket, mathavan2010theoretical}.

\smallskip\noindent\textbf{Billiards Datasets.}
We propose to collect some data capturing billiards games so that various types of data analyses can be conducted on the data for discovering knowledge about the games and players. 
Specifically, we collected the dataset from 9-ball games (i.e., one popular type of billiards sports), where a player needs to hit the object ball with the smallest number among those remaining on the table for each strike and wins the frame if he/she pots the object ball with the number 9.
The dataset covers games of 94 international professional 9-ball tournaments for the last two decades, which were played by 227 professional players. 
In summary, the dataset includes 3,019 records for frames, 6,637 records for turns, and 2,082 records for strikes. More detailed summarization could be found in Table~\ref{tab:data_statics} and description in the supplementary materials~\cite{TR}.
%
% To our best knowledge, this is the first billiards dataset that will be made publicly available.

\begin{table*}[t]
\centering
\setlength{\tabcolsep}{6pt}
\caption{{\revision Summary of the collected billiards dataset (covering 94 tournaments and 227 players).}}
\vspace*{-3mm}
\label{tab:data_statics}
\begin{tabular}{cc|cc|cccc}
\hline
\multicolumn{2}{c|}{Data of Frames (3,019)}                                                                                                                           & \multicolumn{2}{c|}{Data of Turns (6,637)} & \multicolumn{4}{c}{Data of Strikes (2,082)}                                                                                  \\ \hline
1                  & break shot layout                                                                                                                                & 1          & player name                   & 1 & trajectory         & \multirow{2}{*}{6} & \multirow{2}{*}{\begin{tabular}[c]{@{}c@{}}stick top\\ position\end{tabular}}   \\
\multirow{4}{*}{2} & \multirow{4}{*}{\begin{tabular}[c]{@{}c@{}}three performance\\ indicators: clear\\ label, win label and\\ \# of potted balls label\end{tabular}} & 2          & \# of strikes                 & 2 & camera angle       &                    &                                                                                 \\
                  &                                                                                                                                                  & 3          & order of potted balls         & 3 & cushion            & \multirow{3}{*}{7} & \multirow{3}{*}{\begin{tabular}[c]{@{}c@{}}direction\\ on hitting\end{tabular}} \\
                  &                                                                                                                                                  & 4          & type of foul                  & 4 & intersection point &                    &                                                                                 \\
                  &                                                                                                                                                  & 5          & data of strikes               & 5 & cue ball position  &                    &                                                                                 \\ \hline
\end{tabular}
\end{table*}

{\extension{
% On top of the dataset, there are many potential analytic tasks that could be explored. 
In this paper, we focus on the billiards \emph{layouts} data involved in the collected dataset.
A billiards layout consists of the locations and identities of the balls (including the cue ball and the object balls) after a strike during a game. 
A layout example is shown in Figure~\ref{fig:example_intro}.
Intuitively, a billiards layout embeds rich information of a game that could help to inform many aspects of the game, e.g., what's the winning probability of a player, what's the strategy to strike the balls, and is there any historical layout that can be used as a reference point, etc.
We study three tasks on the billiards layouts data, including \emph{prediction}, \emph{generation} and \emph{similarity search}. The first two are covered in our prior study~\cite{zhang2022predicting} while the third one is newly covered in this paper.
% In our prior study~\cite{zhang2022predicting}, we mainly study two tasks, namely prediction and generation, on the billiards \emph{layouts} (an illustration is shown in Figure~\ref{fig:example_intro}). 
%
% A billiards layout consists of the locations and identities of the balls (including the cue ball and the object balls) after a strike during a game. 
% Intuitively, the layout of a break shot embeds rich information of the game.
%
% We have explored three tasks including prediction, generation and similarity search on the collected dataset, with the former two on the break shot layout data in our prior study~\cite{zhang2022predicting}, and the last on the trajectory data of strikes in the supplementary materials~\cite{TR}.
%
% In this paper, we expand this line of research to further study the \emph{similar billiards layout retrieval} (or simply similar layout retrieval) problem.
}}

\smallskip\noindent\textbf{Task 1: Break Shot Layout Prediction.}
% In our prior study~\cite{zhang2022predicting}, the first task we study is a \emph{prediction} task. 
% Intuitively, the layout of a break shot embeds rich information of the game, and 
Given a layout of a game, we predict three aspects of the game: (1) clear or not, it predicts whether the player who performs the break shot will pocket all balls; (2) win or not, it predicts whether the player who performs the break shot will win the game; (3) it predicts how many balls are consecutively potted after the break shot. 
The prediction task is non-trivial, which is mainly due to the characteristics of the data. In particular, the billiards layout data is unlike other existing datasets including a set of points~\cite{zaheer2017deep,skianis2020rep} or a sequence of locations such as trajectory data~\cite{zheng2010geolife}. In a billiards layout, the billiards balls including one cue (white) ball and several object (colored) balls are all located on a rectangular pool table with six pockets. The data embeds rich correlations, e.g., the spatial correlations between the cue ball and object balls, and the correlations between the cue ball (or object balls) and the six pockets, which is unique and important in billiards sports. We carefully extract those unique features that can reflect the spatial correlation of balls and/or pockets. Then, we use Convolutional Neural Network (CNN) to accomplish this task in a supervised learning manner, since CNN is inherently applicable for perceiving the correlations. The proposed model is called BLCNN. {\revision We note that the BLCNN is trained in a supervised learning manner for three distinct prediction tasks: clear, win, and potted balls, where the tasks of predicting (1) clear or not, and (2) win or not, correspond to two binary classification tasks.
}

\smallskip\noindent\textbf{Task 2: Break Shot Layout Generation.}
% In our prior study~\cite{zhang2022predicting}, the second task we study is a \emph{generation} task, which 
This task is to generate realistic yet high-quality (i.e., easy-to-clear) layouts of break shots. For a player, it is critical to gain knowledge about such layouts that (1) he/she can accomplish them (i.e., the layouts are realistic) and (2) he/she would pot many balls and even clear the table given them (i.e., the layouts are of high quality). We explore a data-driven solution to generate the billiards layouts via Generative Adversarial Networks (GANs)~\cite{goodfellow2014generative}, called BLGAN. More specifically, we treat each billiards layout as an ordered sequence, i.e., one cue ball followed by several object balls that remain on the table in ascending order of their numbers. 
Then, a generator is used to generate a sequence of discrete tokens, each representing a discretized location of the billiards table (e.g., a grid cell). 
%Besides, to generate \emph{realistic} layouts that can be accomplished by a player, we collect the real layouts with the clear labels from the dataset to guide the training of the generator. In this way, the layouts are generated with some real layouts but not random ones inputted as seeds, and thus we believe the generated layouts tend to be realistic ones. 
For the discriminator, we collect some real layouts that are predicted to be cleared with high probabilities. These real layouts and the generated layouts together are fed to the discriminator, which tries to discriminate between the real ones and generated ones. %The discriminator provides feedback to encourage the learning of the generator, i.e., to generate layouts that are difficult to be discriminated from those real layouts that are predicted to be cleared with high probabilities. Then, the generated layouts tend to be high-quality ones.

{\extension{\smallskip\noindent\textbf{Task 3: Similar Layout Retrieval.}
% \emph{similar layout retrieval} problem 
This task is to find those billiards layouts from a database that are similar to a billiards layout at hand called \emph{query layout}. 
% Here, a billiards layout consists of the locations and identities of the balls (including the cue ball and the object balls) after a strike during a game. 
%
% Intuitively, a billiards layout embeds rich information of a game that could help to inform many aspects of the game, e.g., what's the winning probability of a player, what's the strategy to strike the balls, and is there any historical layout that can be used as a reference point, etc.
%
% The similar layout retrieval problem can be 
This task can be involved in many scenarios. One scenario is to provide a real-time prediction of the winning result of a game given the latest layout, for which one intuitive solution is to make similar predictions for games with \emph{similar} layouts. This prediction functionality is highly desirable by Web broadcasting platforms of the billiards games such as %{\color{red}
ChalkySticks~\footnote{https://www.chalkysticks.com/tv}. Another scenario is that a billiards coach would like to analyze how well a player performs given different layouts after the first strike (called the break shot). In this scenario, the coach can retrieve those historical layouts that are \emph{similar} to a given one and their performance results (e.g., winning or not) and collect some statistics based on the retrieved layouts. This application can be implemented as a web-based searching engine, putting all layouts at a web server and providing a search functionality to coaches to look for similar layouts for training. A third scenario could be that a billiards fan is watching a billiards game during which, he/she finds some layouts interesting and would like to find if there are \emph{similar} layouts and/or games in the past - note that a game corresponds to a sequence of layouts and if so, how the players play given those layouts. %}

The core challenge of similar billiards layout retrieval is that of measuring the similarity between two billiards layouts, which is non-trivial due to some unique characteristics of billiards layouts.
% since billiards layouts contain some characteristics that make them unique among existing similarity measurements.
First, billiards layouts are order sensitive, e.g., in a billiards 9-ball game, the object balls are numbered from 1 to 9 and players should legally strike the object ball numbered the lowest in the layout first.
Therefore, existing similarity measures on sets (or pointsets), which include (1) pairwise point-matching such as Earth Mover's Distance (EMD)~\cite{rubner2000earth} and Hausdorff distance~\cite{huttenlocher1993comparing} and (2) recently emerging learning-based techniques such as DeepSets~\cite{zaheer2017deep} and RepSet~\cite{skianis2020rep}, all assume order invariance and thus they are not suitable for billiards layouts. 
% In particular, those measures based on (1) pairwise point-matching such as Earth Mover's Distance (EMD)~\cite{rubner2000earth} or Hausdorff distance~\cite{huttenlocher1993comparing}, and (2) recently emerging learning-based techniques such as DeepSets~\cite{zaheer2017deep} or RepSet~\cite{skianis2020rep} all compute the similarity to be invariant in the order of the points in a set.
% Indeed, some measures can treat the unordered sets as ordered sequences, and compute the similarity by identifying the best alignment via dynamic programmings, such as
{\ChengComment{We note that existing similarity measures on sequences and trajectories such as DTW~\cite{yi1998efficient}, Frechet~\cite{alt1995computing},
LCSS~\cite{vlachos2002discovering}, ERP~\cite{chen2004marriage}, EDR~\cite{chen2005robust}, and EDwP~\cite{ranu2015indexing}}}, though order sensitive, are not suitable for billiards layouts either. This is because they would align the balls from two layouts solely based on their locations and totally ignore other information such as the ball numbers.
{\ChengComment{For example, for two billiards layouts with totally different object balls such as Figure~\ref{fig:problem} (c) and (d), it is counter-intuitive to align these different coloured balls to compute the similarity.}}
Second, billiards layout data is with some domain-specific features, e.g., the types and locations of pockets, the factors deciding the difficulty of striking an object ball to a pocket such as the angle formed by the cue ball, a target object ball and a pocket, etc. These features are important for measuring similarities among billiards layouts and should be systematically captured in the similarities.
% e.g., the spatial correlation of balls and/or pockets in the layout, and it remains unexplored to capture them in a similarity measure.
% Thus, a good similarity measure needs to take these features into consideration; however, existing measures fail to capture the similarity in this aspect.

{\nnComment{Another challenge is efficiency. A typical application scenario of similar billiards layout retrieval is to compute the similarity between a query billiards layout and many billiards layouts in the database. However, the existing matching-based methods all involve the procedure of finding an optimal alignment~\cite{yi1998efficient,alt1995computing}, which has at least a quadratic time complexity in terms of the number of balls and would impose a big challenge when performing similar layout retrieval on a huge database}} {\ChengComment{with many layouts and for a big number of queries since the operation of computing the similarity between two layouts needs to be conducted for many pairs of data layouts and query layouts.}}

In this paper, we propose a novel solution called BL2Vec, which overcomes the aforementioned challenges of similar billiards layout retrieval. Regarding the effectiveness, we leverage a deep metric learning model, namely \emph{triplet network}~\cite{kaya2019deep}, to learn a ranking-based metric for measuring the similarities among billiards layouts. More specifically, to capture the ordered nature of the billiards layout data, we treat billiards layouts as ordered sequences, i.e., one cue ball followed by several object balls in the ascending order of their numbers and apply padding when there are missing balls (since some may have been pocketed). To capture those unique characteristics in billiards layouts, we propose to extract features from the balls and pockets and further embed them. Then, we use Convolutional Neural Network (CNN)~\cite{lecun1998gradient} to generate a vector from the embedded features as the representation of the layout since CNN can inherently capture the local relationships between the balls. In addition, we propose a method to generate training data with hard negative mining~\cite{henriques2013beyond} so that BL2Vec is robust to a small shift in a billiards layout and learns the similarity in a self-supervised manner. 
% BL2Vec significantly differs from the existing measures that capture the similarity with some hand-crafted rules/objectives such as pairwise point-matching.
%
{\nnComment{Regarding the efficiency, our method based on BL2Vec computes the similarity between two layouts as the Euclidean distance between their representations (i.e., vectors), which runs in linear time wrt the number of balls in a game.}} 
% existing methods that involve pairwise point-matching have time complexities at least $O(n^2)$, where $n$ denotes the number of balls in the layout. 
When performing a similar layout retrieval on a database, we first learn the representations of billiards layouts once offline. Then, for a given query layout, we learn its representation and perform a \emph{nearest neighbour} query in the representation space (i.e., the vector space), which could be efficiently accomplished with existing methods such as the one in~\cite{bentley1975multidimensional}.
% locality-sensitive hashing~\cite{indyk1998approximate}.
% After embedding the billiards layouts into vectors offline, the similarity between two layouts is captured by the Euclidean distance, which could be computed in constant time and therefore the complexity of searching a billiards layout is linear with the size of the database. The linear complexity makes BL2Vec suitable for large datasets.

% Given that there are no billiards layout datasets that are publicly available, we further collect a billiards layout dataset.
% % , which will be made publicly available.
% %
% %layout
% % A billiards layout corresponds to a set of billiards balls including a cue ball and several object balls, which are located on a billiards table.
% % {\ChengComment{Figure~\ref{fig:example_intro} shows an example of a 9-ball billiards layout.}}
% %
% In particular, the dataset consists of 3,019 billiards (9-ball) layouts after break shots~\footnote{https://en.wikipedia.org/wiki/Glossary\_of\_cue\_sports\_terms}, and covers international professional billiard tournaments including China Open, All Japan and 9-Ball World Championships~\footnote{https://wpapool.com/ranking/} in the last two decades.
% The dataset was collected with a free sports analysis software Kinovea~\footnote{http://www.kinovea.org} and contains rich remarks (e.g., golden break or foul) and labels including \emph{the number of potted balls} after the break shot, \emph{clear or not} and \emph{win or not} based on the layout. 
% This dataset would provide opportunities of conducting various billiards-related analytics tasks.

\smallskip\noindent\textbf{Summary of Contributions.}
% In summary, the paper extends our prior work~\cite{zhang2022predicting}, which studies the prediction and generation tasks (presented in Section~\ref{task:blcnn} and Section~\ref{task:blgan}), with the following new contributions.
The contributions of this paper are summarized as follows, with the first two contributions covered by our prior study~\cite{zhang2022predicting} and the last one newly covered in this extended version.

\begin{enumerate}[leftmargin=*]
    \item We contribute 
% the first 
a billiards sports dataset that includes break shot data (for frames), strike statistics data (for turns) and trajectory data (for strikes).
% which contains 3,019 layouts {\Comment and 2,082 strikes} in recent international professional billiard games. 
The layouts data is a new data type, which is related to yet different from quite a few existing data types including point sets, trajectories, sequences, etc. Specifically, it consists of a point set, is order-sensitive, and is associated with some contexts (e.g., the pockets). 
% To the best of the authors' knowledge, this would be the first dataset of its kind. 
In addition, the dataset has rich contents and label information to support various machine learning tasks. 
Our dataset and codes are publicly accessible via the link~\footnote{\url{https://drive.google.com/drive/folders/1NBqonYLr_cParMMn4xSeE0KTJNhjeYuG?usp=sharing}} and provide an
opportunity for research communities such as machine learning, data mining, computational geometry, computer vision and sports science, to make a significant impact. (Section~\ref{sec:dataset})

    \item 
    We study the prediction and generation tasks on break shot layouts data, which
% The first is to predict some information (e.g., clear or not) that is associated with the game. The second is to generate some layouts which have aforementioned characteristics when performing a break shot for players. 
help better understand the sports and serve different users including coaches, players and fans.
We develop the BLCNN model and the BLGAN model for  the prediction and generation tasks, respectively.
We conduct extensive experiments on the collected 9-ball data, which demonstrate that (1) BLCNN achieves superior classification accuracy over baseline methods; (2) BLGAN has good performance on generating the layouts of high quality (i.e., easily to be cleared) and reality. (Section~\ref{task:blcnn}, Section~\ref{task:blgan}, Section~\ref{subsec:prediction}, and Section~\ref{subsec:generation})

    % \item Extensive empirical evaluations are performed on the collected dataset for the three tasks, demonstrating superior performance over baseline methods.
    
    \item We propose a new problem of searching similar billiards layout for a given query layout from a database of layouts. 
    % To the best of our knowledge, this problem has not been examined before.
    % (Section~\ref{sec:problem}).
    %
    % \item 
    We then develop a deep learning-based similarity measure called BL2Vec for billiards layouts. 
    % To the best of our knowledge, BL2Vec is the first deep learning-based solution for learning similarity between billiards layouts. 
    The similarity not only considers the unique characteristics in a billiards layout but also runs fast in linear time.
    We conduct experiments on BL2Vec with the collected datasets. The effectiveness experiments show that BL2Vec can outperform the best baseline by at least 27\% for similarity search. The efficiency results show that BL2Vec runs up to $420 \times$ faster than the existing methods that involve pairwise point-matching for measuring the similarity between two layouts. We further conduct a user study, which validates BL2Vec with expert knowledge. (Section~\ref{sec:method} and Section~\ref{subsec:similar})
    % Our collected dataset and codes could be accessed via this link~\footnote{https://www.dropbox.com/sh/644aceaolfv4rky/AAAKhpY1yzrbq9-uo4Df3N0Oa?dl=0} (Section~\ref{sec:experiment}).
    
    %\item {\newadd{Our collected dataset and codes could be accessed via this link~\footnote{\url{https://www.dropbox.com/sh/644aceaolfv4rky/AAAKhpY1yzrbq9-uo4Df3N0Oa?dl=0}}.}}
\end{enumerate}
%via the link~\footnote{https://www.dropbox.com/sh/uzszs259ggmyntz/AAAM0zh\_TKmkfdJNaEdCD3cRa?dl=0}
%We will publish the dataset once the paper is accepted, which 
% will be made publicly available.
}}
%The rest of the paper is organized as follows. We review the literature in Section~\ref{sec:related} and give the problem definition in Section~\ref{sec:problem}. We present our BL2Vec model in Section~\ref{sec:method}, {\ChengComment{the dataset in Section~\ref{sec:dataset}}}, and report our experimental results in Section~\ref{sec:experiment}. We finally conclude the paper and discuss some future work in section~\ref{sec:conclusion}.

%% file: related.tex
\section{RELATED WORK}
\label{sec:related}

\textbf{(1) Sports Data Analytics.}
Billiards corresponds to one type of popular sports playing with a cue stick on a pool table. The traditional research in this area addresses the task of training the robotic players such as Deep Green~\cite{lin2004grey} to execute good shots on a physical table and thus win a computer billiards game~\cite{archibald2009analysis, smith2006running}. Recently, Pan et al.~\cite{pan2021can} study the importance of a break shot in predicting the 9-ball game outcomes. Nowadays, it becomes a common practice to deploy some devices such as GPS sensors and cameras to capture some data of sports games, and the collected data proliferates various sports analytic tasks. Specifically, Wang et al.~\cite{wang2019effective,wang2021similar} study the similar soccer game retrieval, which can recommend similar games to sports fans in some sports recommender systems such as ESPN. Decroos et al.~\cite{decroos2018automatic} collect event-stream data from professional soccer matches, and explore the problem of tactics detection, which helps players improve their tactics when they prepare for an upcoming match. Aoki et al.~\cite{aoki2017luck} collect four different sports data including basketball, soccer, volleyball and handball, and analyze the difficulty of predicting the outcome of sports events. Overall, these tasks involve various sports such as basketball, soccer, volleyball and handball.
In our prior study~\cite{zhang2022predicting}, we study the billiards layout prediction and generation tasks, and in this paper, we extend this line of research by studying one additional task of similar billiards layout retrieval.

\smallskip
\noindent\textbf{(2) Sequence Data Prediction.}
The collected billiards data contains the layouts of break shots in real-world 9-ball games, and it can be treated as data of sequences of one cue ball followed by several object balls in the ascending order of their numbers. We review the existing works of sequence data prediction as follows. The common sequence data includes time series data and trajectory data. Time series prediction~\cite{weigend2018time,frank2001time} is a related but different task. It aims to train models to fit historical data and use them to predict future observations of a sequence. Existing methods for time series prediction can be grouped into two categories: classical models such as SVM~\cite{sapankevych2009time} and recent deep learning-based models such as RNN variants~\cite{li2017diffusion} or Transformer variants~\cite{zhou2020informer}. On the other hand, trajectory data corresponds to a sequence of positions to capture the traces of moving objects. The problem of trajectory prediction can be viewed as a sequence generation task of predicting the future trajectories of moving objects based on their past positions, and it is widely used to avoid obstacles for pedestrians~\cite{alahi2016social} and vehicles~\cite{ju2020interaction}. 
Our task differs from these studies mainly in two aspects. First, our billiards data carries the unique characteristics in billiards sports, e.g., it captures the locations of balls on a pool table. Second, our task is to predict some associated information with the layout instead of forecasting a future observation value in a sequence.
%such as time series or trajectory

\smallskip
\noindent\textbf{(3) Sequence Generative Models.}
We review existing generative models~\cite{yu2017seqgan,che2017maximum,fedus2018maskgan,lin2017adversarial,nie2018relgan} related to the task of break shot layout generation as follows. The majority of deep generative models are used to generate text. For example, SeqGAN~\cite{yu2017seqgan} is a typical adversarial text generation model, which builds an unbiased estimator based on the REINFORCE algorithm~\cite{williams1992simple} for the generator, and applies the roll-out policy to obtain the reward from the discriminator. In addition, to deal with the differentiation difficulty when applying GAN to generate sequences of discrete elements, GSGAN~\cite{kusner2016gans} is proposed to use Gumbel-softmax output distributions to train GAN on discrete sequences. Guo et al.~\cite{guo2018long} propose the LeakGAN, which introduces a hierarchical reinforcement learning framework for the generator, and improves the performance of long text generation.
Overall, all of these works consider textual data, e.g., generating some sentences. Our work differs from them mainly in the data, i.e., billiards layout data corresponds to a set of billiards balls that are located on a pool table. Besides, the data is associated with several unique characteristics in billiards sports, e.g., the spatial correlations between the cue (white) ball and object (colored) balls.

{\extension{
\smallskip
\noindent\textbf{(4) Measuring Pointsets Similarity.}
% From the technical perspective, our problem is related to the problem of measuring the similarity between two pointsets. Although the pointsets similarity has been studied extensively, billiards layouts have several characteristics, e.g., a layout is composed of one cue (white) ball and some object (colored) balls on a billiards table with six pockets, which make them unique and differ from existing studies.
%
A billiards layout involves a set containing the locations of the balls, i.e., a pointset. Therefore, it is possible to adopt an existing pointsets similarity for billiards layouts. Existing studies of pointset similarity fall in two categories. 
The first category focuses on finding an optimal alignment of the points between two pointsets~\cite{rubner2000earth, huttenlocher1993comparing}. Earth Mover's Distance (EMD)~\cite{rubner2000earth} and Hausdorff~\cite{huttenlocher1993comparing} are two typical measures in this category. EMD measures the least amount of work needed to transform one set to another. Hausdorff computes the largest distance from a point in one set to the nearest point in the other set. 
% In addition, DTW~\cite{yi1998efficient} and Frechet~\cite{alt1995computing} can also be adapted to measure pointsets similarity, which treats unordered sets as ordered sequences and apply pairwise point-matching methods to find an alignment with a minimal matching cost. %
In general, the time complexity of computing an alignment between two sets is quadratic wrt the number of their points. 
% {\Comment We also consider a straightforward similarity measurement for two billiards layouts, called Peer Matching (PM). It matches the balls peer-to-peer based on their numbers, e.g., cue ball in one layout is matched to the cue ball in another layout, ball 1 in one layout is matched to ball 1 in another layout, and so on. For those mismatched balls with different numbers, we align them in the ascending order of their numbers and match them accordingly.}

The second category aims at designing neural networks for pointsets, which receives growing research attention in recent years~\cite{qi2017pointnet, zaheer2017deep, qi2017pointnet++, skianis2020rep, Arsomngern2021self}. The idea is to learn representations of pointsets in the form of vectors and measure the similarities among pointsets based on the vectors (e.g., the Euclidean distances among the vectors). PointNet~\cite{qi2017pointnet} and DeepSets~\cite{zaheer2017deep} are two classic methods in this category. They transform a pointset into a vector, which guarantees some permutation invariance of points in the set (i.e., a set has unordered nature and therefore it is invariant to permutations of its points).
Specifically, PointNet uses max-pooling layers to aggregate information across the vectors of the points, while DeepSets adds up the vectors. Then, the representation of the set is fed to a standard neural network architecture (e.g., some fully-connected layers with nonlinearities) to produce the output. Further, PointNet++~\cite{qi2017pointnet++} is proposed to capture the local structures of a set by applying PointNet recursively and achieves a more efficient and robust set representation. RepSet~\cite{skianis2020rep} handles the set representation by computing the correspondences between an input set and some generated hidden sets using a bipartite matching algorithm. More recently,  WSSET~\cite{Arsomngern2021self} employs deep metric learning based on EMD to learn set representations and measure the similarity between two pointsets as the Euclidean distance between their corresponding vectors and achieve the state-of-the-art results for measuring pointsets similarity in a self-supervised setting.
%Arsomngern et al.
Nevertheless, these studies target general pointsets, which are not order sensitive and cannot capture unique characteristics of billiards layouts.
% The main difference between our study and these studies is that our problem is on billiards layouts, which are order sensitive for computing the similarity, and with some unique features of billiards.

{\ChengComment{
\smallskip
\noindent\textbf{(5) Measuring Sequences or Structural Similarity.}
We can treat billiards layouts as ordered sequences, i.e., one cue ball followed by several object balls in the ascending order of their numbers. In this case, existing similarity measurements for sequences/trajectories can be used for measuring the similarity between two billiards layouts. Some examples of similarity measurements for sequences/trajectories include DTW~\cite{yi1998efficient}, Frechet~\cite{alt1995computing},
LCSS~\cite{vlachos2002discovering}, ERP~\cite{chen2004marriage}, EDR~\cite{chen2005robust}, and EDwP~\cite{ranu2015indexing}. A more detailed survey on trajectory similarity could be found in~\cite{su2020survey,wang2021survey}. As explained in Section~\ref{sec:introduction}, these measurements are not suitable for billiards layouts since they would align the balls from two layouts solely based on their locations and totally ignore other information such as the ball numbers. In addition, these measurements usually incur high time costs, e.g., each of them has at least a quadratic time complexity in terms of the number of balls.}}
{\nnComment{In addition, some studies take layouts as images for learning the similarity. For example, Patil et al.~\cite{patil2021layoutgmn} measures the structural similarity between layouts such as floorplans, and Manandhar et al.~\cite{manandhar2020learning} measures the similarity between layouts such as Web UI designs. Both of them have specific designs for their data types, which are largely different from our billiards layouts, where we take layouts as a set of locations.
}}

\smallskip
\noindent\textbf{(6) Deep Metric Learning.}
Deep metric learning is proposed to learn a distance function for measuring the similarity between two objects via neural networks. Bromley et al. ~\cite{bromley1993signature} pioneer the classic Siamese network in deep metric learning and employ it for signature verification. Then, the triplet network architecture is proposed for image retrieval, which learns a ranking-based metric and shows superior performance over Siamese network~\cite{kaya2019deep}. %In recent years, the triplet architecture is further used for measuring the similarity of geo-spatial data such as trajectory~\cite{yao2019computing,zhang2020trajectory}.
%urban region~\cite{liu2018efficient}
In this paper, we propose to learn representations of billiards layouts via a deep metric learning method called BL2Vec, which is quite different from those in previous studies. Specifically, our BL2Vec is designed for the spatial layout of billiards based on a triplet architecture of CNNs. The reason why we use the CNN is that it has the inherent ability to capture local spatial correlations of balls in a billiards layout. 
% Moreover, to apply the triplet network on billiards similarity, we design a method to generate training data with hard negative mining~\cite{henriques2013beyond}. 
}}

\smallskip
{\revision
\noindent\textbf{(7) Comparing the Billiards Dataset with Others.} We provide a comparison between our proposed billiards dataset and existing datasets from various sports, such as (1) soccer~\cite{wang2019effective, wang2021similar, decroos2018automatic}, (2) basketball~\cite{sha2016chalkboarding, di2018large}, (3) volleyball~\cite{cui2023sportsmot}, and (4) ice hockey~\cite{pileggi2012snapshot}. For (1), the soccer dataset~\cite{wang2019effective, wang2021similar} comprises 7,500 sequences, each containing tracking data for 11 defense players, 11 attacking players, and a ball. Collected from a professional soccer league, it spans approximately 45 games with over 30 million data points with 10 FPS. The soccer event data~\cite{decroos2018automatic} is sourced from the 2015/2016 English Premier League season, manually annotated during video watching, detailing timestamp, location ($x, y$ coordinates), type (e.g., foul, pass), and involved players. The dataset includes 652,907 events of 39 types, averaging 1,718 events per match, with key types being ``pass'' (368,426), ``out'' (48,046), and ``ball recovery'' (41,448). For (2), the basketball dataset~\cite{wang2019effective, wang2021similar} is produced using STATS' SportVU tracking system, employing 6 cameras to track player locations (including referees and the ball) with 25 FPS. This dataset includes detailed log information, covering actions such as pass, shot, dribble, and turnover. In total, it comprises tracking data from over 1,100 games. For (3), the volleyball dataset is part of SportsMOT~\cite{cui2023sportsmot}. This dataset gathers videos from professional games, including NCAA, Premier League, and the Olympics. Specifically designed for spatio-temporal action localization, it includes typical players’ formations and motion patterns. The dataset comprises a total of 240 video sequences, each in 720P resolution and at a frame rate of 25 FPS. For (4), the ice hockey dataset~\cite{pileggi2012snapshot} is sourced from the National Hockey League (NHL). It comprises 81,158 instances, each represents a shot recorded by NHL during the 2010-2011 season. The collected data for each shot includes details such as the shooter’s name, team, location ($x, y$ coordinates), shot type (wristshot, backhand, tip-in), and associated event (hit, penalty, faceoff), etc.

In comparison to existing datasets, our billiards dataset stands out in three aspects. First, the data is acquired through a tracking system that captures layouts and trajectories, which embed both spatial and temporal features of a billiard game. Consequently, it can be utilized to support a range of quantitative analytics tasks, e.g., leveraging features such as the $x, y$ coordinates. Second, we offer comprehensive label information (such as clear, win, and the number of potted balls), enabling diverse application scenarios for broadcasting platforms and coaching purposes. Third, we have made our dataset publicly available without copyright restrictions, offering an easily accessible opportunity for both research and commercial use. Furthermore, we outline the detailed data collection process in~\cite{TR}. It employs a cost-effective approach, utilizing the free software Kinovea to collect data from publicly available billiards broadcast videos. Given the widespread availability of billiards broadcasts on the internet, this approach holds the potential to enhance dataset collection practices in other sports.

}

%% file: dataset.tex
\section{Billiards Background and Dataset}
\label{sec:dataset}

A billiards game refers to a game playing with a cue stick on a rectangular billiards table, which has six pockets at the four corners and in the middle of each long side. The game involves a cue ball (white) and some object balls (colored), where different billiards games have different numbers of object balls such as 9 object balls in 9-ball game. %, 15 object balls in 8-ball game and 21 object balls in Snooker game.
The game begins with one player's break shot. If a ball is legally pocketed after the break shot, the player keeps his turn; otherwise, the opponent gets a chance to shoot.
%
% The legally pocketed are constrained by the order of balls in the layout,
% The game begins with one player's break shot. 
%Players should obey some rules when playing the game, e.g, a player must strike the cue ball to pocket nine colored balls in an ascending order of their numbers in the 9-ball game, in the way the player can retain his turn; otherwise, the opponent gets a chance to shoot.
% The legally pocketed determines some rules in a game, 
There are usually some rules to follow when striking a ball,
e.g, in a 9-ball game, a player must use the cue ball (white) to strike the object balls (colored) in an ascending order of their numbers.
% %
%The player can retain his turn as long as he calls an object ball legally and make the called ball sink into a pocket.
%
%In case of an illegal shot (i.e., foul), such as sinking the cue ball into a pocket, it offers the opponent "ball in hand", which means the cue ball can be placed at any location on the table.

Our real-world billiards data is extracted from the videos of professional billiards games published on YouTube in the most recent two decades using the software Kinovea (https://www.kinovea.org/). The data collection process by the software Kinovea consists of four steps (i.e., uploading images/videos, adding grids and markers, exporting coordinates and collecting information), which are illustrated in detail in the supplementary materials~\cite{TR}. The dataset covers 227 players and 94 international professional 9-ball tournaments. We collect the billiard dataset for frames, turns and strikes. Table~\ref{tab:data_statics} summarizes the collected dataset (with details included in \cite{TR}). 

%% file: tasks.tex
\section{Layout Prediction with BLCNN}
\label{task:blcnn}
In this section, we introduce how to embed each ball in billiards layouts into a real vector. Then, the vectors are concatenated in an ascending order of ball numbers that should be potted as an embedding of features, which is fed into a classifier based on the architecture of CNN to predict the results. The effectiveness of
prediction depends on the quality of extracted features. The proposed model for billiards layout prediction is called BLCNN.
%and note that our method can be generalized to any other billiards game, including 9-ball, 8-ball and Snooker. 

\begin{figure}%{r}{6.1cm}
\scriptsize{
\centering
\vspace*{-2mm}
\includegraphics[width=0.6\linewidth]{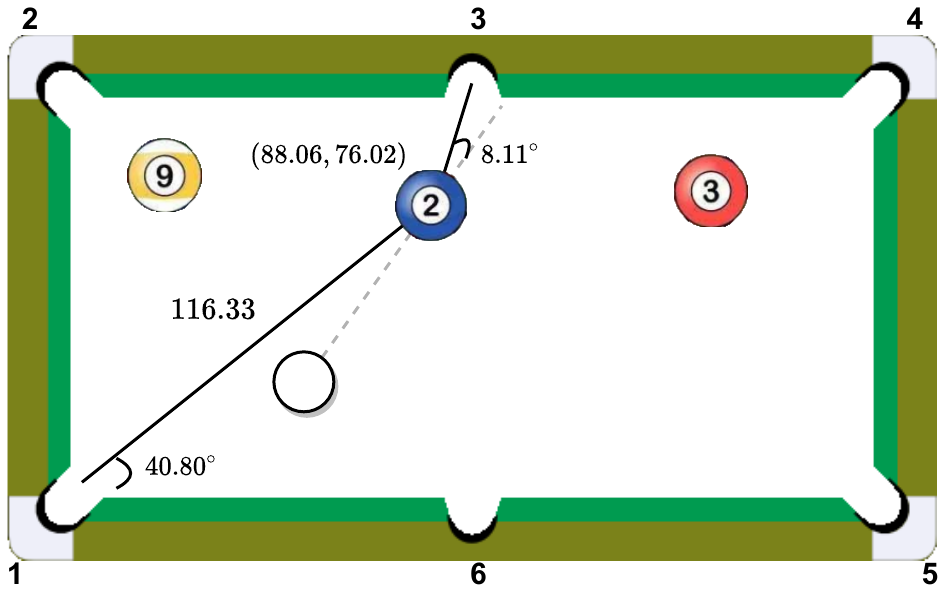}
\caption{Illustration of the BS, BP and BB features of ball 2 (blue) in a layout.}
\label{fig:feature}
\vspace*{-4mm}}
\end{figure}

\smallskip\noindent \textbf{Feature Extraction.} We capture the correlations in a billiard layout and consider the following three kinds of features: the location information of billiards balls on the table, called Ball-Self (BS);
the correlation between balls and pockets, called Ball-Pocket (BP); the correlation between balls, called Ball-Ball (BB). Figure \ref{fig:feature} illustrates these features from a 9-ball billiards layout. 
% The features are unique in billiards sports and differ from existing feature engineering, which is proposed for the data including a set of points~\cite{zaheer2017deep,skianis2020rep} or a sequence of locations~\cite{huang2019grab,zheng2010geolife}.

% \begin{figure*}
% \centering
%   \begin{minipage}{0.6\linewidth}
% 	\includegraphics[width=\linewidth]{figures/feature_ext}
%   \end{minipage}
% \caption{Feature extraction.}
% \label{fig:feature}
% \end{figure*}

\smallskip\noindent (1) Ball-Self (BS). We map the play field of the billiards table into a 200$\times$ 100 coordinate system and set the bottom left corner as the origin. Thus, the range of the x-axis of the coordinate system is [0,200],  the range of the y-axis of the coordinate system is [0,100]. For each ball $b_i$ in the billiards layout, it is related to $(x,y)$ in the coordinate system, which is regarded as its position feature on the pool table. For example, in Figure \ref{fig:feature}, the location of ball 2 (blue) is $(88.06, 76.02)$.

\smallskip\noindent (2) Ball-Pocket (BP). To report features between each ball and each pocket, we set the pocket at the bottom left corner as the start pocket and mark the pockets clockwise from 1 to 6, which is shown in Figure \ref{fig:feature}. For each ball $b_i$, we use four features including cushion angle, distance to pocket, an indicator of occlusion and pocket index to describe it wrt each pocket $p_j (1 \le j \le 6)$ on the table. Thus, there are in total $4 \times 6 = 24 $ features for each ball $b_i$.
(\romannumeral1) Cushion angle. Cushion angle is used to describe the minimum angle between the line from a ball $b_i$ to a pocket $p_j$ (i.e., denoted by $\overline{b_i p_j}$) and its two cushion edges.
(\romannumeral2) Distance to pocket. Distance to pocket is the distance from a ball $b_i$ to a pocket $p_j$ (i.e., $\left\| b_i - p_j \right\|_2$).
(\romannumeral3) Indicator of occlusion. An indicator of occlusion is used to report whether there is a ball on the line from a ball $b_i$ to a pocket $p_j$. If so, the indicator of occlusion is 1; 0 otherwise.
(\romannumeral4) Pocket index. The pocket index is used to distinguish different called pockets, which is from 1 to 6.
Take ball 2 as an example in Figure \ref{fig:feature}, the cushion angle is $40.80^\circ$ between the line $\overline{b_2 p_1}$ and cushion 1-6; the distance from ball 2 to pocket 1 is 116.33; the indicator of occlusion is 1 since cue ball appears in the path from ball 2 to pocket 1; the pocket index is 1 for the above three features. In brief, the feature (\romannumeral1) and (\romannumeral2) capture the feasibility to sink the ball 2 into a called pocket; the feature (\romannumeral3) captures whether there will be collisions or bounces on the path of ball 2 to a called pocket; the feature (\romannumeral4) is to distinguish different called pockets varying from 1 to 6.

\smallskip\noindent (3) Ball-Ball (BB). We consider two features to capture the correlation of balls, including shot angle and pocket index.
(\romannumeral1) shot angle. Shot angle describes the minimum angle between the line of two balls with consecutive numbers on the table (i.e., denoted by $\overline{b_i b_{i+1}}$) and the line from ball $b_{i+1}$ to each pocket $p_j$. In Figure \ref{fig:feature}, the shot angle is $8.11^\circ$ since the minimum angle corresponds to the angle of a line from the cue ball to ball 2 (i.e., $\overline{b_1 b_2}$) and the line from ball 2 to pocket 3.
(\romannumeral2) Pocket index. The pocket index is selected to distinguish the index of the most likely called pocket when performing a strike. In the case of ball 2 in Figure \ref{fig:feature}, the pocket index is 3 since pocket 3 is more likely to be called with the minimum shot angle $8.11^\circ$.

\smallskip\noindent {\revision \textbf{Discussion on Feature Selection and Engineering.} We discuss the feature selection for BS, BP, and BB below. In examining a billiard layout, we recognize two crucial elements: balls and pockets. The integration of these two elements allows us to capture essential features that describe a layout, namely BS, BP, and BB. In particular, a layout corresponds to a collection of billiards balls, and the BS feature captures fundamental locations of the balls within this layout. The BP feature represents the relationships between balls and pockets, indicating, for instance, the difficulty of potting a ball with an occlusion indicator. Further, the BB feature captures the interactions between balls. In a billiards game, a player uses the cue ball to strike the object ball, leading to the transfer of the shot angle between two consecutive object balls to calculate the angle between the cue ball and one of the object balls. 

To engineer the features, we utilize the free software Kinovea, as detailed in~\cite{TR}. The play field of a billiard table is calibrated into a $200*100$ coordinate system along the x-axis and y-axis, respectively. We incorporate a ``perspective grid'' onto the table and utilize markers provided by Kinovea to record the coordinates of the balls. Additionally, the coordinates of six pockets are known in the given coordinate system, encompassing four corner positions and two middle positions along the x-axis. Based on these coordinates, we calculate the BS, BP, and BB features according to the previously introduced definitions.
}

\begin{figure}
\centering
  \begin{minipage}{\linewidth}
	\includegraphics[width=\linewidth]{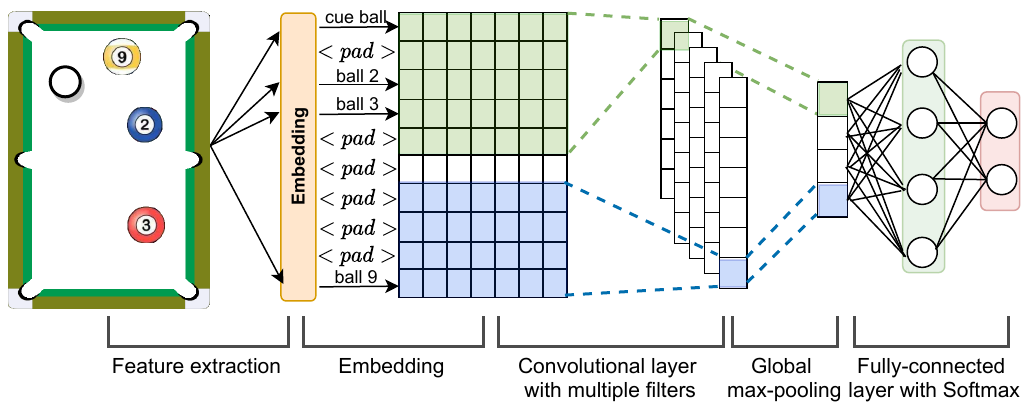}
  \end{minipage}
\caption{The model architecture of BLCNN. The colours (i.e., green and blue) denote the convolutional filters with different sizes to obtain feature maps.
%Ball 2, ball 3 and ball 9 are left on the table after break shot. For each ball, the BS, BB and BP features are extracted and embedded into real vectors. Further, the embedded features are concatenated and fed to a convolutional layer with multiple filters, where the colours (i.e., green and blue) mean the convolutional filters with different sizes to obtain feature maps, and a global max-pooling layer is then employed to capture the representation of the whole layout, and a fully-connected layer is employed to produce the output.
}
\label{fig:blcnn}
\vspace{-2mm}
\end{figure}

\smallskip\noindent \textbf{CNN-based Classifier.}
BLCNN for the prediction task is a CNN-based architecture, which is illustrated in Figure~\ref{fig:blcnn}. We consider three labels in the task including clear or not, win or nor and the number of potted balls after the break shot based on a given billiards layout.
% For the generation task, the GAN-based architecture is illustrated in Figure~\ref{fig:BLGAN}. We generate the billiards layouts by exploring existing layouts which are cleared with a data-driven solution, and it could help players to explore more layouts as the targets of a break shot in the game.

%(a figure depicting the architecture could be found in the supplementary materials for illustration).

%as shown in Figure~\ref{fig:blcnn}. It includes four basic layers: (1) embedding layer, (2) convolutional layer, (3) global max-pooling layer and (4) fully-connected layer.
%
In particular, for each billiards ball, we transform its extracted features (i.e., BS, BB and BP) into real vectors via an embedding layer, then the vectors are concatenated into a long vector as the embedding of the ball. To achieve the embedding, we granulate these features into the tokens. Next, the ball embeddings are further concatenated in the ascending order of their numbers and we apply padding when there are missing balls on the table. Thus, we get an embedding for the features of billiards balls in a matrix form, called embedded features.
Note that we embed these features via an embedding layer instead of using the feature values directly because the feature values restrict the embeddings in a low dimensional space, which makes it difficult for the model to be further optimized.
Further, the embedded features are fed to a convolutional layer with multiple filters of varying sizes to obtain feature maps, and a global max-pooling layer is then employed to capture the most important feature for each feature map, and those important features are further aggregated into a vector, which corresponds to a representation of the billiards layout.
The layout representation is then fed to a standard neural network architecture, i.e., the fully-connected layer, with softmax nonlinearity to produce the output. We train the BLCNN in a supervised learning manner, with the cross-entropy loss for the prediction of three types of labels (i.e., clear, win and potted balls).

{\revision The novelty of the layout prediction task lies in the comprehensive methodology adopted by the proposed BLCNN model to effectively capture and exploit correlations inherent in billiards layouts. This involves three key aspects. (1) Real Vector Embedding of Billiards Balls: the BLCNN model employs a unique approach by embedding each billiards ball into a real vector, specifically utilizing features such as BS, BP, and BB. This selection of features reflects a systematic understanding of billiard layouts, recognizing the essential elements of balls and pockets. By integrating information about both balls and pockets, the model gains the ability to grasp fundamental spatial locations, relationships, and interactions within the layout. (2) Cost-effective Feature Engineering: our feature engineering is cost-effective, utilizing the freely available software Kinovea. This not only allows for the efficient extraction of features from widely accessible billiards broadcasts on the internet but also holds the potential to contribute to improved feature engineering practices in other research studies. (3) Innovative Billiard Format and CNN-based classifier: we introduce a novel format where ball numbers are arranged in ascending order to represent the billiards layout. This layout representation is complemented by the proposed BLCNN model featuring a CNN-based classifier. This design is tailored specifically for predicting outcomes in the context of billiards, and leveraging the hierarchical structure inherent in the ascending order of ball numbers.
}

\section{Layout Generation with BLGAN}
\label{task:blgan}
We first explain that the layout generation task is useful by referring to some existing literature of billiards sports~\cite{smith2006running,archibald2009analysis}, which aim to find high-quality break shots so as to achieve desirable layouts. For example, in~\cite{smith2006running}, it models a layout (i.e., the positions of balls in a billiards table) as a state and how to execute a shot as an action. Each action is controlled by five continuous parameters including the direction of hitting, etc. It aims to search for a shot that would generate a desirable layout. It is also worth mentioning that our dataset includes sufficient information on the the strike level (including the break shots), such as the hitting position, stick top position, direction on hitting, which we believe will facilitate more in-depth research along this line.

\smallskip\noindent\textbf{Challenges.}
Generative Adversarial Network (GAN)~\cite{goodfellow2014generative} is a promising framework to generate new data based on existing data. 
% It consists of a discriminative model and a generative model. {\Comment We design a GAN-based architecture (called BLGAN) to generate the synthetic break shot layouts which are \emph{feasible} and of \emph{high quality} to be cleared in a billiards game. Specifically, we organize billiards layouts as ordered sequences, i.e., one cue ball followed by object balls that remain on the table in ascending order of their numbers. In this way, BLGAN to generate billiards layouts is like GAN to generate a sequence of discrete tokens~\cite{yu2017seqgan,kusner2016gans}. 
A natural idea is to treat a billiards layout as an ordered sequence, i.e., one cue ball followed by object balls that remain on the table in ascending order of their numbers, and apply GAN to generate sequences of balls (and their locations).
Nevertheless, it has the following challenges.
First, when the physical space of the billiards balls (i.e., the billiards table) is discretized (e.g., as a 15 $\times$ 15 grid), applying GAN to generate the balls and their locations (as discrete tokens) would suffer from a classical issue of indifferentiability~\cite{kusner2016gans}, since the samples from a distribution of discrete tokens are not differentiable with respect to the distribution parameters.
Second, in a break shot layout, some balls may have been potted already, and in this case, the player who performs the break shot can keep his turn and thus  have the chance to clear the table given the layout. Yet a straightforward GAN does not provide a mechanism of deciding some ball/balls to be omitted from the generated layouts, i,e., meaning some of the balls are potted already.
% It should be explored in the generation.
% Third, recall that our target is to generate layouts that are \emph{realistic} and \emph{high-quality} instead of some random layouts. Therefore, how to serve the target by assessing the quality of the generated layouts (i.e., whether this layout can really be cleared) should also be explored in this task.

\begin{figure}
%\vspace{-4mm}
\small{
\centering
  \begin{minipage}{\linewidth}
	\includegraphics[width=\linewidth]{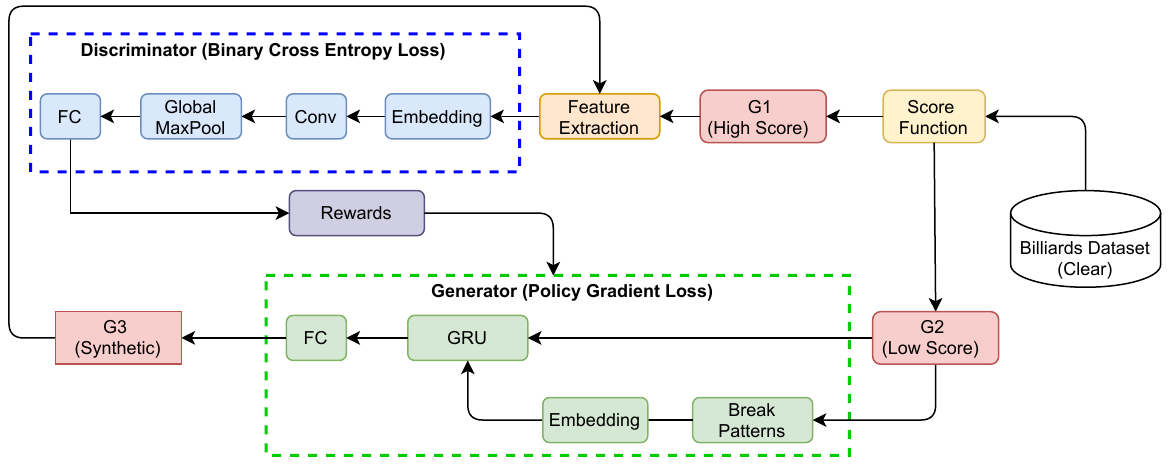}
  \end{minipage}
%\vspace{-2mm}
%\setlength{\abovecaptionskip}{1pt}
\caption{%The model architecture of BLGAN, where FC denotes a fully-connected layer.
The model architecture of BLGAN. The score function is to identify two groups of layouts with high and low scores (i.e., $G_1$ and $G_2$) from the dataset. In the discriminator, it follows the architecture in BLCNN, and provides the rewards outputted via a fully-connected (FC) layer for training the generator. In the generator, the break patterns are extracted from $G_2$, each pattern corresponds to a token, which is fed into an embedding layer to obtain a latent vector as the initial hidden vector, to generate the $G_3$ via GRU followed by a FC layer.
}
\label{fig:BLGAN}}
\vspace{-2.5mm}
\end{figure}

\smallskip\noindent \textbf{Overview.} To serve the target of generating the layouts that are more likely to be cleared, we collect all layouts with clear labels from the dataset. We employ a \emph{score function} (more details would be discussed later) to score the quality of a layout, sort the layouts in descending order of their scores and divide the billiards layouts equally into two groups without overlapping (i.e., one group ($G_1$) represents layouts with high scores (high-score), the other group ($G_2$) represents layouts with low scores (low-score)).
% Then, we equally divide the layouts into two groups. The first group includes the layouts with higher scores, denoted by $G_1$, and the second group includes the layouts with lower scores, denoted by $G_2$. 
$G_2$ is fed to the \emph{generator} in BLGAN to guide the new layout generation denoted by $G_3$. 
% {\ChengComment The intuition of the generator is }
% {\Comment The intuition behind the generator is to generate layouts $G_3$ with higher scores from those sequences (i.e., the layouts $G_2$) with lower scores.}
%
Also, we train a \emph{discriminator} to provide guidance for improving the quality of billiards layouts the generator generates by feeding the examples of $G_1$ (i.e., the real layouts with high scores) and the examples of $G_3$ (i.e., the generated layouts). 
%A figure depicting the architecture could be found in the supplementary materials for illustration.
The GAN-based architecture is illustrated in Figure~\ref{fig:BLGAN}.
%We introduce the detailed designs next.
%We generate the billiards layouts by exploring existing layouts which are cleared with a data-driven solution, and it could help players to explore more layouts as the targets of a break shot in the game.

%\vspace{-8mm}

\smallskip\noindent \textbf{Score Function.}
%{\color{blue}In order to evaluate the performance of BLGAN and identify two groups of the layouts with high and low scores (i.e., G1 and G2), we propose to use a score function to score the billiards layouts, i.e., evaluating the quality of the layouts whether they can be cleared.}
%
% Based on the empirical study, BLCNN has shown good performance on the clear prediction task, which corresponds to a score function to give the scores of layouts. 
 We use the BLCNN model to compute the score of a layout since BLCNN model can predict whether a layout will be cleared or not, which indicates the quality of the layout. BLCNN involves a fully-connected layer with the softmax function and thus it computes scores from 0 to 1. {\revision The rationale for employing the BLCNN model as the score function can be clarified in three aspects. First, in the generation task, all layouts with clear labels are utilized. Within this context, we categorize layouts into two distinct groups: $G_1$ (high-score) and $G_2$ (low-score). The objective is to guide the generator to produce $G_3$ (the generated layouts) in a manner similar to $G_1$, aiming to confuse the discriminator in its identification task. The BLCNN naturally serves as an effective scoring function because it is trained to capture layout features and classify clear labels (0 or 1) using the associated scores. Second, since both $G_1$ and $G_2$ are grouped based on clear layouts, additional supervised information, such as win labels or potted balls labels, becomes irrelevant in providing a scoring function. Within these groups, all layouts are associated with a win, and all the balls on the table are considered potted. Third, the BLCNN plays a role in data preprocessing by preparing $G_1$ and $G_2$ to guide the training of the discriminator and generator in BLGAN. It remains inactive during the training process of BLGAN, which allows BLGAN to undergo an end-to-end training procedure independently.
 }

{\extension{
\begin{figure*}[h]
  %\hspace*{-.3cm}
  \centering
  \begin{tabular}{c c}
    \begin{minipage}{0.38\linewidth}
    \includegraphics[width=\linewidth]{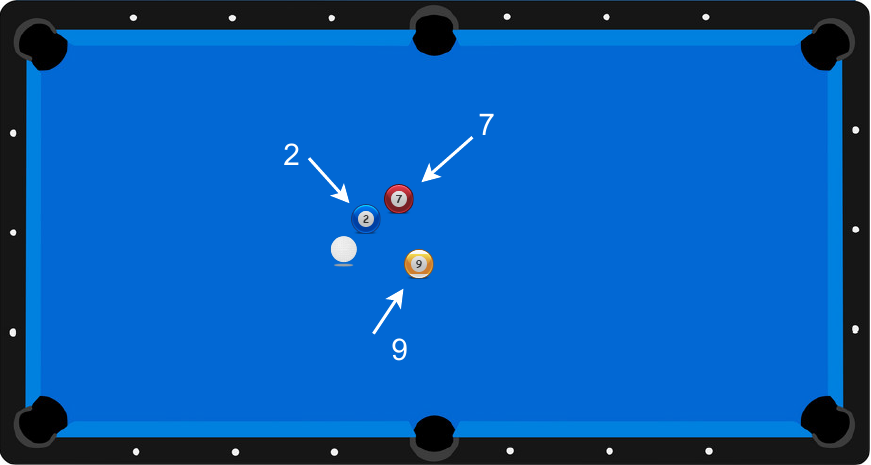}
    \end{minipage}
    &
    \begin{minipage}{0.38\linewidth}
      \includegraphics[width=\linewidth]{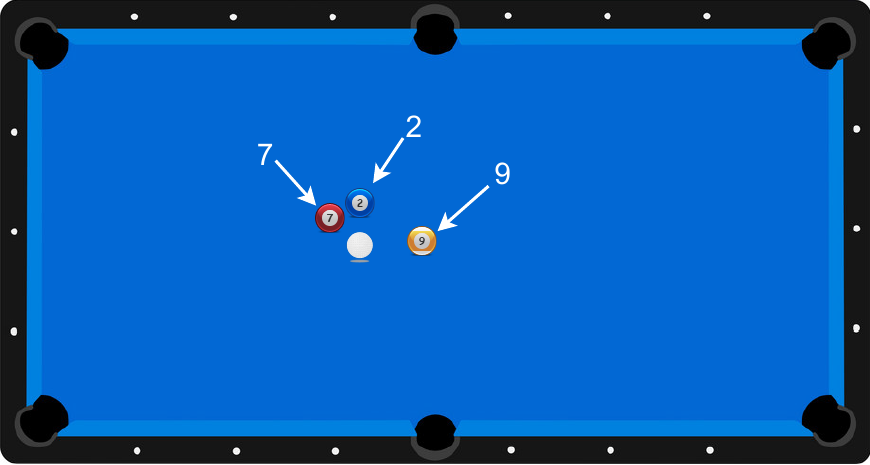}
    \end{minipage}
    \\
    (a) $\mathcal{B}_q$ (cue ball and ball 2,7,9)
    &
    (b) $\mathcal{B}_1$ (cue ball and ball 2,7,9)
    \\\\
    \begin{minipage}{0.38\linewidth}
    \includegraphics[width=\linewidth]{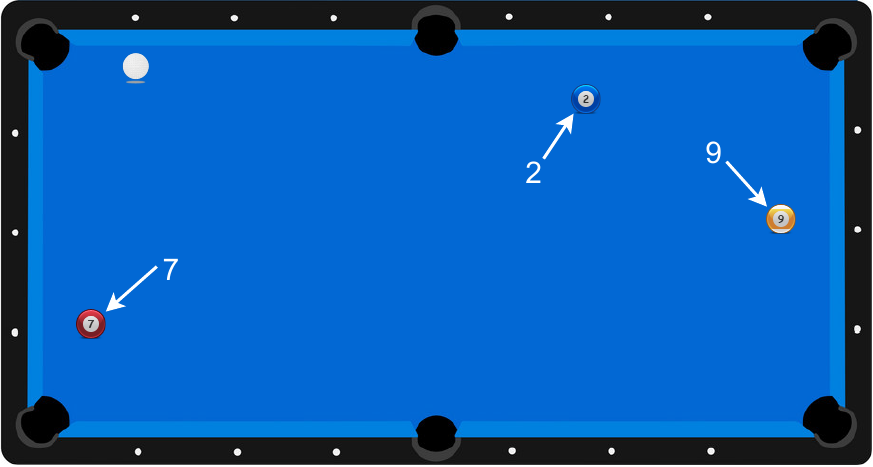}
    \end{minipage}
    &
    \begin{minipage}{0.38\linewidth}
      \includegraphics[width=\linewidth]{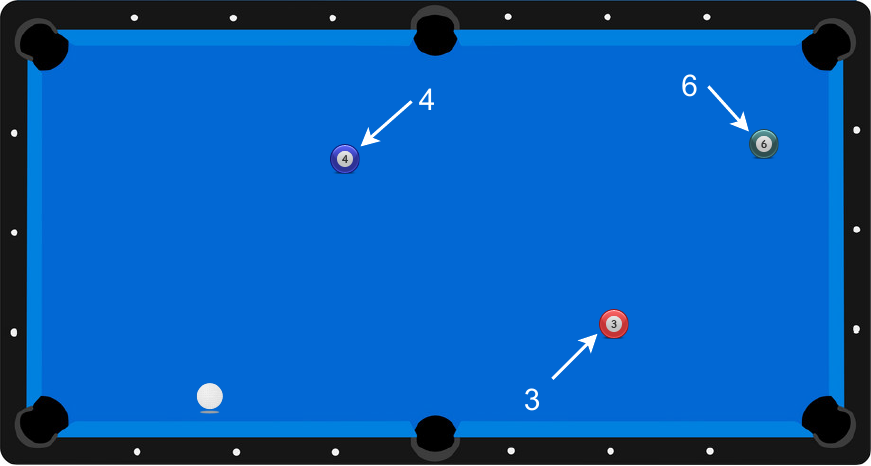}
    \end{minipage}
    \\
    (c) $\mathcal{B}_2$ (cue ball and ball 2,7,9)
    &
    (d) $\mathcal{B}_3$ (cue ball and ball 3,4,6)
  \end{tabular}
  \vspace{-2mm}
  \caption{An example of illustrating the billiards layout similarity, where $sim(\mathcal{B}_q, \mathcal{B}_1) > sim(\mathcal{B}_q, \mathcal{B}_2) > sim(\mathcal{B}_q, \mathcal{B}_3)$.}
  \label{fig:problem}
  \vspace*{-3mm}
\end{figure*}
}}

\smallskip\noindent \textbf{Generator.} To overcome the issue of indifferentiability, we follow an existing strategy~\cite{yu2017seqgan,bachman2015data,bahdanau2016actor} to model the sequence generation procedure as a sequential decision making process, which is optimized by the REINFORCE algorithm via Policy Gradient~\cite{williams1992simple} and thus it naturally bypasses the differentiation difficulty for discrete tokens.
In particular, we train a generative model $G$ to generate a sequence $C_{1:T} = (c_1, c_2, ...c_t, ....c_T)$, $c_t \in \gamma$, where $c_t$ denotes the location token of a ball, and $\gamma$ corresponds to the vocabulary of all location tokens on the billiards table.
At time step $t$, the state $s$ corresponds to the sequence of generated tokens so far, i.e., $(c_1, c_2, ...c_{t-1})$; and the action $a$ is to select the next location token (i.e., $c_t$) based on the state $s$. To obtain the reward $r$, the existing study~\cite{yu2017seqgan} adopts the roll-out policy by sampling the unknown tokens in a sequence to obtain an immediate reward from the discriminator. 
In this paper, we adopt a simpler strategy, i.e., we collect a reward only after an entire layout is generated and fed into the discriminator - recall the discriminator would return a score as the reward signal, and it is shared to all steps when the entire layout has not been generated.
Besides, we collect a set of patterns, each corresponding to a set of balls of a layout, which appear in the real dataset. Each such pattern is called a \emph{\underline{break pattern}} and is encoded using one-hot vectors. For example, consider a break shot layout in a 9-ball game with the cue ball and ball 1, 2, 3, 5, 7, 8, 9 on the pool table. The break pattern of this layout is denoted by $(4,6)$ since ball 4 and 6 are potted already and thus missing from the layout. When generating a layout, we randomly sample a known break pattern and feed its vector to the generator as an initial hidden vector to guide the generation task. This is to handle the second challenge.

\smallskip\noindent \textbf{Discriminator.}
By empirical findings, the CNN-based model has shown good performance to capture the correlations for the billiards layouts, e.g., it achieves the accuracy around 90\% on the previous prediction tasks. Thus, we adopt the CNN-based architecture in the discriminator, whose main function is to discriminate the generated layouts (i.e., $G_3$) and high-score real layouts (i.e., $G_1$), and provide a reward signal to guide the learning of the generator, i.e. it encourages the generator to generate the layouts of high-score (i.e., $G_3$) from those low-score real layouts (i.e., $G_2$).
% , which targets the third challenge.
%
The optimization of the discriminator is to minimize Binary Cross-Entropy (BCE) between $G_1$ and $G_3$.

%% file: method.tex
{\extension{
  
\section{Layout Retrieval with BL2Vec}
\label{sec:method}

% In this paper, our goal is to design such $sim(\cdot)$ function for measuring the similarity among billiards layouts. We formulate the problem as follows.

% \subsection{Problem Statement} 
% \smallskip\noindent\textbf{Problem Statement.}
% Given a database of billiards layouts $\mathcal{D}$, we aim to map each billiards layout $\mathcal{B} \in \mathcal{D}$ to a low-dimensional space to capture similarities among billiards layouts. 
% More specifically, for any two given billiards layouts $\mathcal{B}_1$ and $\mathcal{B}_2$, we map them to two $K$-dimensional vectors $v_1$ and $v_2$ via deep metric learning. The learned vectors should be similarity preserving, i.e., if two billiards layouts are similar, the Euclidean distance between their vectors would be small.

{\extension{
\begin{figure}
\hspace*{1.5cm}
\centering
\begin{tabular}{c}
  \begin{minipage}{\linewidth}
	\includegraphics[width=0.8\linewidth]{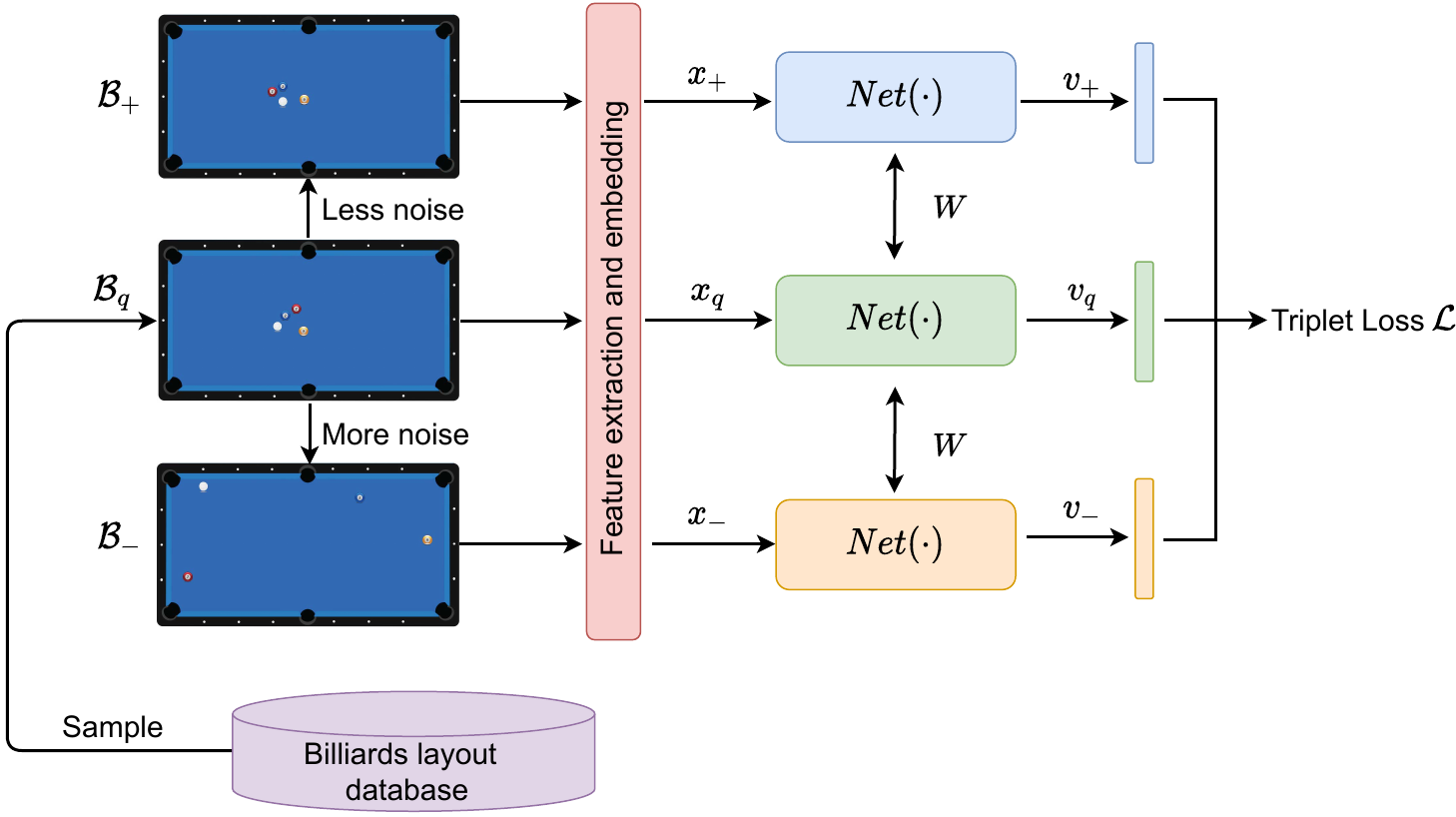}
  \end{minipage}
\end{tabular}
\vspace{-2mm}
\caption{The model architecture of BL2Vec.}
\label{fig:bl2vec}
\vspace*{-2mm}
\end{figure}
}}

\subsection{{Overview of BL2Vec}}
\label{subsec:triplet}

Given two billiards layouts, it is hard to tell if they are similar or dissimilar. However, it could be noticed from the example in Figure~\ref{fig:problem} that given a query layout and two candidate layouts, it is much easier to distinguish which one is more similar to the query layout. Specifically, in Figure~\ref{fig:problem}, $\mathcal{B}_q$ denotes a query layout, which contains a cue ball and three object balls 2,7,9. There are three different candidate layouts $\mathcal{B}_1$, $\mathcal{B}_2$ and $\mathcal{B}_3$. Among these candidate layouts, $\mathcal{B}_3$ is the most dissimilar to $\mathcal{B}_q$ since $\mathcal{B}_3$ contains totally different object balls (i.e., ball 3,4,6). Both $\mathcal{B}_1$ and $\mathcal{B}_2$ contain the same balls as $\mathcal{B}_q$; however, it is clear that the locations of the balls in $\mathcal{B}_1$ are more similar to those in $\mathcal{B}_q$ since these balls are located near to each other as they are in $\mathcal{B}_q$, while the balls in $\mathcal{B}_2$ are scattered. Therefore, while we do not know the exact values of these similarities, we tend to know a ranking of them (i.e., $sim(\mathcal{B}_q, \mathcal{B}_1) > sim(\mathcal{B}_q, \mathcal{B}_2) > sim(\mathcal{B}_q, \mathcal{B}_3)$). 
Here, $sim(, )$ defines the similarity between two layouts.
 
Motivated by this, we propose a deep metric learning model based on \emph{triplet network}~\cite{hoffer2015deep} to learn a similarity measure for billiards layouts. We call the model \emph{BL2Vec} and show its overview in Figure~\ref{fig:bl2vec}. 
Specifically, we randomly sample a billiards layout from the dataset as an anchor layout $\mathcal{B}_q$. We then generate a positive sample $\mathcal{B}_+$ and a negative sample $\mathcal{B}_-$ by injecting less and more noise to $\mathcal{B}_q$, respectively. This is based on the intuition that a billiards layout, when injected with less noise, should be more similar to the original layout. 
We then extract features $x_q$ (resp. $x_+$ and $x_-$) from the generated layouts $\mathcal{B}_q$ (resp. $\mathcal{B}_+$ and $\mathcal{B}_-$) and fed them to a triplet network. The triplet network consists of three instances of a shared feedforward neural network, denoted as $Net(\cdot)$, and outputs two Euclidean distances as follows: $d_+ = ||Net(x_q)-Net(x_+)||_2$ and $d_- =||Net(x_q)-Net(x_-)||_2$,
where $Net(x_q)$ denotes the embedded representation of $x_q$ via the network (i.e., a vector). We adopt the following loss for the triplet network:
\begin{equation}\label{triplet}
  \mathcal{L}(x_q,x_+,x_-) = \max\{d_+-d_-+\delta,0\},
\end{equation}
where $\delta$ is a margin between the positive and negative distances.
By optimizing this loss, it learns to correctly distinguish $\mathcal{B}_+$ and $\mathcal{B}_-$.

{\revision The triplet network in BL2Vec introduces technical innovations in three aspects. (1) Incorporation of Ranking-based Intuition: the triplet network aligns with the ranking-based intuition introduced in BL2Vec. Instead of requiring an exact similarity measure, the model learns to rank layouts based on their similarity to a query layout, which is often more intuitive and applicable to real-world scenarios. (2) Learning Layout Similarities: Unlike conventional methods (e.g., EMD~\cite{rubner2000earth}, Hausdorff~\cite{huttenlocher1993comparing}) that rely on established similarity measures for pointsets, the triplet loss function serves as a mechanism for the model to learn a similarity measure for billiards layouts. By optimizing the network's parameters based on the triplet loss, the model learns to embed layouts in a way that aligns with human intuition regarding layout similarity. (3) Training Stability and Convergence: the use of the triplet network helps address the challenge of training stability when dealing with layout similarity. Existing methods (e.g., DeepSets~\cite{zaheer2017deep}, RepSet~\cite{skianis2020rep}) that directly embed layouts might suffer from inherent convergence issues. The triplet network, by focusing on the relative ranking of layouts with a designed margin $\delta$, offers a stable training framework.
}

Next, we present the details of BL2Vec and analyze the time complexity of computing the similarity between two layouts with BL2Vec.

\subsection{Positive and Negative Layouts Generation}
\label{subsec:sample}

\noindent \textbf{Positive billiards layout $\mathcal{B}_+$.} We generate $\mathcal{B}_+$ by conducting two operations on $\mathcal{B}_q$, including (1) randomly shifting the locations of the balls to some extent and (2) randomly removing some balls and adding them back at random locations. We control the generation process with two parameters, namely a noise rate and a drop rate and ensure that no balls overlap. More specifically, setting the noise rate as $\alpha$ means moving each ball along the $x$ axis and the $y$ axis towards a random direction and by a random distance, which is bounded by $\alpha$ times the billiards table's length and width, respectively. Setting the drop rate as $\beta$
means randomly deleting $\beta\cdot n'$ balls rounded with a ceiling function and then adding them at random locations on the billiards table, where $n'$ is the number of balls in the current layout. The two parameters of noise rate and shift rate would be studied in experiments.

\noindent \textbf{Negative billiards layout $\mathcal{B}_-$.} 
One intuitive idea is to randomly sample a billiards layout that is not the same as $\mathcal{B}_q$ to be $\mathcal{B}_-$ from the dataset. However, it suffers from the issue that the randomly sampled $\mathcal{B}_-$ would be generally very dissimilar to $\mathcal{B}_q$, and thus distinguishing $\mathcal{B}_+$ and $\mathcal{B}_-$ becomes trivial. In this case, the model would converge with inferior performance. To address this issue, we propose to generate the $\mathcal{B}_-$ based on $\mathcal{B}_q$ in the same way as we generate $\mathcal{B}_+$, but with more noise (i.e., larger noise rate and drop rate). In summary, both $\mathcal{B}_+$ and $\mathcal{B}_-$ are noisy versions of $\mathcal{B}_q$, where $\mathcal{B}_+$ is more similar to $\mathcal{B}_q$ since it contains less noise than $\mathcal{B}_-$.

\noindent {\revision \textbf{Discussion on Training the Positive and Negative for Retrieval.} We discuss the relationship between the training of positive and negative samples and its impact on the retrieval task within the BL2Vec model. This training strategy is designed to tackle the challenges in determining the similarity between two billiards layouts for retrieval. The task of discerning the similarity or dissimilarity between two billiards layouts is inherently complex and lacks a robust measurement. However, as illustrated in Figure~\ref{fig:problem}, distinguishing which one is more similar to the query layout among two candidate layouts is relatively straightforward. Motivated by this observation, we explore a deep metric learning model based on the triplet network. This model involves creating a positive sample and a negative sample from an anchor layout. It is trained to encode layouts into vectors, aligning the similarity between these vectors with the intuition that the positive sample should be more similar to the query layout than the negative one.

In the execution of the layout retrieval task, we employ the trained BL2Vec to encode all layouts in the database and the query into vectors. Subsequently, we calculate the similarities between these vectors to identify the layout in the database that is closest to the query. This process establishes an effective retrieval mechanism based on learned layout similarities.
}

\subsection{Billiards Layout Feature Extraction and Embedding}
We extract the features Ball-Self (BS), Ball-Pocket (BP) and Ball-Ball (BB) from a billiards layout as presented in Section~\ref{task:blcnn}, and then map the continuous BS, BP and BB features into discrete tokens with some predefined granularities (e.g., the granularity of cell size is set to $15 \times 15$ for BS, $15^\circ$ and 10 are used to partition the angle and distance ranges for BP and BB, respectively). 
We then obtain a vector for each layout by embedding the tokens as one-hot vectors and concatenating the vectors in an ascending order of the ball numbers, denoted by $x \in \mathbb{R}^{n \times K'}$, where $K'$ is the dimension of the embedded space and $n$ denotes the total number of balls in a billiards game to which the layout belongs, such as $n=10$ for 9-ball layouts.

\subsection{Model Architecture of $Net(\cdot)$}
\label{subsec:net}
We adopt a convolutional neural network (CNN) to instantiate the $Net(\cdot)$ , since the CNN is inherently appropriate for perceiving the layout, and preserves the spatial correlation of the balls that form the layout.

To feed a layout $\mathcal{B}$ into $Net(\cdot)$, we use its embedded features $x \in \mathbb{R}^{n \times K'}$.
Let $x_i$ be a $K'$-dimensional embedding vector corresponding to the ball with the number $i$ in the layout.
Then, a billiards layout could be represented as $x_{1:n} = x_1 \oplus x_2 \oplus ... \oplus x_n$, where $\oplus$ is the concatenation operator. We denote by $x_{i:i+j}$ the concatenation of $x_i, x_{i+1},...,x_{i+j}$. A convolution operation~\cite{zhang2015sensitivity}, which involves a \emph{filter} $\bm{W} \in \mathbb{R}^{h \times K'}$, is applied to a window of $h$ balls to extract a new feature. Specifically, a feature $c_i$ is extracted from a window of balls $x_{i:i+h-1}$ as follows.
$$
c_i = \sum_{l=i}^{i+h-1}\sum_{r=1}^{K'} \bm{W}_{l-i+1,r} \cdot x_{l,r} + \bm{b},
$$
where $b \in \mathbb{R}$ denotes a bias term, {\nnComment{and $r$ denotes each dimension in an embedded feature $x_i$ ($1 \leq r \leq K'$).
}}
Then, the filter is applied to each possible window of balls in the layout $\{x_{1:h}, x_{2:h+1},..., x_{n-h+1:n}\}$ to produce a feature map as $c = \left[c_1, c_2,..., c_{n-h+1}\right]$, where $c \in \mathbb{R}^{n-h+1}$. We then apply a global max-pooling layer over the feature map and take the maximum value 
$$
\hat{c} = \max\limits_{1 \le i \le n-h+1}\{c_i\}
$$ as the feature corresponding to this particular filter.
The intuition is that it can capture the most important feature (i.e., one with the highest value) for each feature map, and deals with variable lengths of different feature maps.

Note that each feature $\hat{c}$ is extracted from each convolution filter, and we use multiple filters with varying widths to obtain multiple features. Then, these features are concatenated to be a $K$-dimensional vector $v$ as the representation of a billiards layout.

\begin{algorithm}[t]
\caption{Training of the BL2Vec Model}
\label{framework}
\begin{algorithmic}[1]
        \Require
        $\mathcal{D}$: the database of billiards layouts need to be embedded;
        $K$ (resp. $K'$): the embedding dimension of each billiards layout (resp. ball)
        \Ensure
        The trained BL2Vec model
        \Repeat
        \State sample an anchor $\mathcal{B}_q$ from $\mathcal{D}$, and generate $\mathcal{B}_+$ and $\mathcal{B}_-$ by adding noises;
        \State extract features and obtain tokens $(\mathcal{T}_q, \mathcal{T}_+, \mathcal{T}_-)$;
        \State embed the tokens and obtain the embedded features $(x_q, x_+, x_-)$;
        \State get $(v_q, v_+, v_-)$ from $Net(\cdot)$;
        \State optimize the triplet loss $\mathcal{L}$ according to Equation~\ref{triplet};
        \Until{No improvement on validation set}
\end{algorithmic}
\end{algorithm}

\subsection{Model Training}
\label{subsec:bl2vec}
The detailed procedure of training the BL2Vec is presented in Algorithm~\ref{framework}. 
During the iterative training process (lines 1-7).
It first samples a billiards layout $\mathcal{B}_q$ from the database $\mathcal{D}$ as an anchor and generates its positive version $\mathcal{B}_+$ and negative version $\mathcal{B}_-$ by adding noises (line 2). It then extracts the BS, BP and BB features from the layouts, and map them into discrete tokens $(\mathcal{T}_q, \mathcal{T}_+, \mathcal{T}_-)$ (line 3). These tokens are further fed into the embedding layer of the model to get the embedded features $(x_q, x_+, x_-)$ (line 4). Next, it obtains the embedding vectors $(v_q, v_+, v_-)$ from $Net(\cdot)$ (line 5).
Finally, it computes the triplet loss $\mathcal{L}$ according to Equation~\ref{triplet} and optimizes the loss via some optimizers such as Adam stochastic gradient descent (line 6).

Once the model has been trained, {\ChengComment{we can embed each billiard layout in a database to a vector as a pre-processing step. When a query layout comes, we embed the query layout to a vector, perform a \emph{$k$ nearest neighbor} ($k$NN)~\cite{bentley1975multidimensional} query to search $k$ data vectors that have the smallest Euclidean distances to the query vector, and then return the billiards layouts corresponding to the found data vectors.
}}

\subsection{Time Complexity Analysis}
We analyze the time complexity of computing the similarity between two billiards layouts based on BL2Vec.
Specifically, it includes two parts, namely the embedding part and the distance computation part. For embedding, the time complexity is $O(n)$, where $n$ denotes the total number of balls in a billiards game to which the layout belongs. We explain as follows:
(1) It takes $O(n')$ to extract features from a layout and map the features to the corresponding token, where $n'$ denotes the number of balls in the current layout, and it is bounded by $n$.
(2) It takes $O(n)$ to get the embedded features with padding the missing balls.
(3) It takes roughly $O(n)$ to map the embedded features to the target $K$-dimensional vector representation via the classical convolutions~\cite{he2015convolutional}. %More specifically, the complexity in a classic convolutional unit~\cite{he2015convolutional} is $O(M_{length}*M_{width}*K_{length}*K_{width}*C_{in}*C_{out})$, where $M_{length}$ (resp. $M_{width}$) denotes the length (resp. width) of the feature map, $K_{length}$ (resp. $K_{width}$) denotes the length (resp. width) of the filter, $C_{in}$ is the number of input channels of the convolutional layer and $C_{out}$ is the number of filters. $K_{length}$, $K_{width}$, $C_{in}$ and $C_{out}$ are set to constants in the model. For the feature map, $M_{width}=1$ and $M_{length}$ is bounded by $n$.
For distance computation, it costs $O(K)$ which is obvious. Hence, the overall time complexity is $O(n + K)$.
We note that existing matching-based methods~\cite{yi1998efficient,alt1995computing} have the time complexity at least $O(n^2)$ and the operator of computing similarity is usually conducted highly frequently. As a result, the similar billiard layout retrieval based on BL2Vec is significantly faster than that based on existing methods (e.g., up to 420x speed-up in our experiments). 

{\ChengComment{Finally, we analyze the time complexity of the similar billiards layout retrieval for a query layout based on BL2Vec. Suppose there are $N$ billiards layouts in a database. The time complexity of computing the vectors of the $N$ billiards layouts is $O(N\cdot (n+K))$, where $n$ is the number of balls in a billiards game and $K$ is the number of dimensions of a vector. Note that this cost is one-time cost and shared by all queries. 
% With the vectors of $N$ billiards layouts computed as a pre-processing step, the time complexity is dominated by that 
The time complexity of answering each similar layout retrieval query is dominated by that of conducting a $k$NN query, for which we can leverage a rich literature. For example, the time complexity is $O(k\cdot \log (N))$ if the algorithm in~\cite{bentley1975multidimensional} is adopted for the $k$NN query.}}
% Given a billiards layout database, the billiards layout embeddings only need to be computed once. For the similarity search task, we only need to embed a new billiards layout (query) and perform the search based on the distance of embeddings. Thus, the search complexity is linear with the size of the database.
}}

%% file: experiments.tex
\section{EXPERIMENTS}
\label{sec:experiment}

\subsection{Evaluation on the Billiards Layout Prediction Task}
\label{subsec:prediction}
We evaluate the BLCNN for the prediction task on the collected dataset, including (1) an effectiveness study to evaluate the accuracy of the classification on three tasks (i.e., clear, win and potted ball), (2) a parameter study to evaluate the effect of cell size, and (3) an ablation study to evaluate the importance of each component of extracted features in BLCNN. It demonstrates the superior performance of BLCNN (e.g., with the accuracy of 89.69\%, 86.56\% and 80.94\% for the clear, win and potted ball tasks, respectively). Detailed results and explanations can be found in~\cite{zhang2022predicting}.

\subsection{Evaluation on the Billiards Layout Generation Task}
\label{subsec:generation}
We further evaluate the BLGAN for the generation task. We first evaluate (1) the quality and reality of the generated layouts by BLGAN. Overall, we observe BLGAN has good performance on generating the layouts of high quality (e.g., the scores are all-around 0.9 for four frequent break patterns), and the generated layouts are in general quite similar to real ones (e.g., it exceeds the similarity of more than half of the real layouts). We also conduct (2) a user study to show the quality of the generated layouts. We observe BLGAN generates the layouts that can be easily cleared with 82.5\% votes from users. We present the detailed results and explanations for BLGAN in~\cite{zhang2022predicting}.

{\extension{\subsection{Evaluation on the Billiards Layout Retrieval Task}
\label{subsec:similar}

\subsubsection{Experimental Setup}
\label{sec:setup}

\begin{figure*}[ht]
\centering
  %\hspace*{3mm}
  \begin{minipage}{0.7\linewidth}%5.5
    \includegraphics[width=\linewidth]{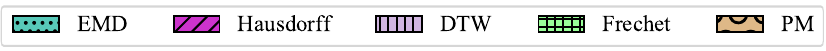}
  \end{minipage}
  \\
  \begin{minipage}{0.63\linewidth}%5.5
    \includegraphics[width=\linewidth]{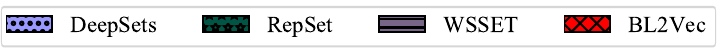}
  \end{minipage}
  \begin{tabular}{c c}
   \begin{minipage}{0.47\linewidth}%5.5
    \includegraphics[width=\linewidth]{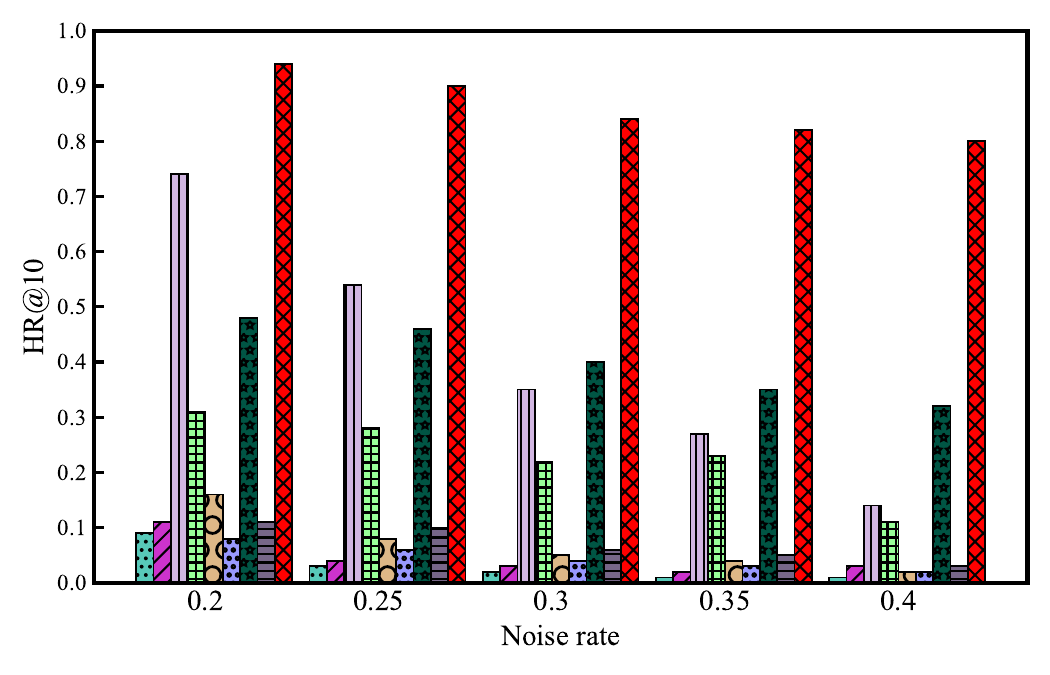}
    \end{minipage}
    &
    \begin{minipage}{0.47\linewidth}
    \includegraphics[width=\linewidth]{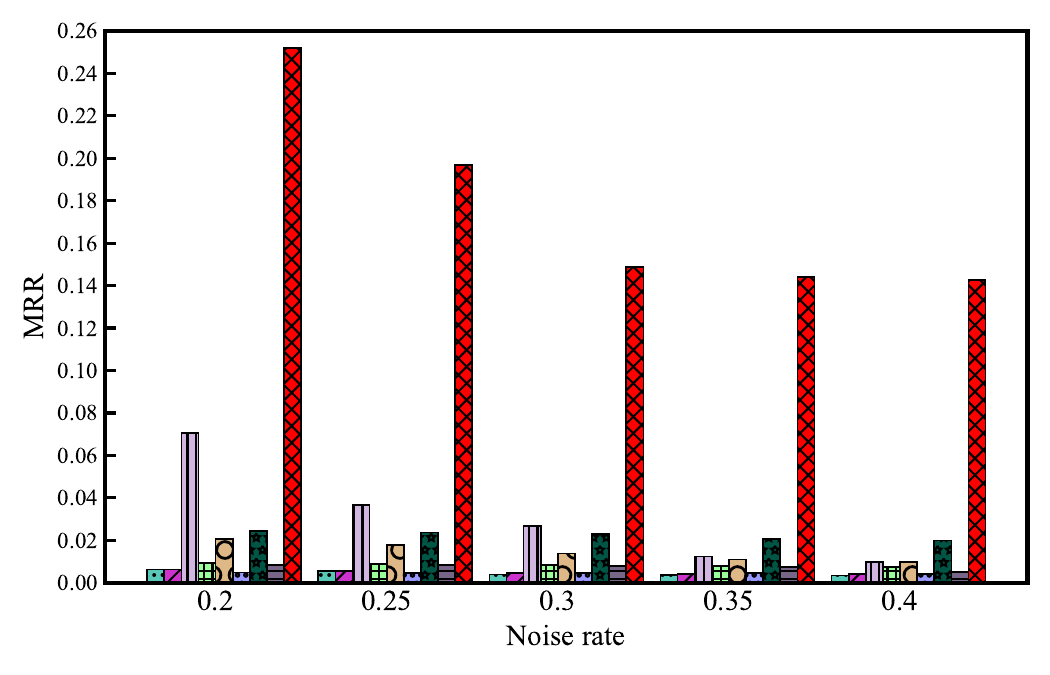}
    \end{minipage}
    \\
    (a) Varying noise rate (HR@10)%\hspace{-4mm}
    &
    (b) Varying noise rate (MRR)%\hspace{-4mm}
    \\
    \begin{minipage}{0.47\linewidth}
    \includegraphics[width=\linewidth]{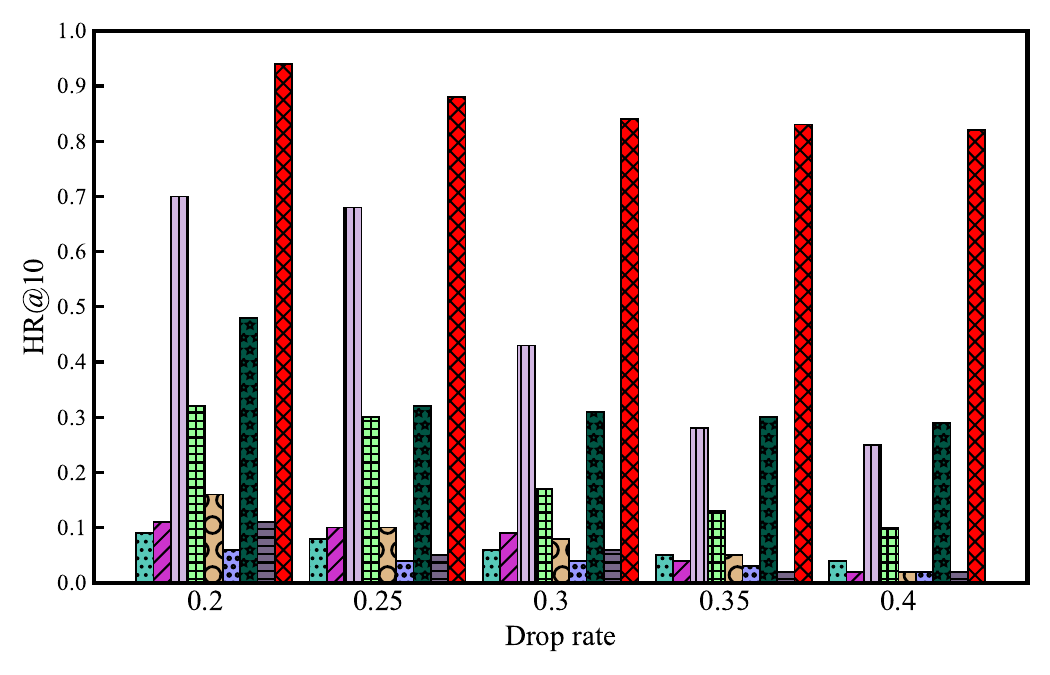}
    \end{minipage}
    &
    \begin{minipage}{0.47\linewidth}
    \includegraphics[width=\linewidth]{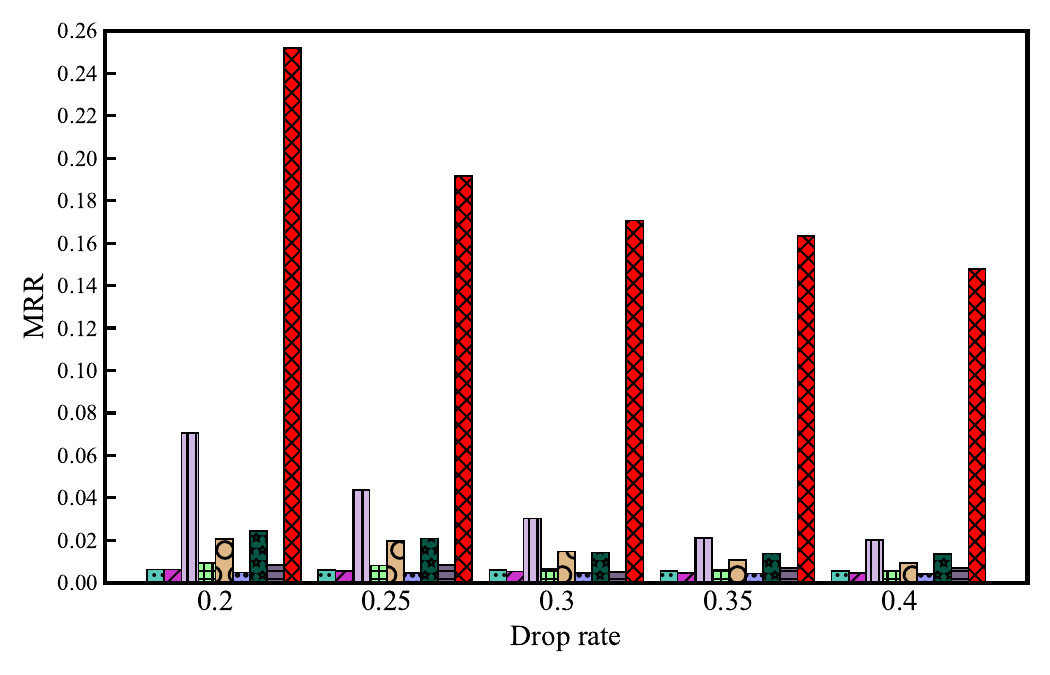}
    \end{minipage}
    \\
    (c) Varying drop rate (HR@10)%\hspace{-4mm}
    &
    (d) Varying drop rate (MRR)%\hspace{-4mm}
\end{tabular}
\vspace{-3mm}
{\extension{\caption{Effectiveness evaluation (similarity search).}
\label{fig:effectiveness_similarity}
}}
\vspace*{-3mm}
\end{figure*}

\noindent \textbf{Baseline.}
To evaluate the effectiveness and efficiency of our BL2Vec, we compare it with the following baselines: %(1) EMD~\cite{rubner2000earth}, (2) Hausdorff~\cite{huttenlocher1993comparing}, (3) DTW~\cite{yi1998efficient}, (4) Frechet~\cite{alt1995computing}, (5) Peer Matching (PM), (6)DeepSets~\cite{zaheer2017deep}, (7) RepSet~\cite{skianis2020rep} and (8) WSSET~\cite{Arsomngern2021self} as discussed in Section~\ref{sec:related}.
%\begin{itemize}
\\
$\bullet$
%\item 
EMD~\cite{rubner2000earth}. The Earth Mover's Distance (EMD) is based on the idea of a transportation problem that measures the least amount of work needed to transform one set to another. %a well-known measurement for pointsets  Besides, EMD is widely used for various applications such as document classification, image retrieval, etc.
\\
$\bullet$
Hausdorff~\cite{huttenlocher1993comparing}. Hausdorff distance is based on the idea of point-matching and computes the largest distance among all the distances from a point in one set to the nearest point in the other set.
\\
$\bullet$
DTW~\cite{yi1998efficient} and Frechet~\cite{alt1995computing}. DTW and Frechet are two widely used similarity measurements for 
sequence data.
%trajectory/time series analysis. Specifically, Frechet is
%a metric, namely the distance is symmetric and satisfies the triangle inequality; however, DTW is not a metric, since it does not satisfy the triangle inequality.
%
To apply the measurements to billiards layouts, we model each layout as a sequence of locations, which corresponds to one cue ball followed by several object balls in ascending order of ball numbers.
\\
$\bullet$
Peer Matching (PM). We adopt a straightforward solution for measuring the similarity between two billiards layouts as the average of the distances between their balls matched based on the numbers, e.g., ball 1 in one layout is matched to the ball 1 in another layout, and so on.
For those mismatched balls with different numbers, we align them in the ascending order of their numbers and match them accordingly.
%}
\\
$\bullet$
DeepSets~\cite{zaheer2017deep}. DeepSets is used to generate pointsets representation in an unsupervised fashion. The main idea of DeepSets is to transform vectors
of points in a set into a new representation of the set. It aggregates the representations of vectors and then passes them to fully-connected layers to get the representation.
\\
$\bullet$
RepSet~\cite{skianis2020rep}. RepSet is also used to generate the representations of pointsets. It adapts a bipartite matching idea by computing the correspondences between an input set and some generated hidden sets to get the set representation. Besides, RepSet captures the unordered nature of a set and can generate the same output for all possible permutations of points in the set.
\\
$\bullet$
WSSET~\cite{Arsomngern2021self}. WSSET achieves the state-of-the-art performance on pointsets embedding, which is an approximate version of EMD based on the deep metric learning framework and uses the EMD as a metric for inferring the positive and negative pairs. It generates the representations of pointsets, and these generated representations can be used for similarity search, classification tasks, etc.
%\end{itemize}

\noindent \textbf{Parameter Setting.}
% We set the cell size to be 15. 
Intuitively, with the smaller cell size, it generates more tokens and provides a higher resolution of the pool table. However, with the smaller size, it reduces the robustness against the noise for the similarity search task. The parameter is set to 15 because it provides better performance in general via empirical studies (with the results of its effect shown later on), and thus %The results and detailed description are included in the technical report~\cite{TR} due to the page limit), % (with the results of its effect shown later on) 
we have 99 unique tokens for the BS features. For BP and BB features, {\ICDEComment{we vary the parameter of angle (resp. distance) from $5^\circ$ to $20^\circ$ (resp. 5 to 20), and since the results are similar, we use the setting of $15^\circ$ and 10 for partitioning the angle range and distance range, respectively.}}
For each feature, we embed it to a 10-dimensional vector, and thus the representation dimension of the ball is $10*(1+4*6+2)=270$ (i.e., $K'=270$). Recall that for each ball, there is one token in the BS feature, four tokens in the BP feature with totally six pockets, and two tokens in the BB feature.We use a convolutional layer as $Net(\cdot)$ with 3 output channels. For each, it has 52 filters with varying sizes from $(1,K')$ to $(7,K')$. Note that larger filters help capture the correlations between the balls. A ReLU function is applied after the convolutional layer and then a global max-pooling layer is employed to generate a 156-dimensional representation of the billiards layout (i.e., $K=3 \times 52 = 156$). Additionally, in the training process, we randomly sample 70\% billiards layouts to generate training tuples and adopt Adam stochastic gradient descent with an initial learning rate of $0.00001$, and the remaining layouts are used for testing. To avoid overfitting, we employ a L2 regularization term with $\lambda=0.001$ for $Net(\cdot)$. The margin in the triplet loss is set to $\delta = 1.0$ based on empirical findings.For generating each $\mathcal{B}_+$ (resp. $\mathcal{B}_-$), we set both the noise rate and the drop rate to 0.2 by default. For parameters of baselines, we follow their settings in the original papers.

\noindent \textbf{Evaluation Metrics.}
We consider the following three tasks to evaluate the effectiveness.
(1) Similarity search, we report the Hitting Ratio for Top-10 ($HR@10$) and Mean Reciprocal Rank ($MRR$). In particular, for a query billiards layout $\mathcal{B}_q$, if its positive version $\mathcal{B}_+$ is in the top-10 returned billiards layouts, then $HR@10$ is defined as $HR@10 = 1$, otherwise $HR@10 = 0$.
$MRR$ is defined as $MRR = \frac{1}{rank}$, where $rank$ denotes the rank of $\mathcal{B}_+$ in the database for its query billiards layout $\mathcal{B}_q$. The average results of $HR@10$ and $MRR$ over all queries are reported in the experiments. A higher $HR@10$ or $MRR$ indicates a better result.
(2) Classification, we report the average classification accuracy for the label of ``clear'' and ``not clear''. 
% "clear" or "not clear".
Classification accuracy is a widely used metric in the classification task. A higher accuracy indicates a better result.
(3) Clustering, we report two widely used metrics Adjusted Rand Index (ARI) and Adjusted Mutual Information (AMI) for clustering. %{\color{blue} ~\cite{rand1971objective} ~\cite{vinh2010information}
They measure the correlation between the predicted result and the ground truth. The ARI and AMI values lie in the range $[-1, 1]$. For ease of understanding, we normalize the values in the range of $[0,1]$.
%as a percentage value. %{\color{blue}
A higher ARI or AMI indicates a higher correspondence to the ground-truth.%}

\noindent \textbf{Evaluation Platform.}
All the methods are implemented in Python 3.8. The implementation of $BL2Vec$ is based on PyTorch 1.8.0. The experiments are conducted on a server with 10-cores of Intel(R) Core(TM) i9-9820X CPU @ 3.30GHz 64.0GB RAM and one Nvidia GeForce RTX 2080 GPU. The datasets and codes can be downloaded via the link~\footnote{\url{https://www.dropbox.com/sh/644aceaolfv4rky/AAAKhpY1yzrbq9-uo4Df3N0Oa?dl=0}}.

\begin{table}[]
\centering
{\extension{\caption {Effectiveness with classification and clustering.}\label{tab:tasks}}}
\vspace*{-3mm}
\begin{tabular}{c|c|l|l|l|c|lcl}
\hline
\multirow{2}{*}{Tasks} & \multicolumn{4}{c|}{Classification} & \multicolumn{4}{c}{Clustering}                     \\ \cline{2-9}
                       & \multicolumn{4}{c|}{Acc (\%)}       & \multicolumn{2}{c}{ARI (\%)} & \multicolumn{2}{c}{AMI (\%)} \\ \hline
EMD                    & \multicolumn{4}{c|}{66.33}               & \multicolumn{2}{c}{52.15}    & \multicolumn{2}{c}{55.81}    \\
Hausdorff              & \multicolumn{4}{c|}{59.17}               & \multicolumn{2}{c}{51.68}    & \multicolumn{2}{c}{54.98}    \\
DTW                    & \multicolumn{4}{c|}{59.33}               & \multicolumn{2}{c}{50.12}    & \multicolumn{2}{c}{50.13}    \\
Frechet                & \multicolumn{4}{c|}{59.16}               & \multicolumn{2}{c}{50.16}    & \multicolumn{2}{c}{50.22}    \\
PM                        & \multicolumn{4}{c|}{54.24}               & \multicolumn{2}{c}{50.02}    & \multicolumn{2}{c}{50.32}    \\
DeepSets               & \multicolumn{4}{c|}{54.23}               & \multicolumn{2}{c}{49.50}    & \multicolumn{2}{c}{50.41}    \\
RepSet                 & \multicolumn{4}{c|}{58.99}               & \multicolumn{2}{c}{50.54}    & \multicolumn{2}{c}{50.72}    \\
WSSET                  & \multicolumn{4}{c|}{56.62}               & \multicolumn{2}{c}{49.12}    & \multicolumn{2}{c}{49.78}    \\
BL2Vec                 & \multicolumn{4}{c|}{\textbf{72.62}}               & \multicolumn{2}{c}{\textbf{61.94}}    & \multicolumn{2}{c}{\textbf{63.26}}    \\ \hline
\end{tabular}
\vspace*{-4mm}
\end{table}

\subsubsection{Experimental Results}
\label{sec:effectiveness}

\textbf{(1) Effectiveness evaluation (similarity search).}
The lack of ground truth makes it a challenging problem to evaluate the similarity. To overcome this problem, we follow recent studies~\cite{Arsomngern2021self} which propose to use the $HR@10$ and $MRR$ to quantify the similarity evaluation via self-similarity comparison. Specifically, we randomly select 100 billiards layouts to form the query set (denoted as $Q$) and 500 billiards layouts as the target database (denoted as $D$), For each billiards layout $ \mathcal{B}_q \in Q$, we create its positive version denoted as $\mathcal{B}_+$ by introducing two types of noises. As discussed in Section~\ref{subsec:sample}, (1) we shift each ball along a random direction by a random distance, which is controlled by a noise rate; (2) we randomly delete some balls controlled by a drop rate and then add them at random locations. Then for each $\mathcal{B}_q$, we compute the rank of $\mathcal{B}_+$ in the database $D \cup \mathcal{B}_+$ using BL2Vec and other baselines. Ideally, $\mathcal{B}_+$ should be ranked at the top since $\mathcal{B}_+$ is generated from the $\mathcal{B}_q$.

Figure~\ref{fig:effectiveness_similarity} (a) and (b) show the results of varying the noise rate from 0.2 to 0.4 when the drop rate is fixed to 0.2 in terms of
$HR@10$ and $MRR$. We can see that $BL2Vec$ is significantly better than the baselines in terms of $HR@10$ and $MRR$. For example, when applying 40\% noise, BL2Vec outperforms RepSet and DTW (the two best baselines) by 1.5 times and 4.7 times in terms of $HR@10$, respectively, and by more than 6.2 times and 13.3 times in terms of $MRR$, respectively. EMD and DeepSets perform the worst in most of the cases because EMD is based on the idea of point-matching and affected significantly by the noise;
DeepSets is developed for a set that has an unordered nature and therefore it is not suitable for measuring the similarity among billiards layouts, which are order sensitive. %In addition, compared with baselines, our BL2Vec is more robust to noise. For example, when the noise rate increases from 0.2 to 0.4, the $HR@10$ of BL2Vec only decreases by 15\%, while the values of DTW decrease by 80\%.
Figure \ref{fig:effectiveness_similarity} (c) and (d) show the results of varying the drop rate from 0.2 to 0.4 when the noise rate is fixed to 0.2 in terms of
$HR@10$ and $MRR$. BL2Vec still performs the best in terms of both metrics. For example, it outperforms RepSet, which performs best in baselines when the drop rate is 0.4, by more than 1.8 times and 9.6 times in terms of $HR@10$ and $MRR$. %We also note that the performance of BL2Vec gets worse with a larger drop rate. This is because the drop rate may change the form of their relative locations, which affects the similarity.

\noindent \textbf{(2) Effectiveness evaluation (classification).}
Table \ref{tab:tasks} shows the results of different methods in the classification task for predicting the labels of ``clear" and ``not clear". Recall that ``clear" means the player who performs the break shot pocketed all balls and thus win the game, and ``not clear" means the otherwise case.
We train a k-nearest neighbour classifier (kNN) with $k = 10$ for this task and report the average accuracy on 500 billiards layouts. For BL2Vec, we fix the noise rate and drop rate to be 0.2 by default, and use the representations of billiards layouts for computing the distance in the KNN classifier, we do the same for DeepSets, RepSet and WSSET. For EMD, Hausdorff, DTW, Frechet and PM, we use their distances for the KNN classifier directly.
We observe BL2Vec outperforms all the baselines. For example, it outperforms EMD (the best baseline) by 9.5\%. The result indicates that the representation learned from BL2Vec is more useful for learning the target task.

\noindent \textbf{(3) Effectiveness evaluation (clustering).}
We further show the clustering task, which is to group billiards layouts belonging to the same family (i.e., ``clear" or ``not clear") into the same cluster. We use the K-means algorithm with $k = 2$ and report ARI and AMI on the
500 billiards layouts. The results are shown in Table~\ref{tab:tasks}. According to the results, BL2Vec also performs best.
For example, it outperforms EMD (the best baseline) by 18.8\% (resp. 13.3\%) in terms of ARI (resp. AMI). In addition, we observe other baselines have similar ARI and AMI, e.g., the values are between 49\% and 50\%. Similar to the classification task, the results indicate adopting the embeddings of BL2Vec for billiards layout clustering will yield superior performance.

\noindent \textbf{(4) Effectiveness evaluation (parameter study).}
\label{sec:parameter}
We next evaluate the effect of the cell size on the effectiveness of similarity search, classification and clustering for BL2Vec.
Intuitively with the smaller cell size, it generates more tokens and provides a higher resolution of the pool table. However, with the smaller size, it reduces the robustness against the noise for the similarity search task. In Table \ref{tab:parameter}, we report the performance for the similarity search, classification and clustering. We notice that the performance of the similarity search becomes better as the cell size grows and this is in line with our intuition. In addition, we observe the smallest cell size of 10 leads to the worst performance, especially for classification and clustering, because the smallest cell size produces more tokens, which makes the model more difficult to train. In our experiments, we set the cell size to 15 because it provides better performance in general.

\begin{table}[]
\centering
{\extension{\caption {The effect of the cell size.}\label{tab:parameter}}}
%\vspace*{-2mm}
\centering
\setlength{\tabcolsep}{1.8pt}
\begin{tabular}{c|c|cc|c|l|l|l|c|lcl}
\hline
\multirow{2}{*}{Cell size} & \multirow{2}{*}{\#tokens} & \multicolumn{2}{c|}{Similarity} & \multicolumn{4}{c|}{Classification} & \multicolumn{4}{c}{Clustering}                     \\ \cline{3-12}
                           &                           & HR@10           & MRR           & \multicolumn{4}{c|}{Acc (\%)}       & \multicolumn{2}{c}{ARI (\%)} & \multicolumn{2}{c}{AMI (\%)} \\ \hline
10                         &    200                    &  0.92         & 0.240              & \multicolumn{4}{c|}{66.67}               & \multicolumn{2}{c}{52.25}    & \multicolumn{2}{c}{55.80}    \\
15                         &    98                     &  0.94         & 0.252         & \multicolumn{4}{c|}{\textbf{72.62}}               & \multicolumn{2}{c}{\textbf{61.94}}    & \multicolumn{2}{c}{\textbf{63.26}}    \\
20                         &    50                     & 0.95          &0.259          & \multicolumn{4}{c|}{67.33}               & \multicolumn{2}{c}{56.96}    & \multicolumn{2}{c}{58.25}    \\
25                         &    32                     &  0.96          &0.303              & \multicolumn{4}{c|}{70.14}               & \multicolumn{2}{c}{55.21}    & \multicolumn{2}{c}{57.37}    \\
30                         &    28                     &  \textbf{0.96}          &\textbf{0.306}            & \multicolumn{4}{c|}{68.72}               & \multicolumn{2}{c}{53.96}    & \multicolumn{2}{c}{56.28}    \\ \hline

\end{tabular}
%\vspace*{-3mm}
\end{table}

\begin{table}[]
\centering
{\extension{\caption {Effectiveness evaluation with ablation study.}\label{tab:ablation}}}
%\vspace*{-3mm}
%\footnotesize{
\setlength{\tabcolsep}{3pt}
\begin{tabular}{c|cc|c|l|l|l|c|lcl}
\hline
\multirow{2}{*}{Model} & \multicolumn{2}{c|}{Similarity} & \multicolumn{4}{c|}{Classification} & \multicolumn{4}{c}{Clustering}                     \\ \cline{2-11}
                       & HR@10           & MRR           & \multicolumn{4}{c|}{Acc (\%)}       & \multicolumn{2}{c}{ARI (\%)} & \multicolumn{2}{c}{AMI (\%)} \\ \hline
BL2Vec                 & \textbf{0.94}            & \textbf{0.252}              & \multicolumn{4}{c|}{\textbf{72.62}}               & \multicolumn{2}{c}{\textbf{61.94}}    & \multicolumn{2}{c}{\textbf{63.26}}    \\
w/o BS                 & 0.92            & 0.207              & \multicolumn{4}{c|}{62.42}               & \multicolumn{2}{c}{53.50}    & \multicolumn{2}{c}{59.25}    \\
w/o BP                 & 0.11            & 0.009              & \multicolumn{4}{c|}{51.08}               & \multicolumn{2}{c}{49.27}    & \multicolumn{2}{c}{50.18}    \\
w/o BB                 & 0.91            & 0.197              & \multicolumn{4}{c|}{69.52}               & \multicolumn{2}{c}{52.45}    & \multicolumn{2}{c}{61.73}    \\ 
w/o $Net(\cdot)$       &  0.40          & 0.078             & \multicolumn{4}{c|}{41.76}               & \multicolumn{2}{c}{50.38}    & \multicolumn{2}{c}{50.67}   \\ \hline
\end{tabular}%}
%\vspace*{-3mm}
\end{table}

\noindent \textbf{(5) Effectiveness evaluation (ablation study).}
\label{sec:ablation}
To show the importance of each component of the extracted features in our BL2Vec model, including Ball-self (BS), Ball-Pocket (BP) and Ball-Ball (BB), and the ability of CNN for capturing spatial features in $Net(\cdot)$, we conduct an ablation experiment. We denote our BL2Vec model without BS, BP, BB and $Net(\cdot)$ as w/o BS, w/o BP, w/o BB and w/o $Net(\cdot)$, respectively. Table~\ref{tab:ablation} compares the different versions for similarity search, classification and clustering tasks. Overall, all these components help improve the effectiveness of the BL2Vec model and enable the model to achieve superior performance than baselines. A detailed discussion on the effectiveness of each component is given as follows.
(1) BS features can effectively capture the spatial information of each ball; if the features are omitted, the classification performance drops significantly by around 14\%. %{\color{blue}
(2) BP features capture the correlations between the ball and each pocket, and these are the unique characteristics in the billiards layout data. In addition, BP features are associated with the most tokens (i.e., $4 \times 6 =24$) in the total of 27 tokens to represent each ball as discussed in Section~\ref{task:blcnn}, and therefore the BP features contribute the most for all of the tasks. As expected, we observe that if the BP features are omitted, the performance of similarity search drops by 88\% (resp. 96\%) in terms of HR@10 (resp. MRR); that of classification drops by 30\% in terms of accuracy; that of clustering drops by 20\% (resp. 21\%) in terms of ARI (resp. AMI). (3) BB features capture the correlation between balls in the layout. The features also affect the overall performance. For example, if the BB features are omitted, the clustering performance drops a lot by around 15\% (resp. 2\%) in terms of ARI (resp. AMI). (4) The CNN in $Net(\cdot)$ can well capture the correlation of those spatial features (i.e., BS, BP and BB), and simply using the average of feature embeddings cannot lead to the superior performance, e.g., similarity search drops by 57.4\% (resp. 69.05\%) in terms of HR@10 (resp. MRR).

\begin{figure}[ht]
%\footnotesize
%\vspace*{-1mm}
\centering
\begin{tabular}{c c}
%   \begin{minipage}{8.5cm}
%     \includegraphics[width=17.2cm]{figures/running_time_legend.pdf}
%   \end{minipage}
%   &
    \\
   \begin{minipage}{0.4\linewidth}
    \includegraphics[width=\linewidth]{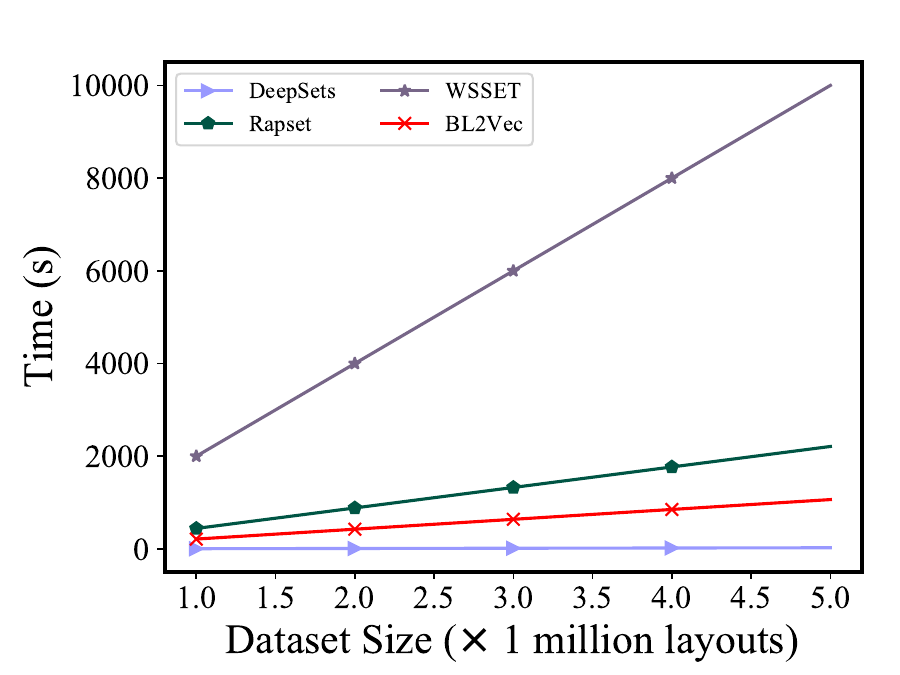}
    \end{minipage}
    &
    \begin{minipage}{0.4\linewidth}
    \includegraphics[width=\linewidth]{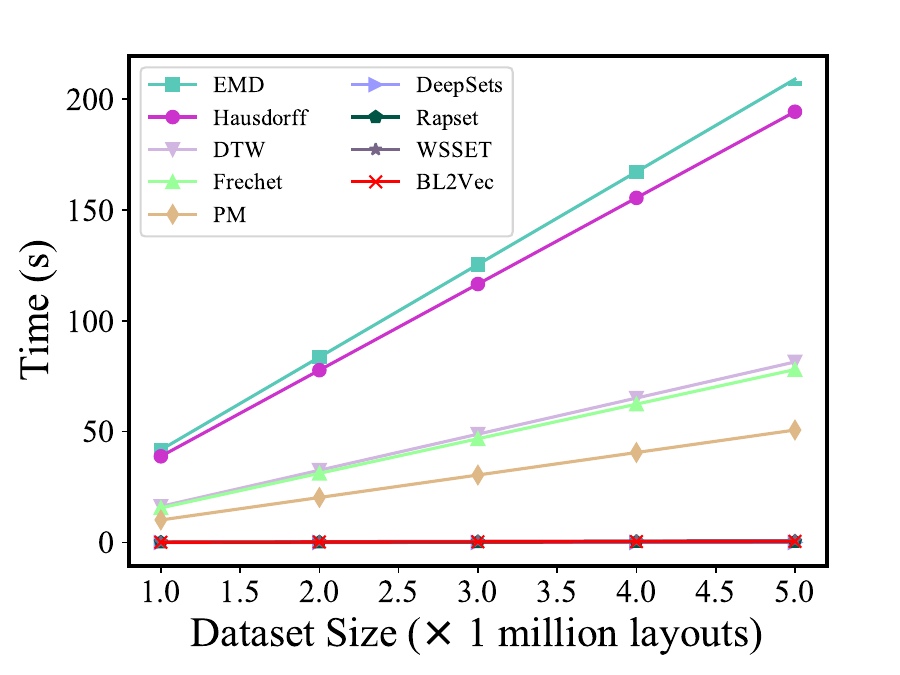}
    \end{minipage}
    \\
    (a) Embedding scalability 
    &
    (b) Search scalability
\end{tabular}
%\vspace*{-2mm}
{\extension{\caption{Scalability test.}\label{fig:efficiency}
}}
\vspace*{-2mm}
\end{figure}
\noindent \textbf{(6) Scalability test.}
\label{sec:efficiency}
%In this part, we evaluate the efficiency of different methods for the billiards layout retrieval over 100 queries. 
To test scalability, we generate more layouts by adding noise on the original dataset.
{\ICDEComment{Figure \ref{fig:efficiency}(a) reports the running time of embedding the database from 1 million layouts to 5 million layouts for the learning-based methods, i.e., DeepSets, WSSET, Rapset and BL2Vec.
}}
Figure \ref{fig:efficiency}(b) reports the average running time over 10 queries for computing the similarity between a query billiards layout and the billiards layouts when we vary the size of the target database from 1 million layouts to 5 million layouts. 

{\ICDEComment{For embedding time, we notice all learning-based methods have their running times linearly increase with the dataset size, which is consistent with their linear time complexity for embedding layouts. We observe DeepSets model is fastest for the embedding, because the model is light via embedding each point into a vector, and aggregating all embeddings together to an overall embedding of the pointsets, 
which can be accomplished very quickly. Overall, BL2Vec runs comparably fast for the layout embedding, compared with other learning-based  methods.
}}

For searching time, EMD and Hausdorff perform extremely slow, which match their quadratic time complexity.
We observe the running time of DTW and Frechet is smaller than that of EMD and Hausdorff though all of them have the same time complexity. This is because DTW and Frechet compute the similarity via dynamic programming to find an
optimal pairwise point-matching.
%
%{\color{blue}
PM is faster than DTW and Frechet because it matches the balls peer-to-peer based on their numbers and the complexity is linear.%}
In addition, DeepSets, WSSET, Rapset and BL2Vec have a similar running time.
Specifically, they run up to 420 times faster than EMD and Hausdorff, 160 times faster than DTW and Frechet and 100 times faster than PM. Moreover, we notice that DeepSets, WSSET, Rapset and BL2Vec scale linearly with the dataset size and the disparity between them increases as the size of the target database grows. This is because all of the learning-based methods have linear time complexity to compute the similarity by embedding layouts to vectors, and the embedding process can also be done offline, which is consistent with their time complexities.
%the discussion in Section~\ref{subsec:complexity}.

\begin{figure}
\centering
\begin{tabular}{c c}
   \begin{minipage}{0.4\linewidth}
    \includegraphics[width=\linewidth]{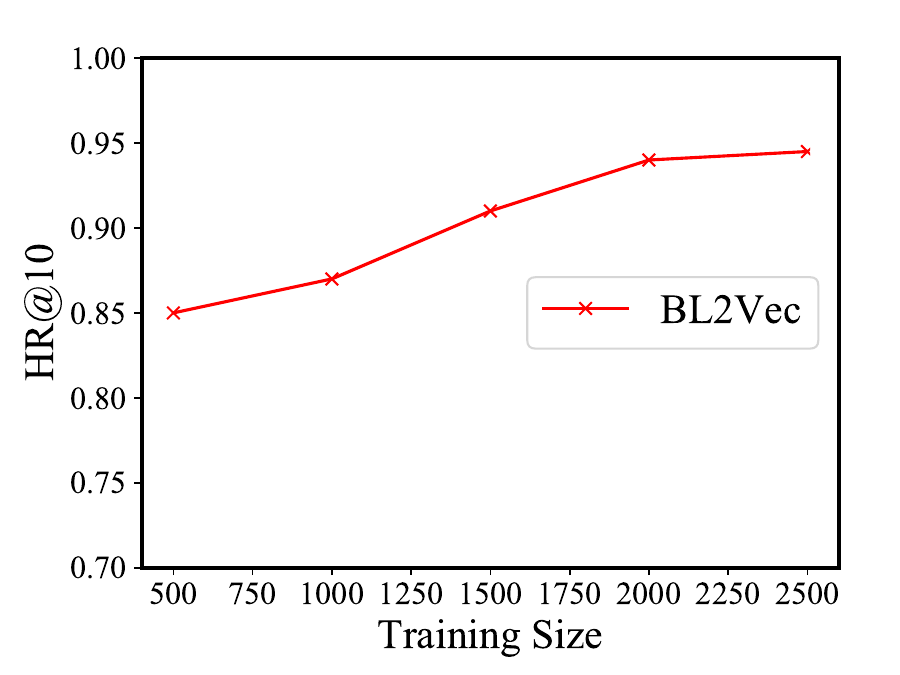}
    \end{minipage}
    &
    \begin{minipage}{0.4\linewidth}
    \includegraphics[width=\linewidth]{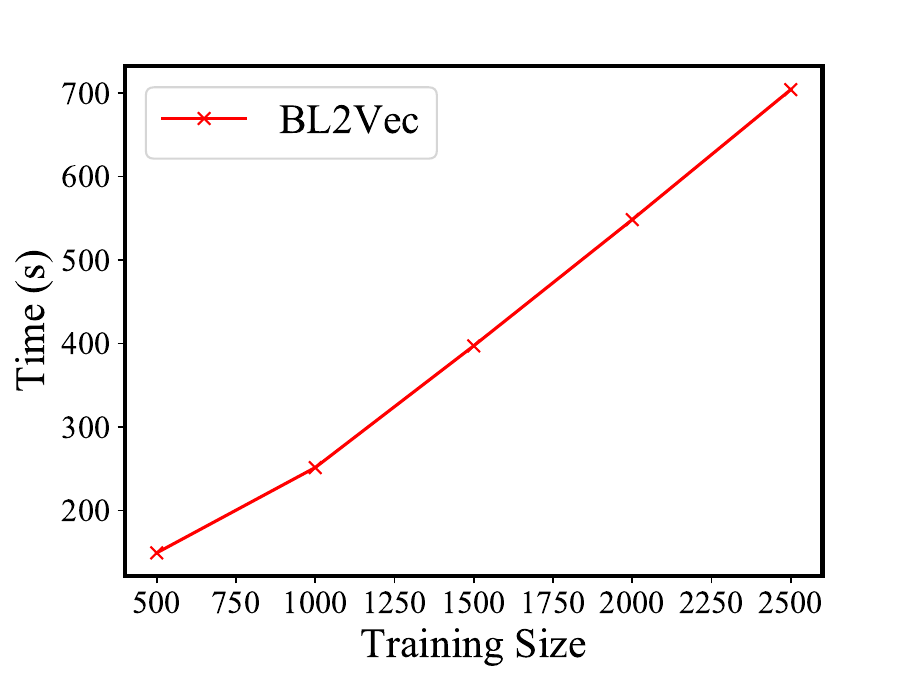}
    \end{minipage}
    \\
    (a) HR@10
    &
    (b) Time cost 
\end{tabular}
%\vspace*{-2mm}
{\extension{\caption{Training cost.}
\label{fig:traing}}}
\vspace*{-2mm}
\end{figure}

\noindent \textbf{(7) Training time.}
\label{sec:training}
{\ICDEComment{The training time of BL2Vec is shown in Figure~\ref{fig:traing}. In particular, we study how the number of training samples affects the model performance for similarity search (i.e., $HR@10$) and the training time cost. We randomly 5 training sets from the dataset, including 500, 1,000, 1,500, 2,000, 2,500 billiards layouts, respectively. For each training set, we report
its training cost per epoch and the corresponding effectiveness on 
the testing set with 500 layouts, where we follow the default setup in Section~\ref{sec:setup}. 
Based on the results, we observe the effectiveness improves as the training size increases, and it approximately converges when the training size exceeds 2,000. For the training time, it increases almost linearly with the training size as expected. We use the setting of 70\% (i.e., 2,000) billiards layouts for training, since it provides a reasonable trade-off between effectiveness and training cost.}}

\begin{figure}[ht]
    \centering
    \includegraphics[width=0.5\linewidth]{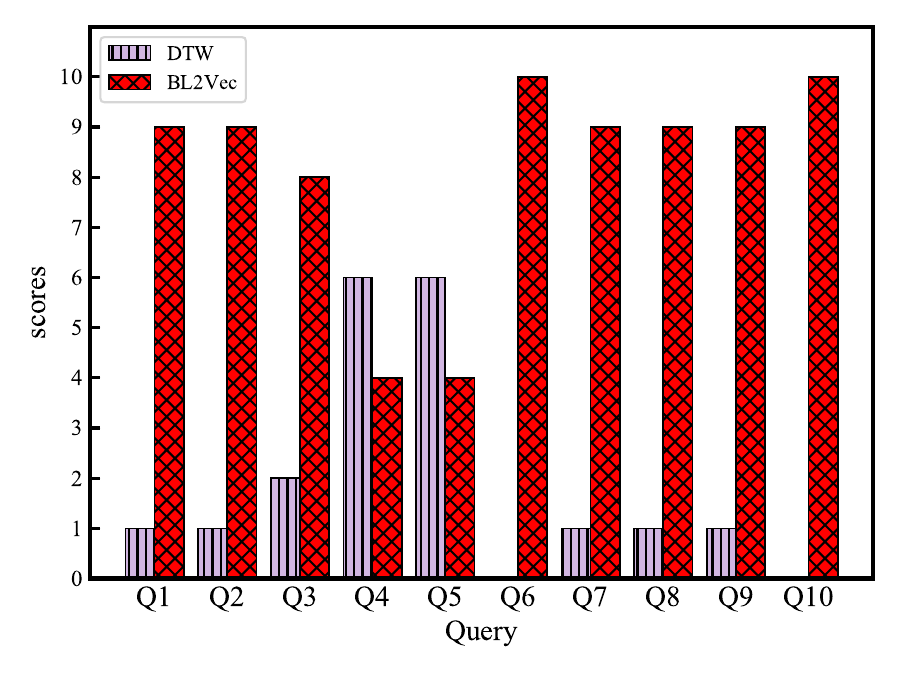}
%\vspace*{-4mm}
{\extension{\caption{User study.}
\label{fig:userstudy}}}
%\vspace*{-4mm}
\end{figure}

\begin{figure*}
\centering
\begin{tabular}{c c c c c}
  \begin{minipage}{0.18\linewidth}
  \includegraphics[width=\linewidth]{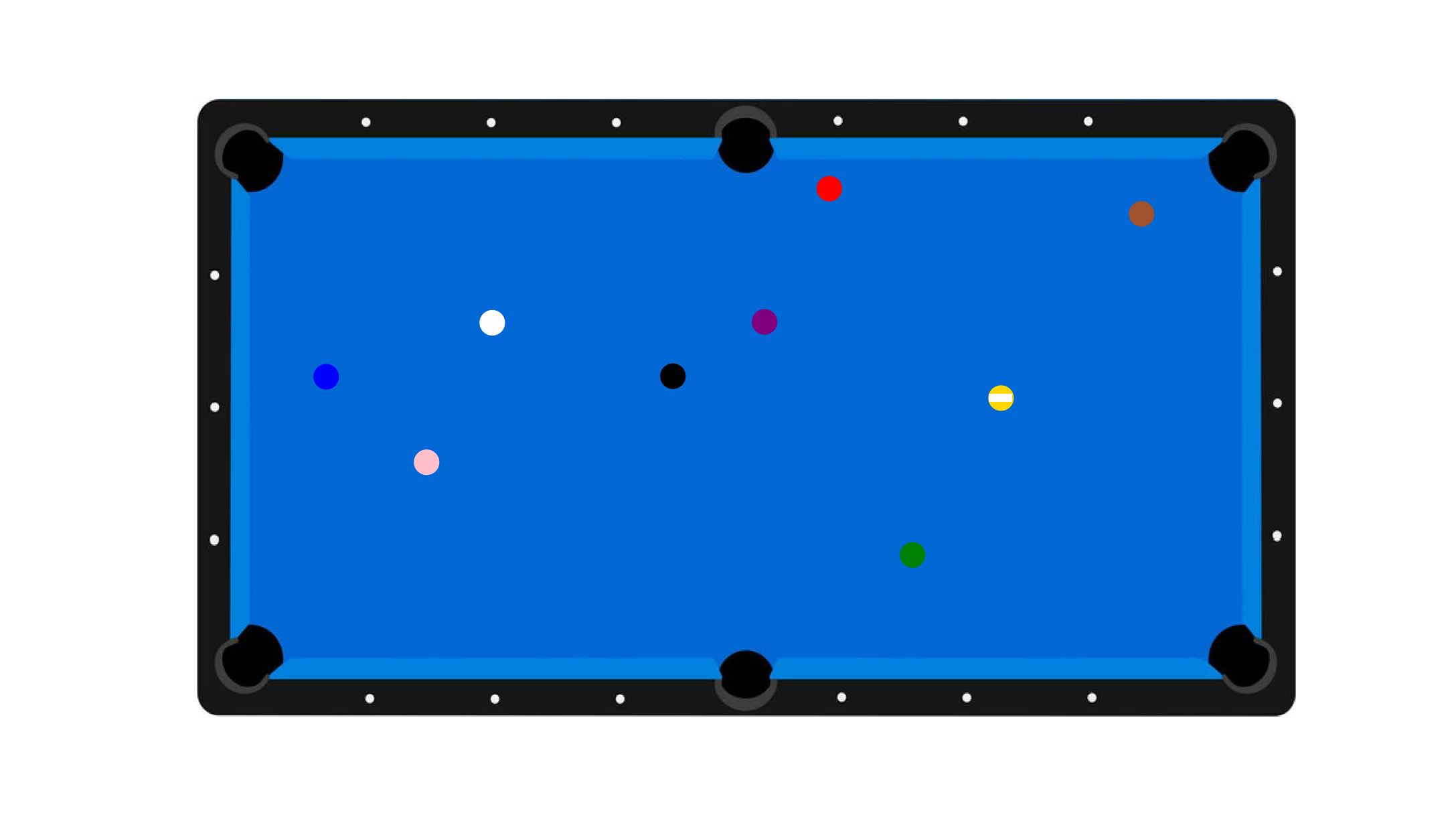}%
  \end{minipage}
  &
  \begin{minipage}{0.18\linewidth}
    \includegraphics[width=\linewidth]{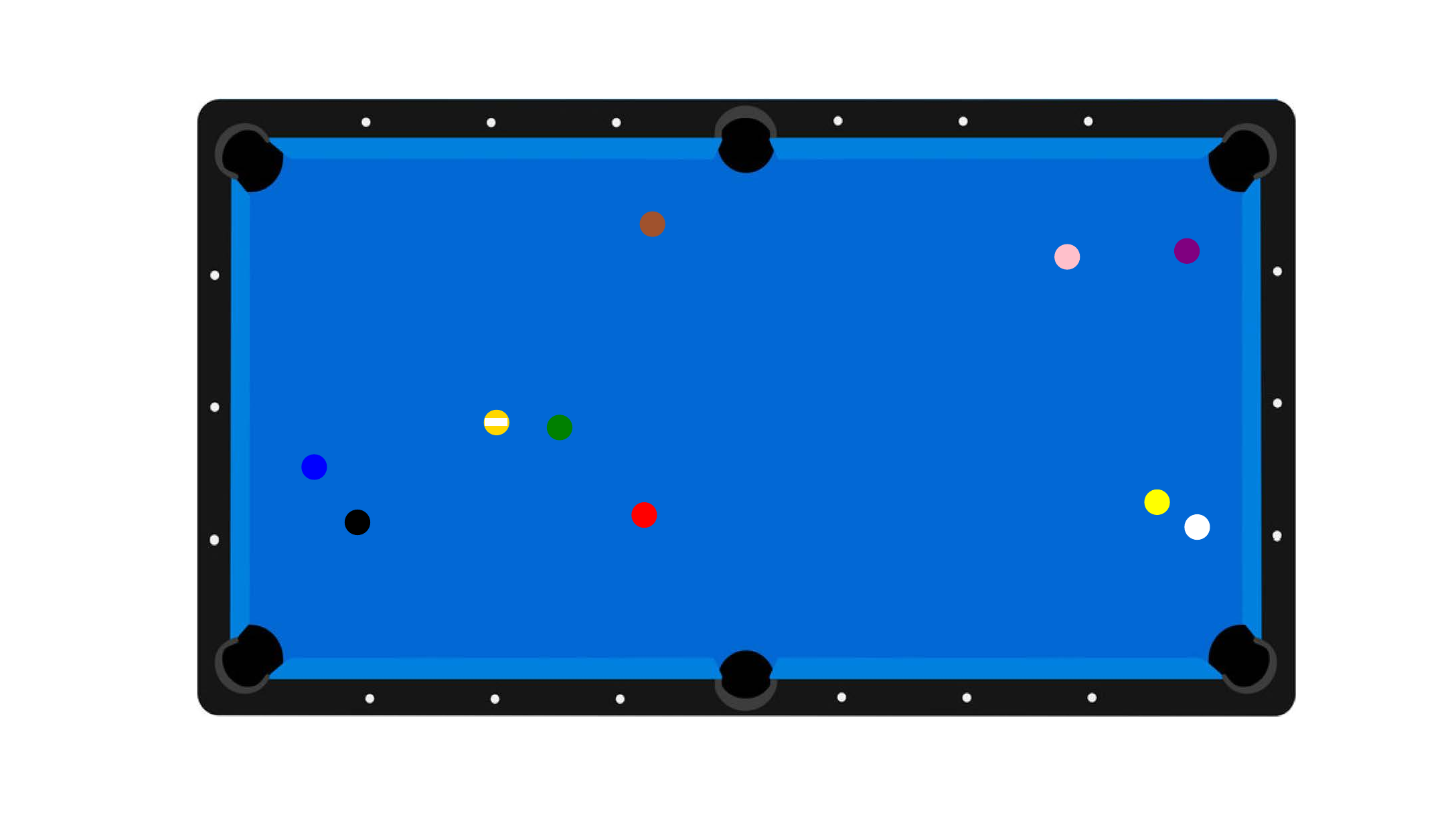}%
  \end{minipage}
  &
  \begin{minipage}{0.18\linewidth}
  \includegraphics[width=\linewidth]{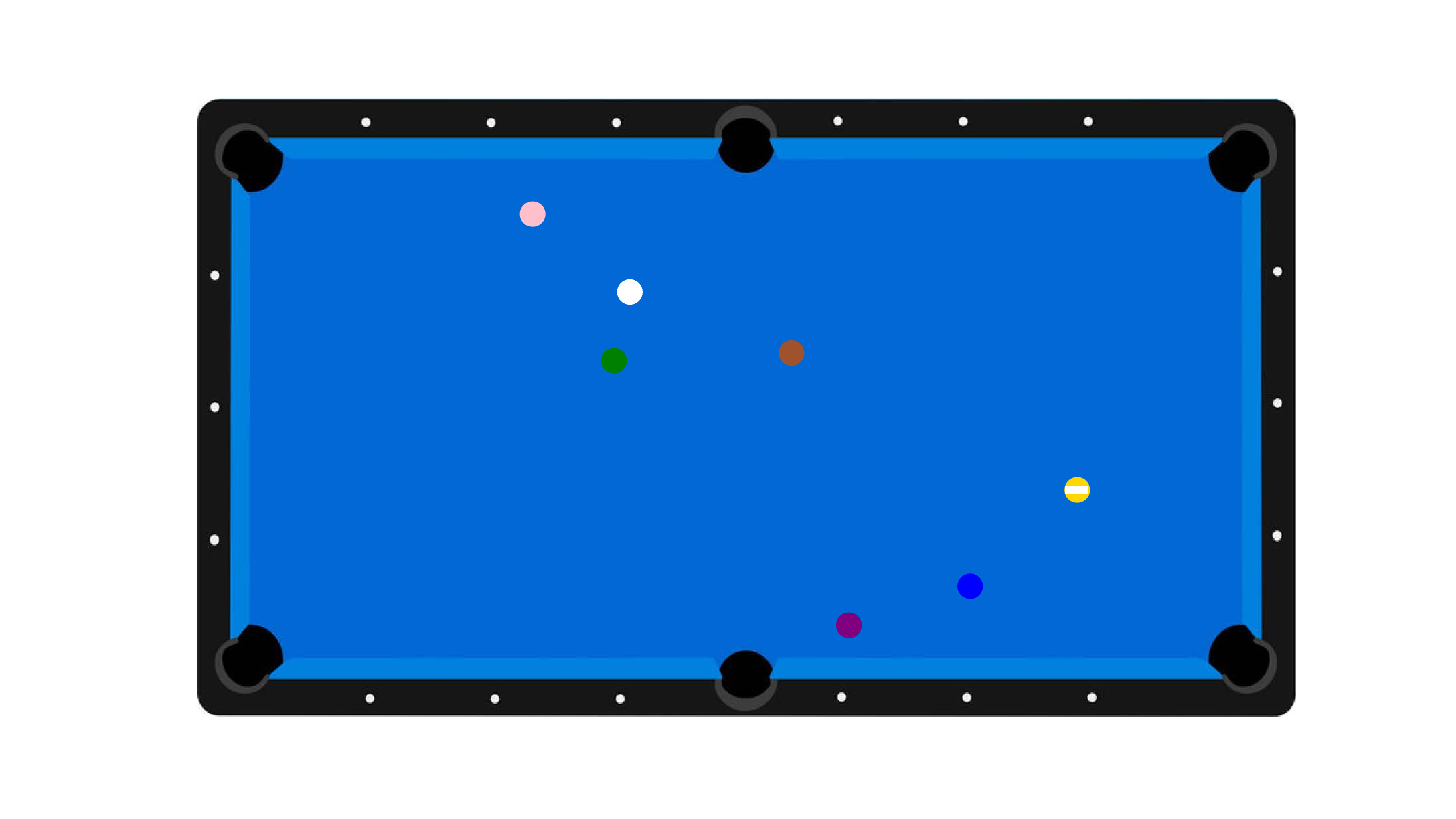}%
  \end{minipage}
  &
  \begin{minipage}{0.18\linewidth}
    \includegraphics[width=\linewidth]{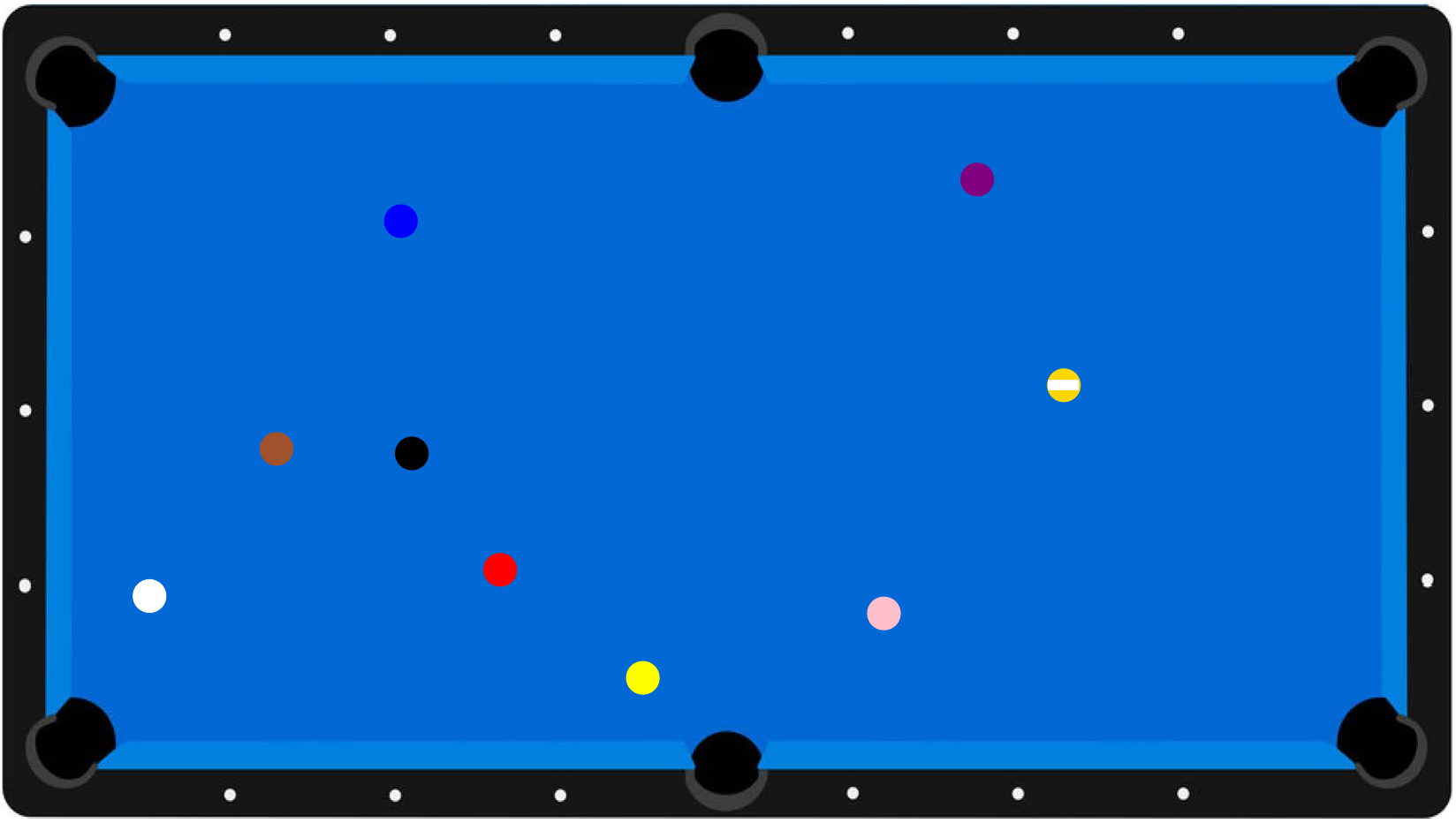}%
  \end{minipage}
  &
  \begin{minipage}{0.18\linewidth}
    \includegraphics[width=\linewidth]{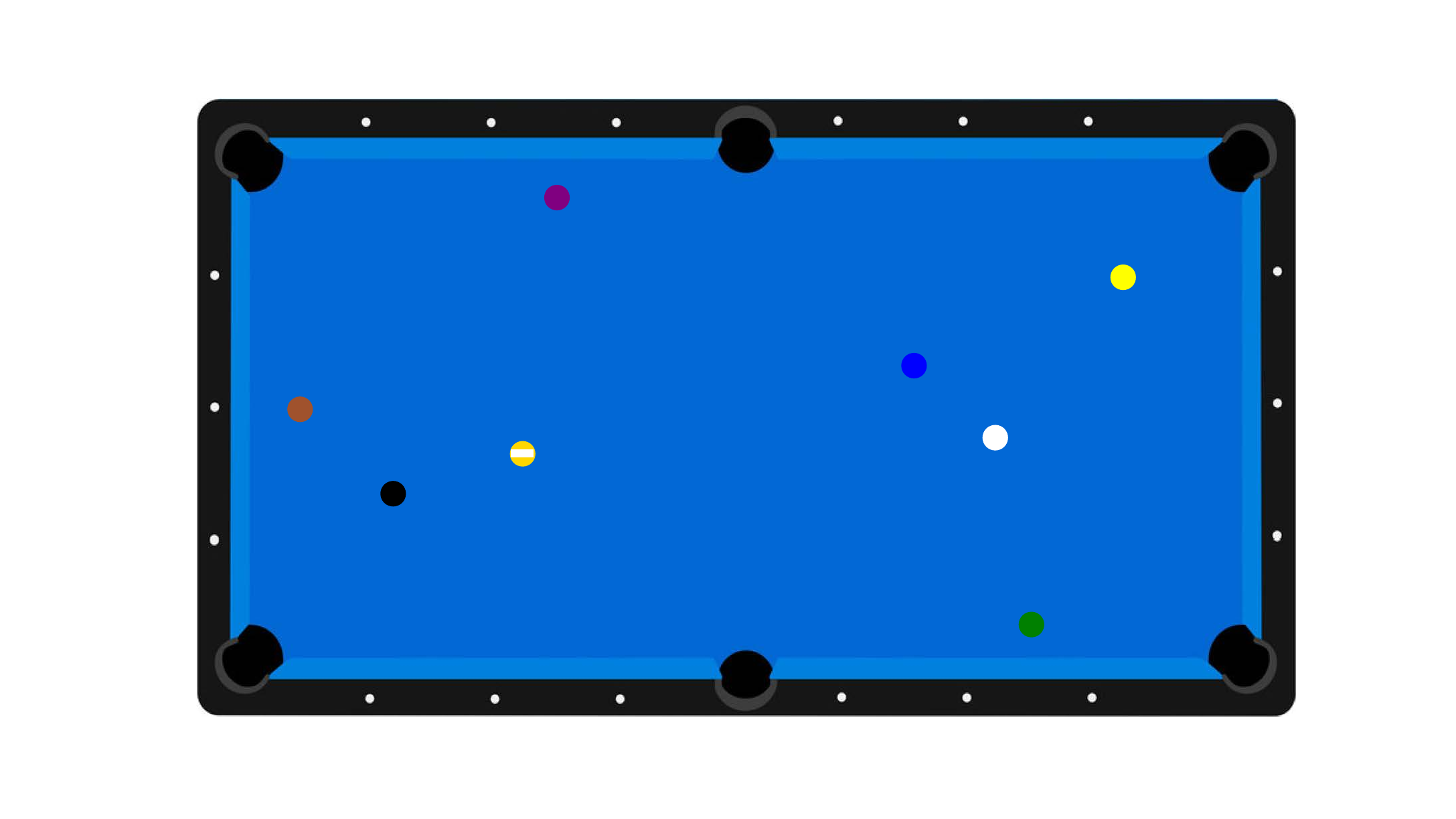}%
  \end{minipage}
  \\
  Q1
  &
  Q2
  &
  Q3
  &
  Q4
  &
  Q5
  \\
  \begin{minipage}{0.18\linewidth}
  \includegraphics[width=\linewidth]{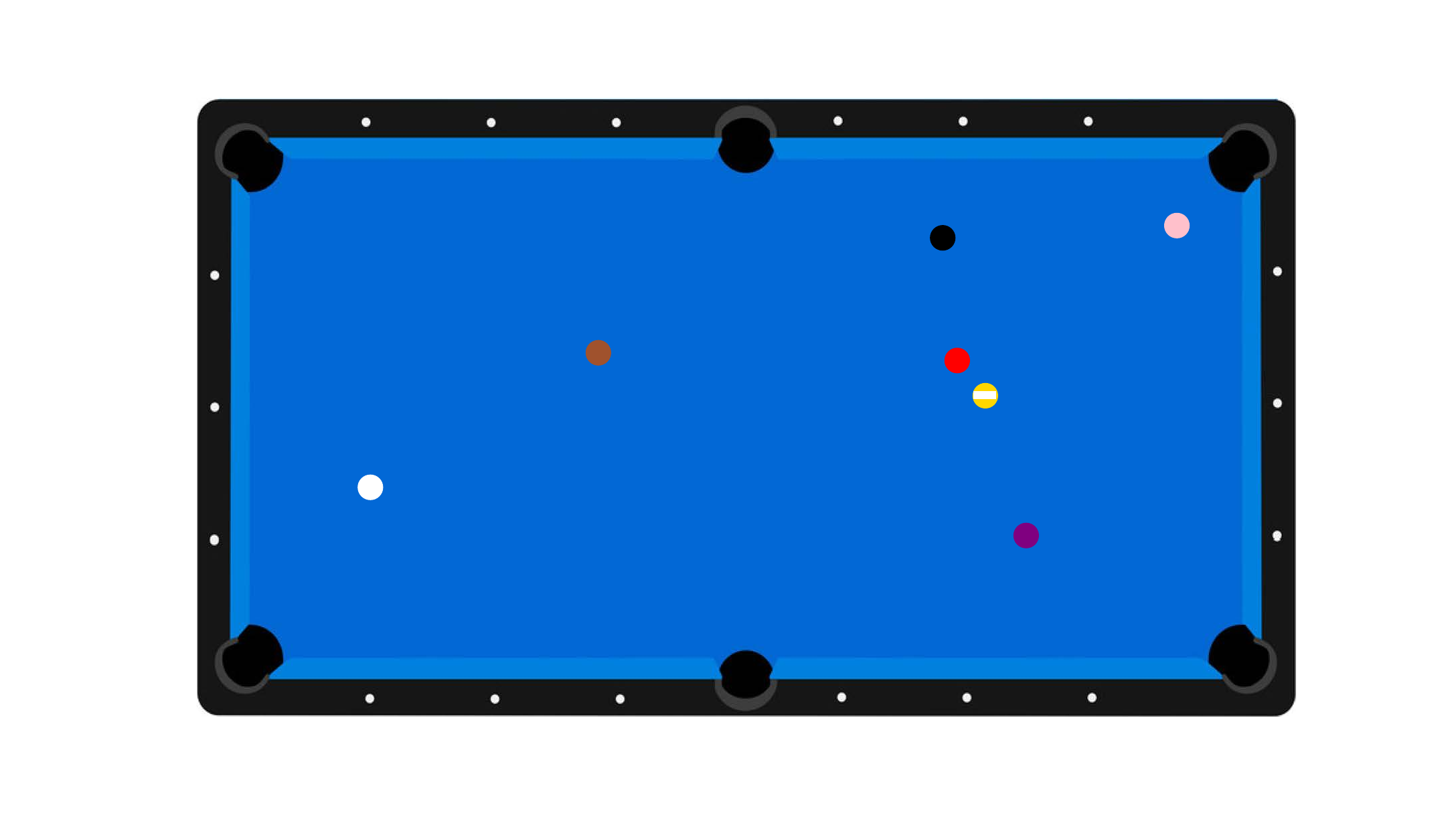}%
  \end{minipage}
  &
  \begin{minipage}{0.18\linewidth}
    \includegraphics[width=\linewidth]{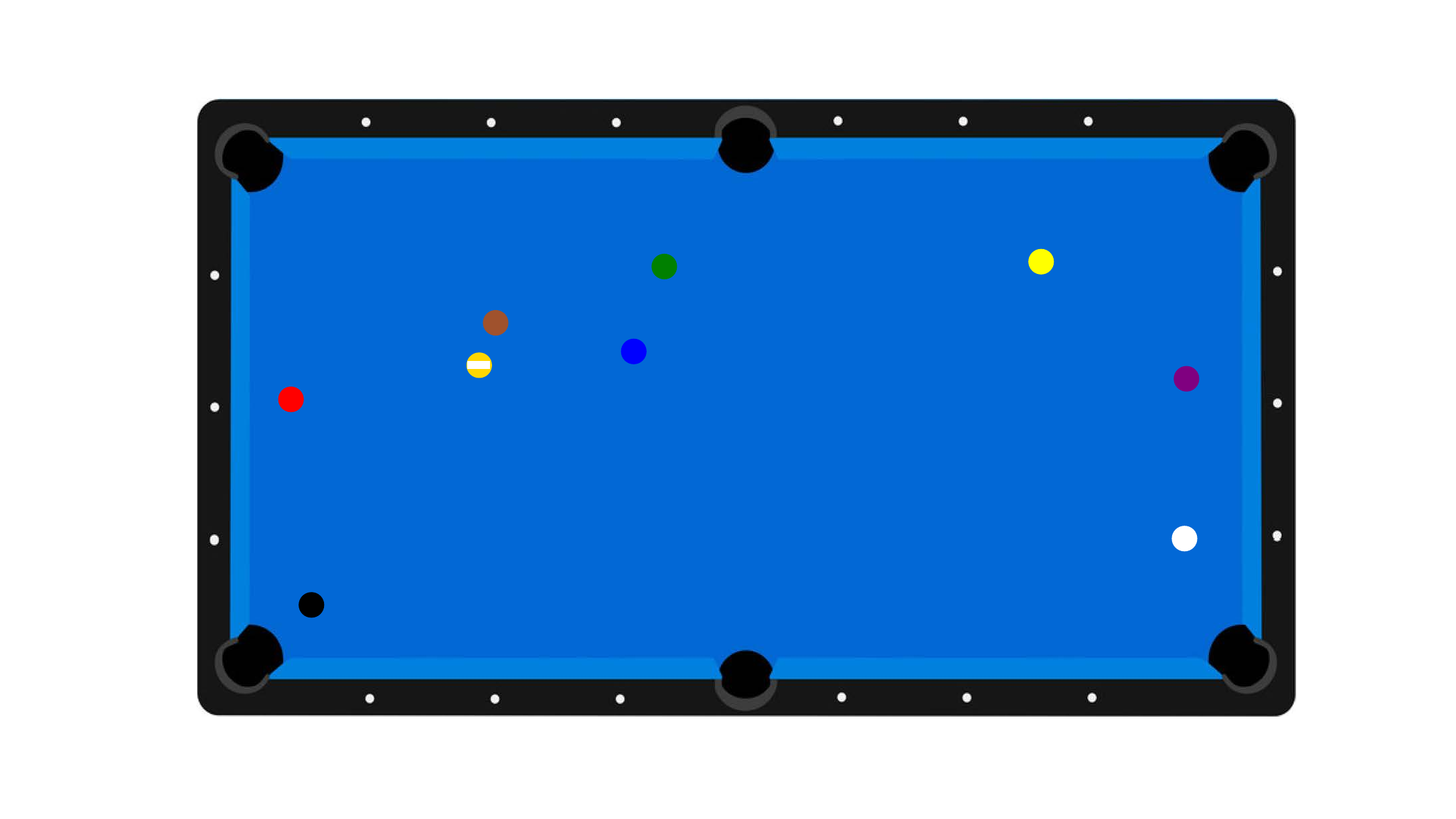}%
  \end{minipage}
  &
  \begin{minipage}{0.18\linewidth}
  \includegraphics[width=\linewidth]{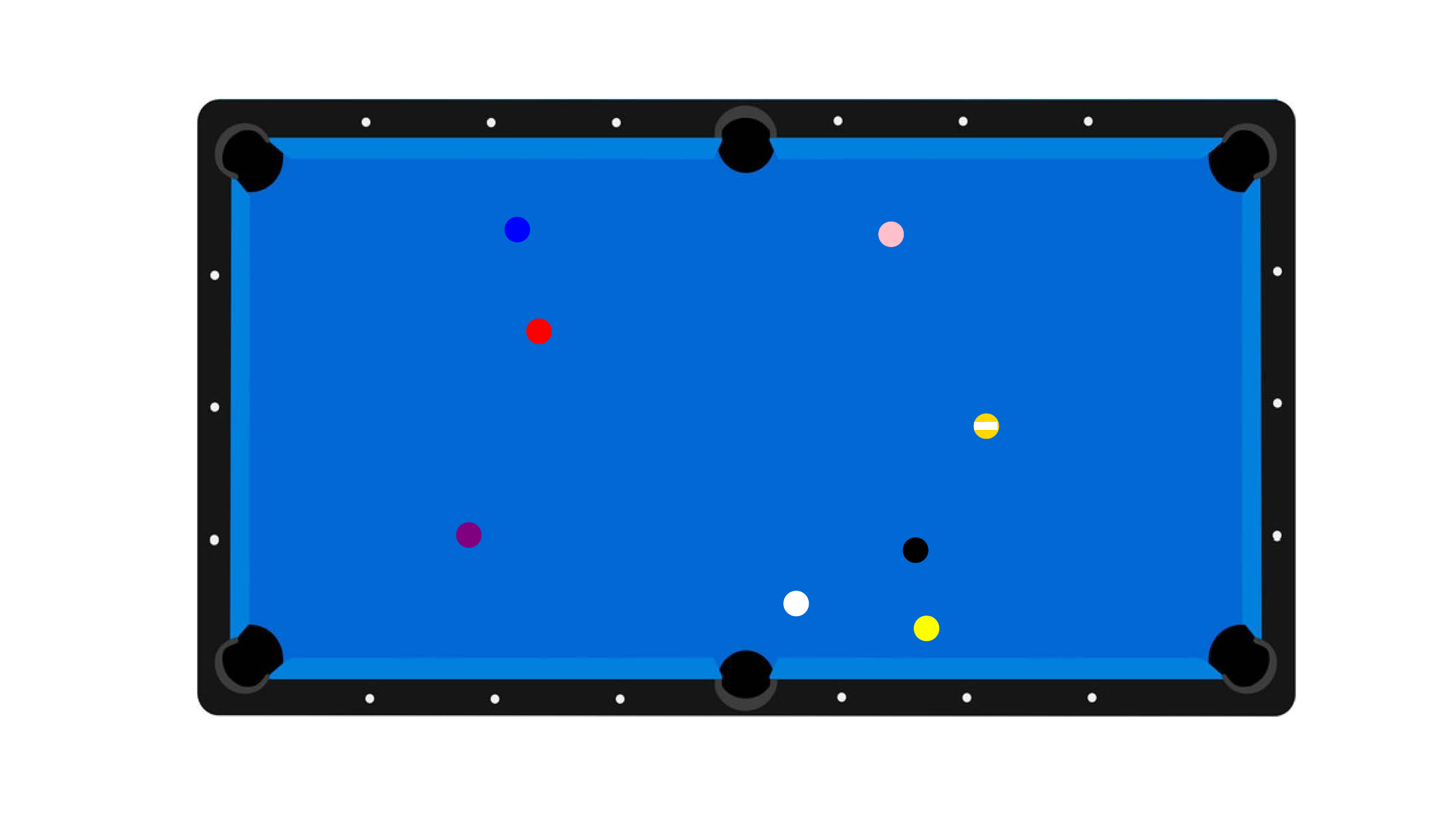}%
  \end{minipage}
  &
  \begin{minipage}{0.18\linewidth}
    \includegraphics[width=\linewidth]{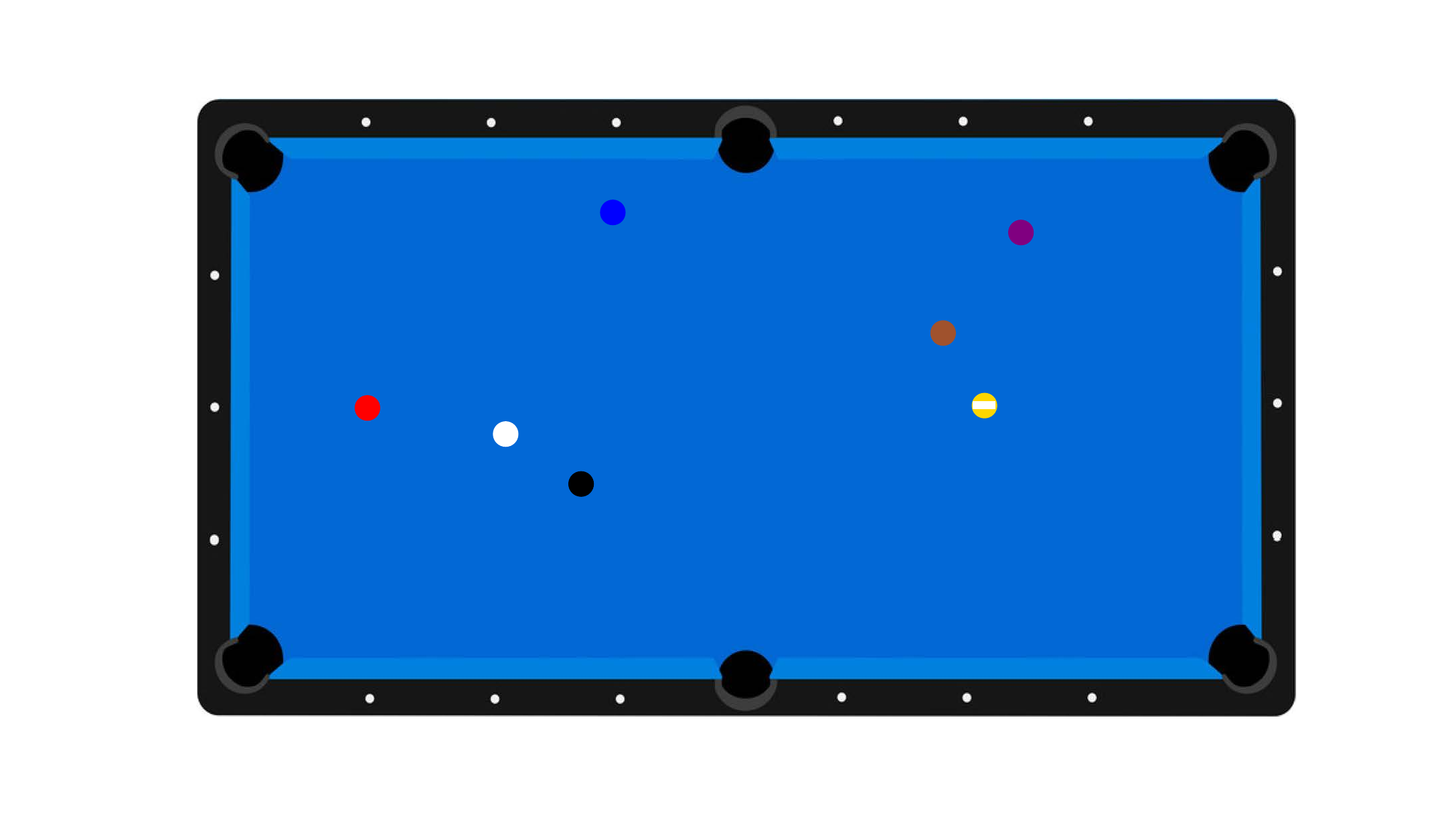}%
  \end{minipage}
  &
  \begin{minipage}{0.18\linewidth}
    \includegraphics[width=\linewidth]{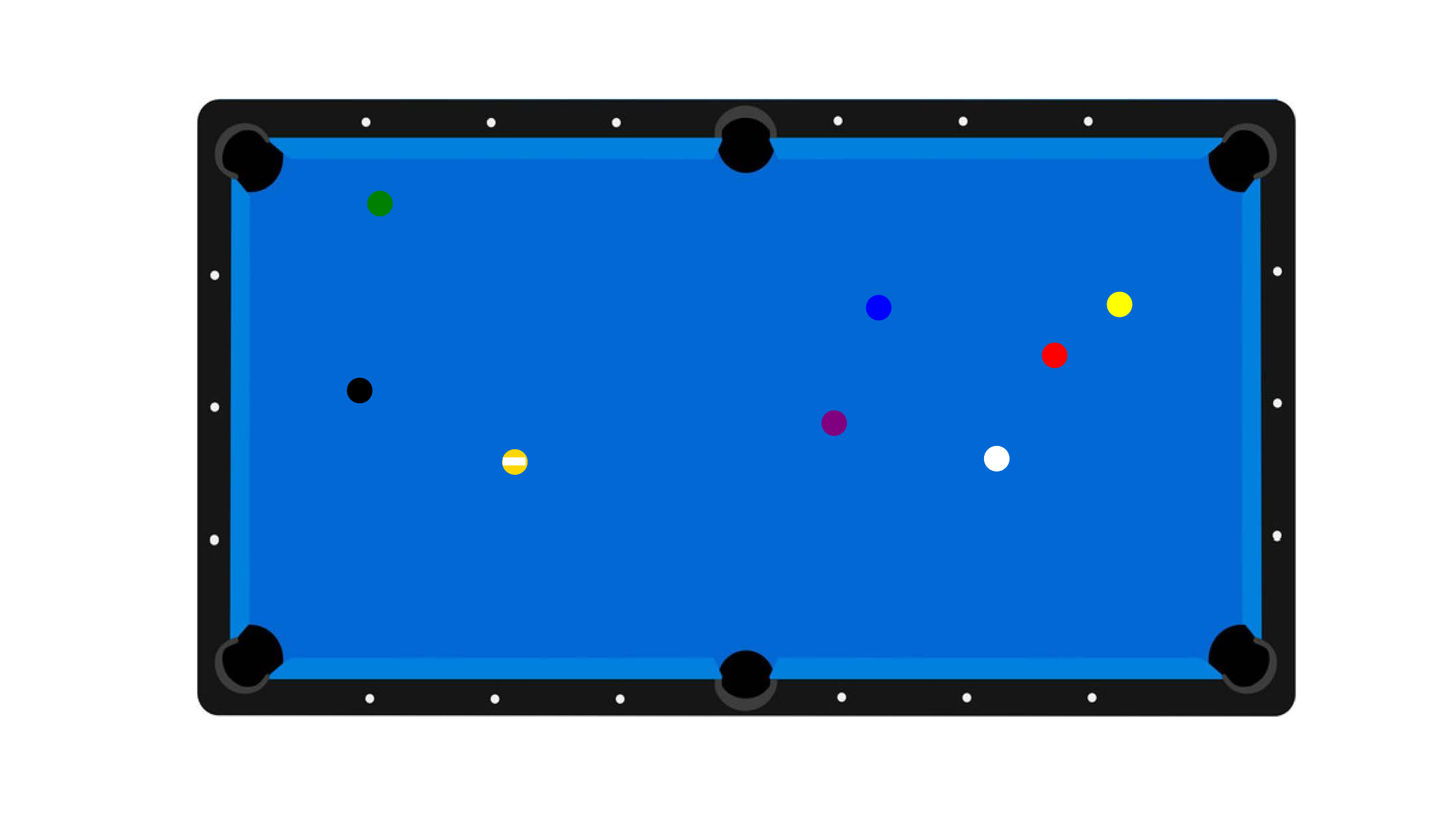}%
  \end{minipage}
  \\
  DTW 1 (1/10)
  &
  DTW 2 (2/10)
  &
  DTW 3 (1/10)
  &
  DTW 4 (6/10)
  &
  DTW 5 (6/10)
  \\
  \begin{minipage}{0.18\linewidth}
  \includegraphics[width=\linewidth]{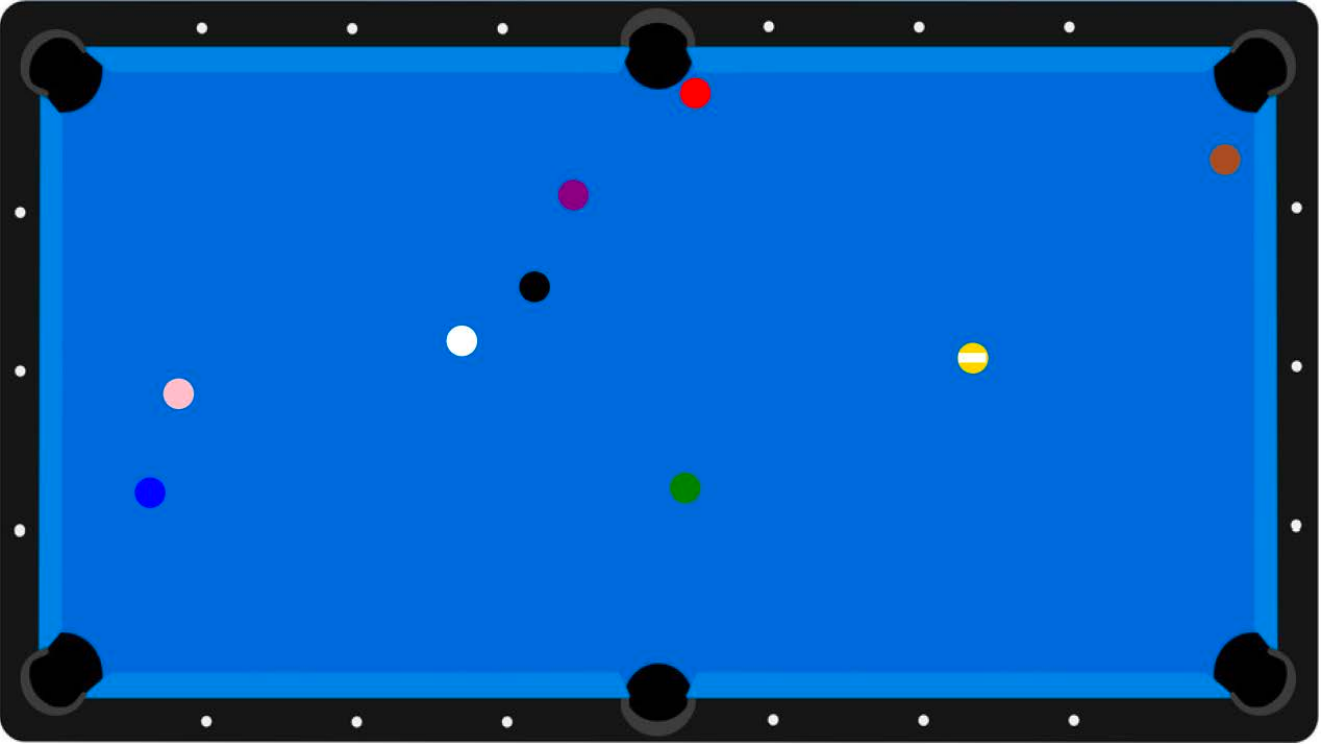}%
  \end{minipage}
  &
  \begin{minipage}{0.18\linewidth}
    \includegraphics[width=\linewidth]{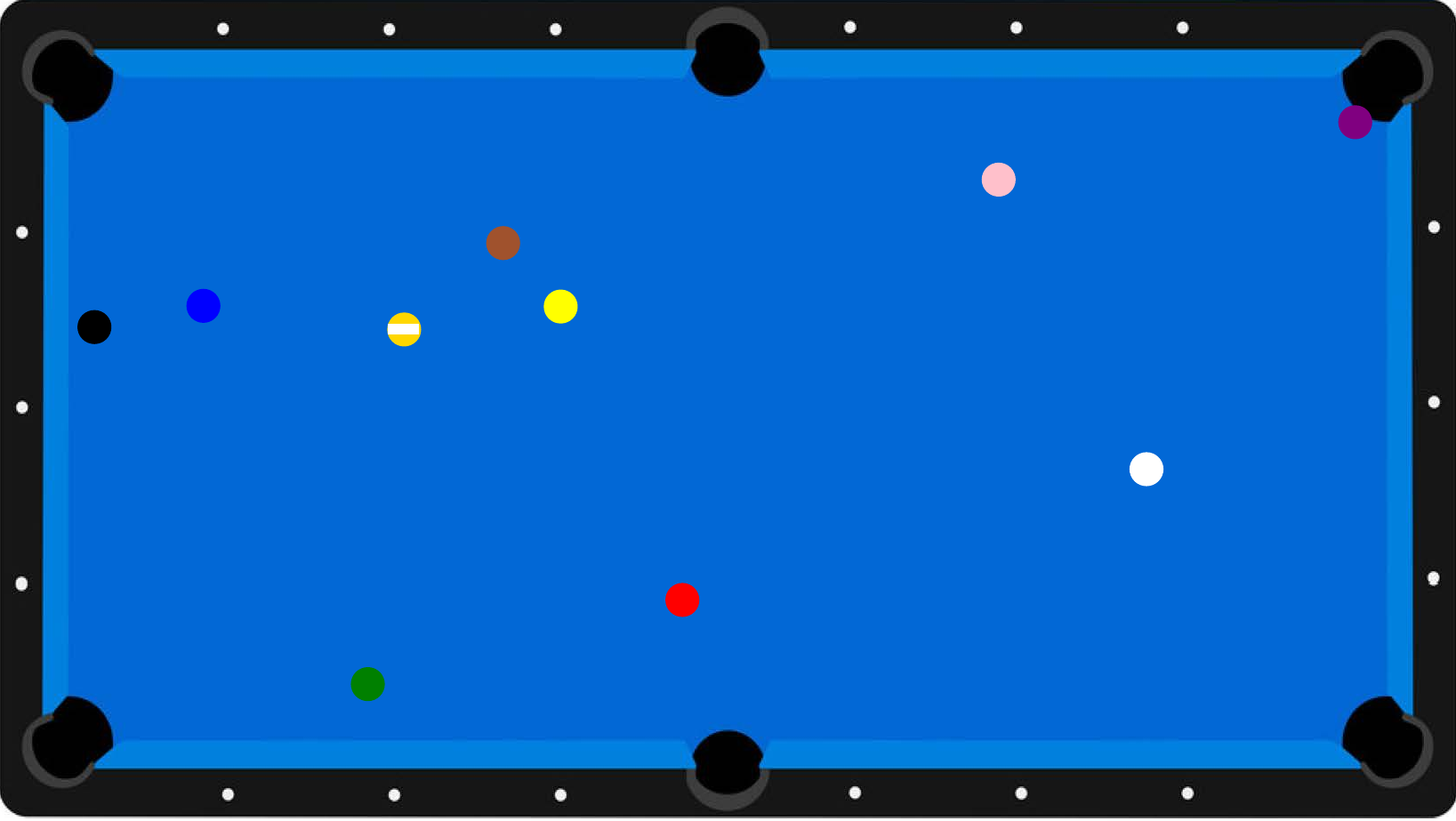}%
  \end{minipage}
  &
  \begin{minipage}{0.18\linewidth}
  \includegraphics[width=\linewidth]{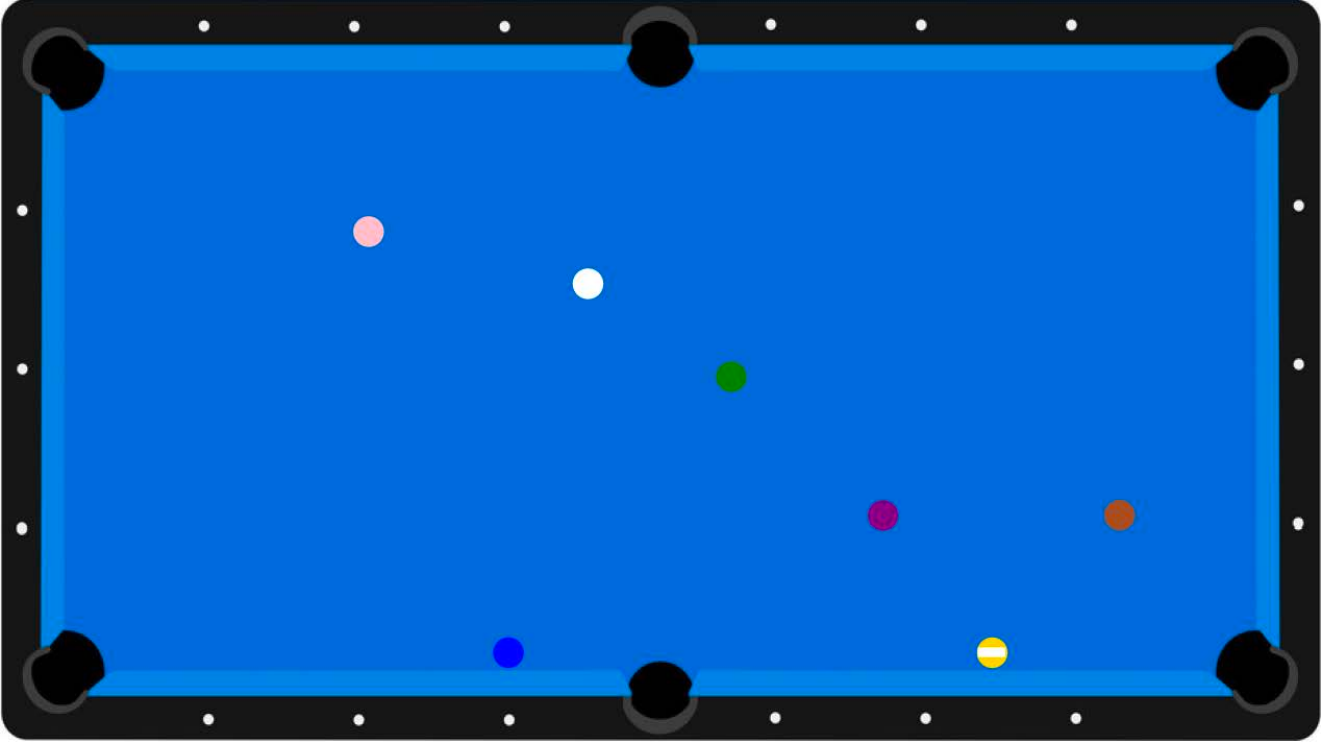}%
  \end{minipage}
  &
  \begin{minipage}{0.18\linewidth}
    \includegraphics[width=\linewidth]{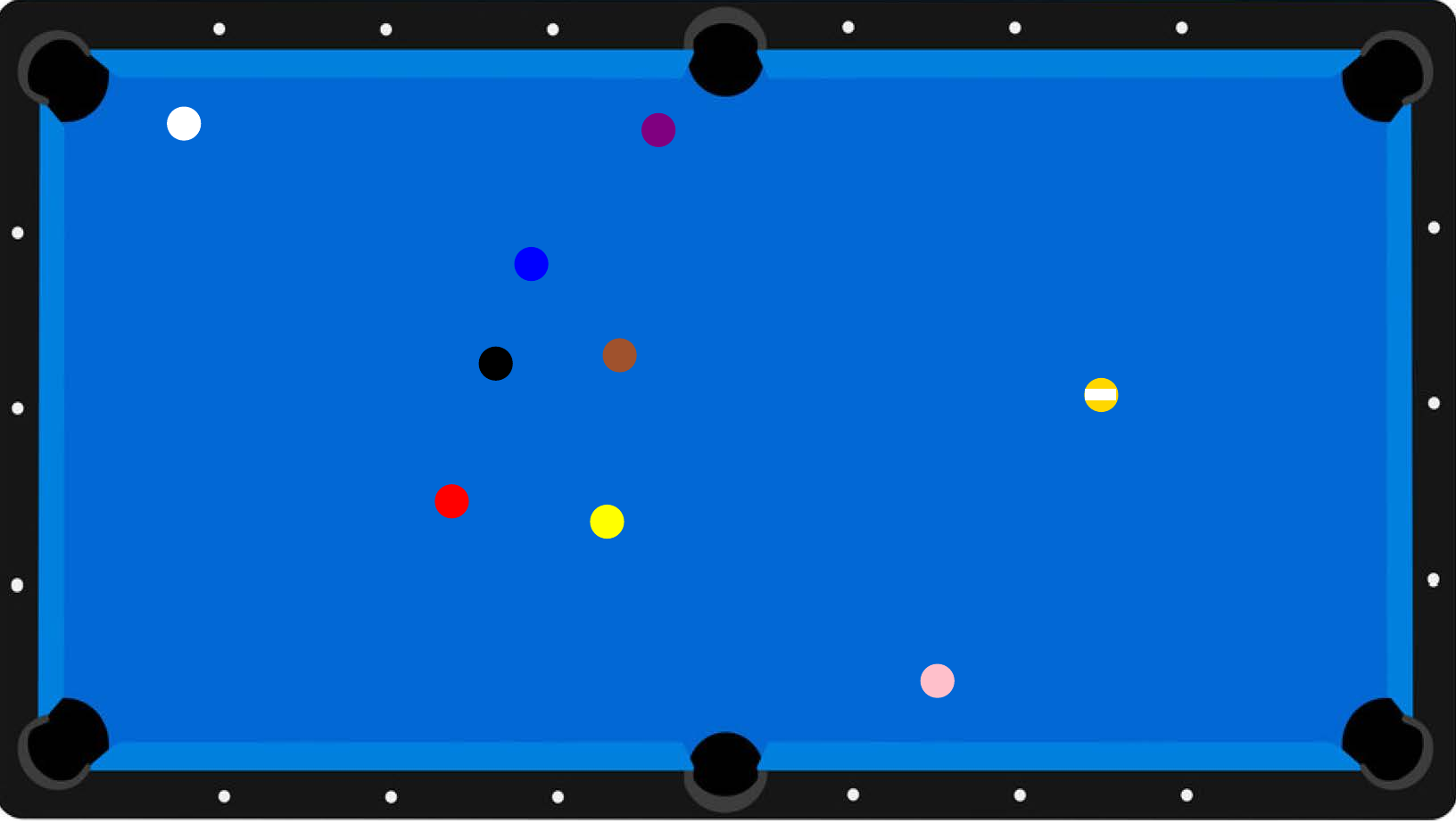}%
  \end{minipage}
  &
  \begin{minipage}{0.18\linewidth}
    \includegraphics[width=\linewidth]{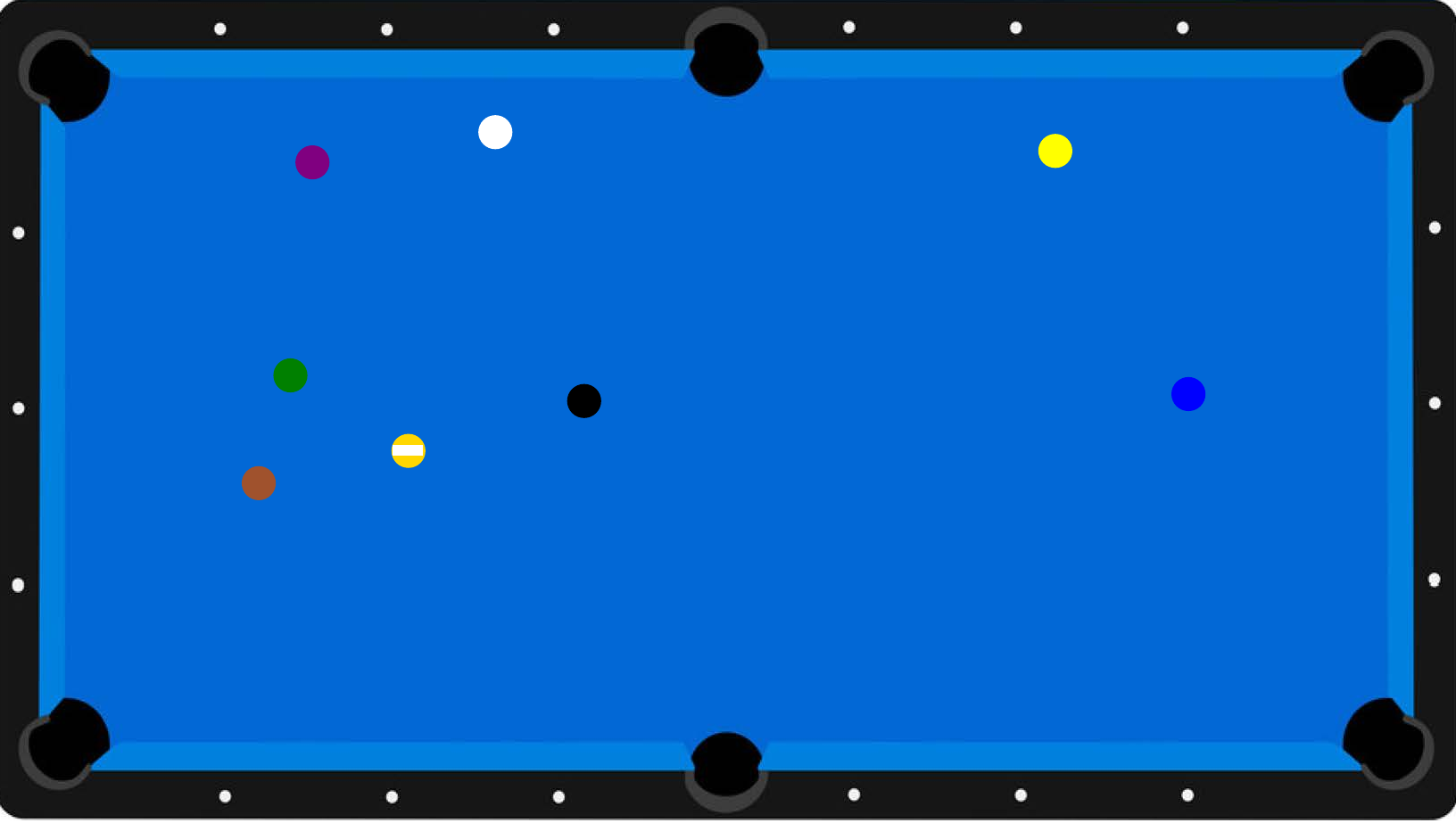}%
  \end{minipage}
  \\
  BL2Vec 1 (9/10)
  &
  BL2Vec 2 (8/10)
  &
  BL2Vec 3 (9/10)
  &
  BL2Vec 4 (4/10)
  &
  BL2Vec 5 (4/10)
\end{tabular}
%\vspace*{-3mm}
{\extension{\caption{Top-1 billiards layouts for Q1-Q5 returned by DTW and BL2Vec, (1/10) means the 1 user supports this result among the total of 10 users.}
%\vspace{-4mm}
\label{q1toq5}}}
%\vspace*{-4mm}
\end{figure*}

\begin{figure*}
\centering
\begin{tabular}{c c c c c}
 \begin{minipage}{0.18\linewidth}
 \includegraphics[width=\linewidth]{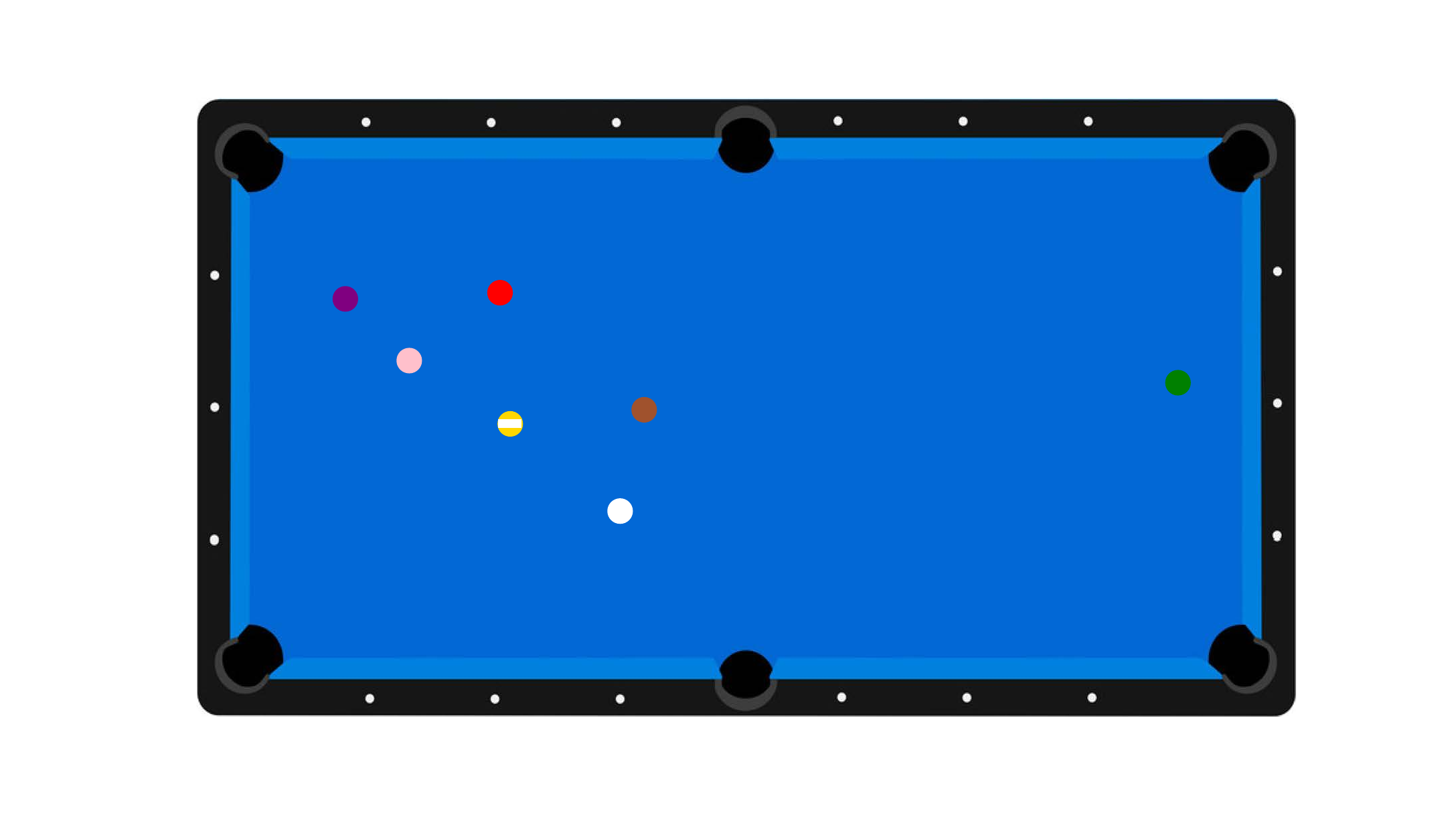}%
 \end{minipage}
 &
 \begin{minipage}{0.18\linewidth}
    \includegraphics[width=\linewidth]{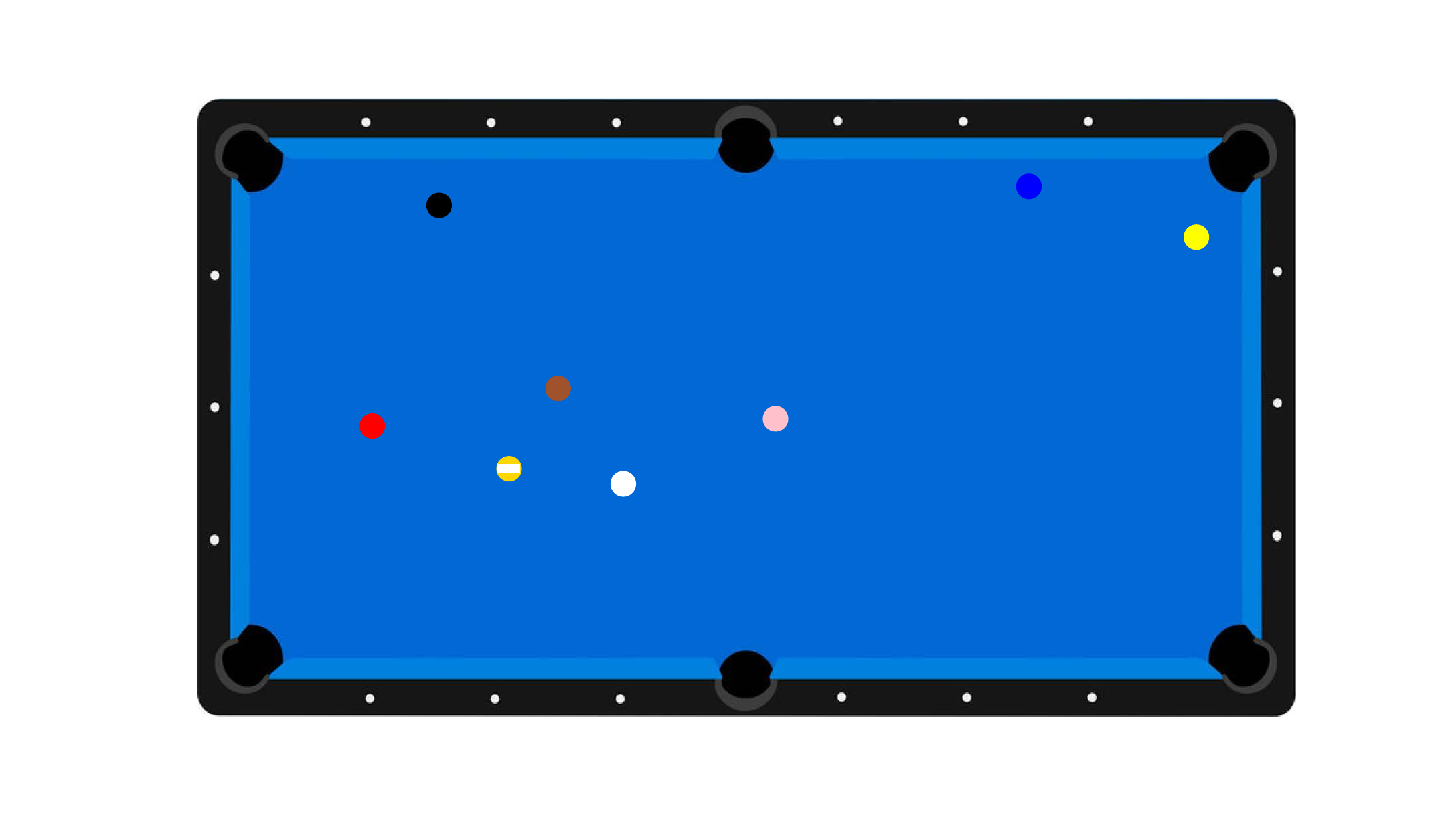}%
 \end{minipage}
 &
 \begin{minipage}{0.18\linewidth}
 \includegraphics[width=\linewidth]{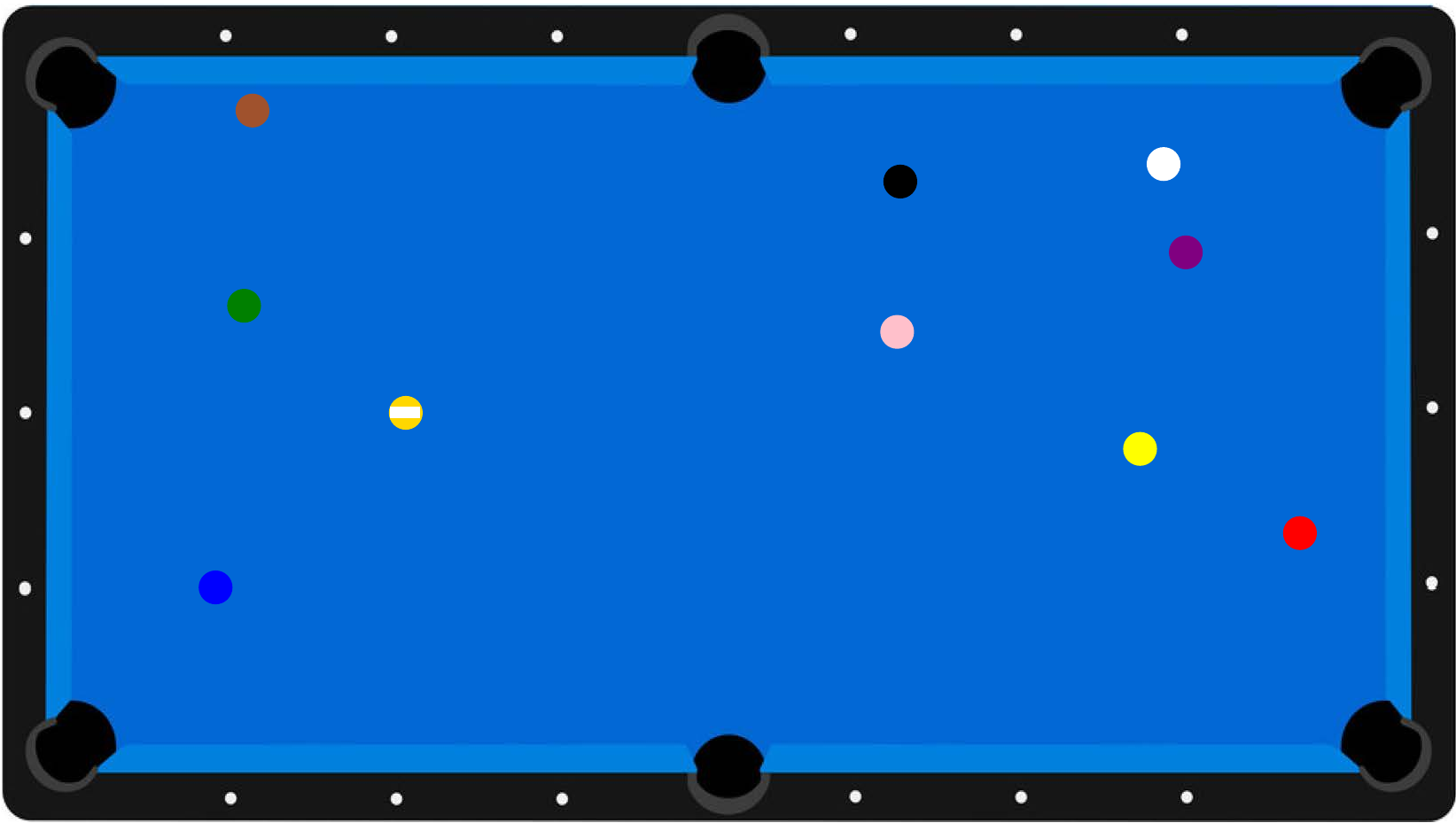}%
 \end{minipage}
 &
 \begin{minipage}{0.18\linewidth}
    \includegraphics[width=\linewidth]{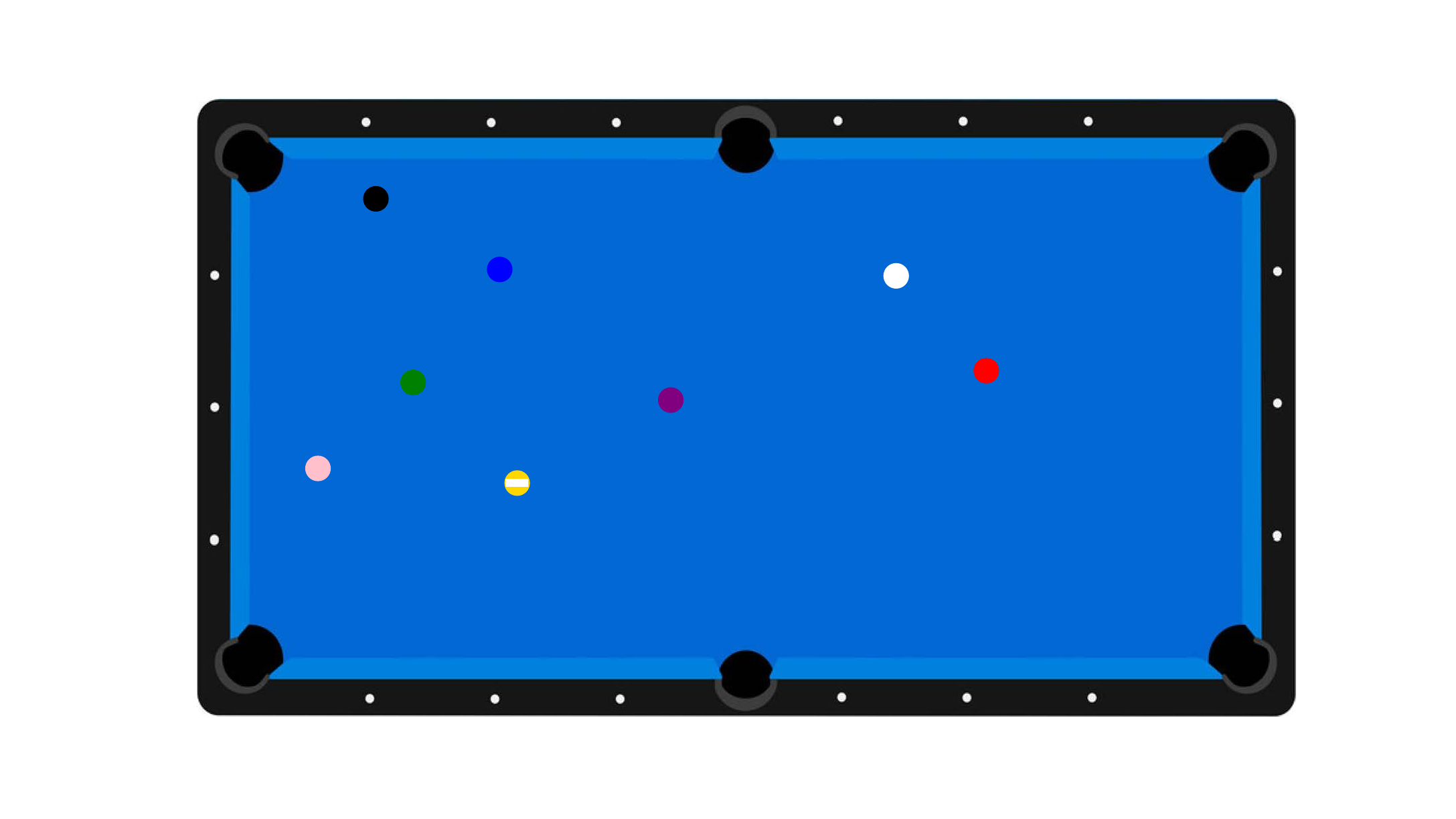}%
 \end{minipage}
 &
 \begin{minipage}{0.18\linewidth}
    \includegraphics[width=\linewidth]{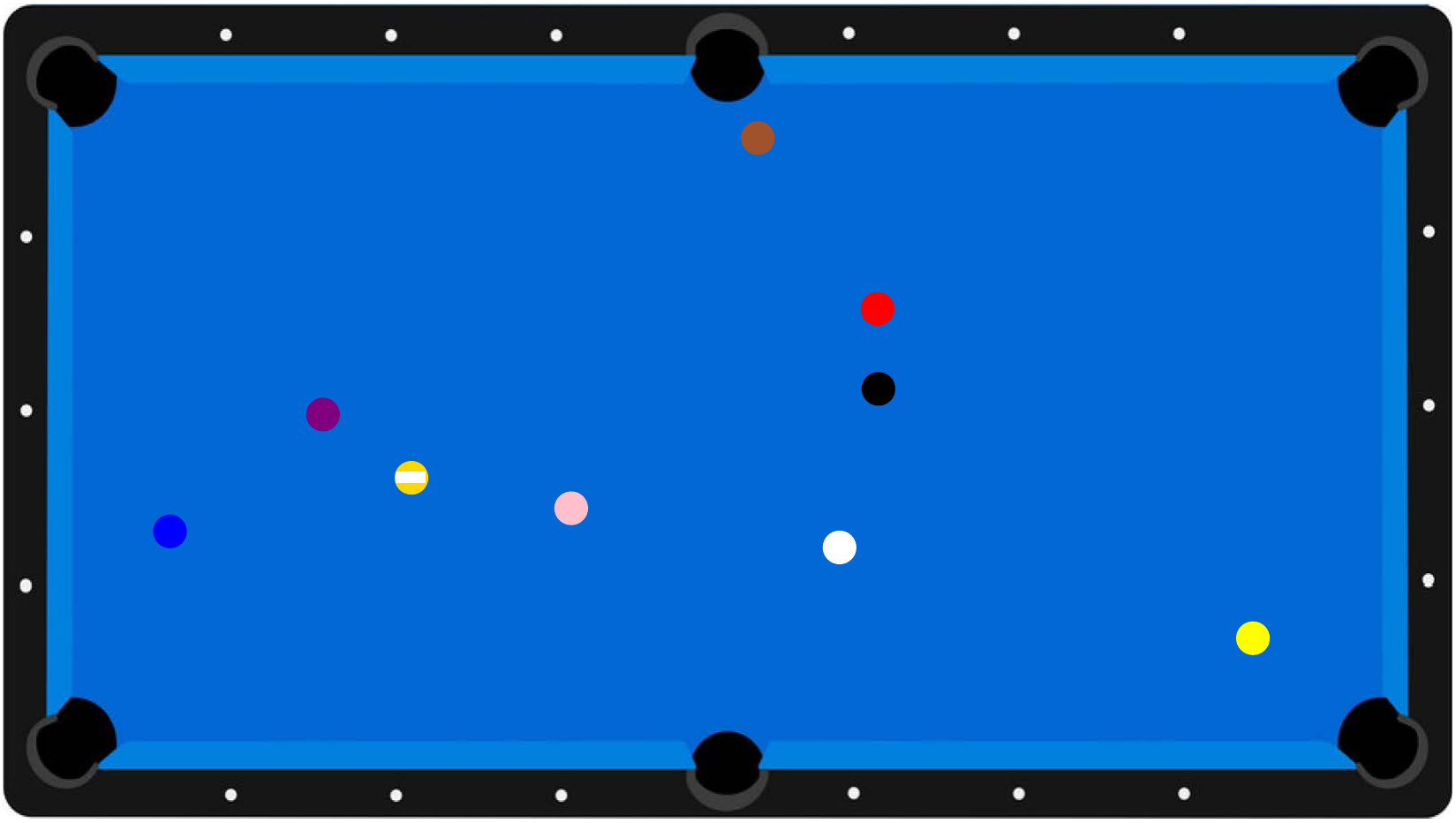}%
 \end{minipage}
 \\
 Q6
 &
 Q7
 &
 Q8
 &
 Q9
 &
 Q10
 \\
  \begin{minipage}{0.18\linewidth}
    \includegraphics[width=\linewidth]{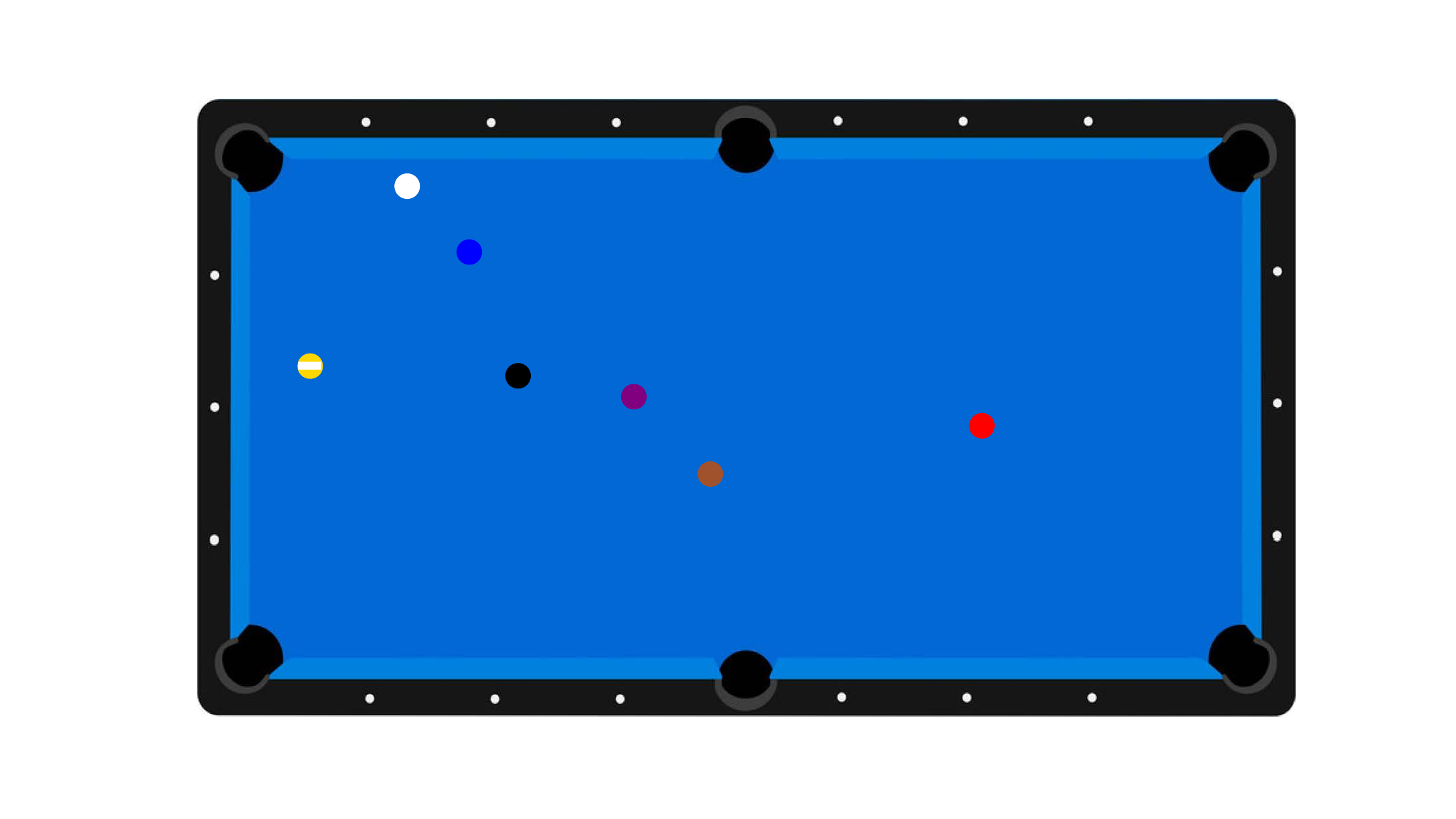}%-eps-converted-to
 \end{minipage}
 &
 \begin{minipage}{0.18\linewidth}
 \includegraphics[width=\linewidth]{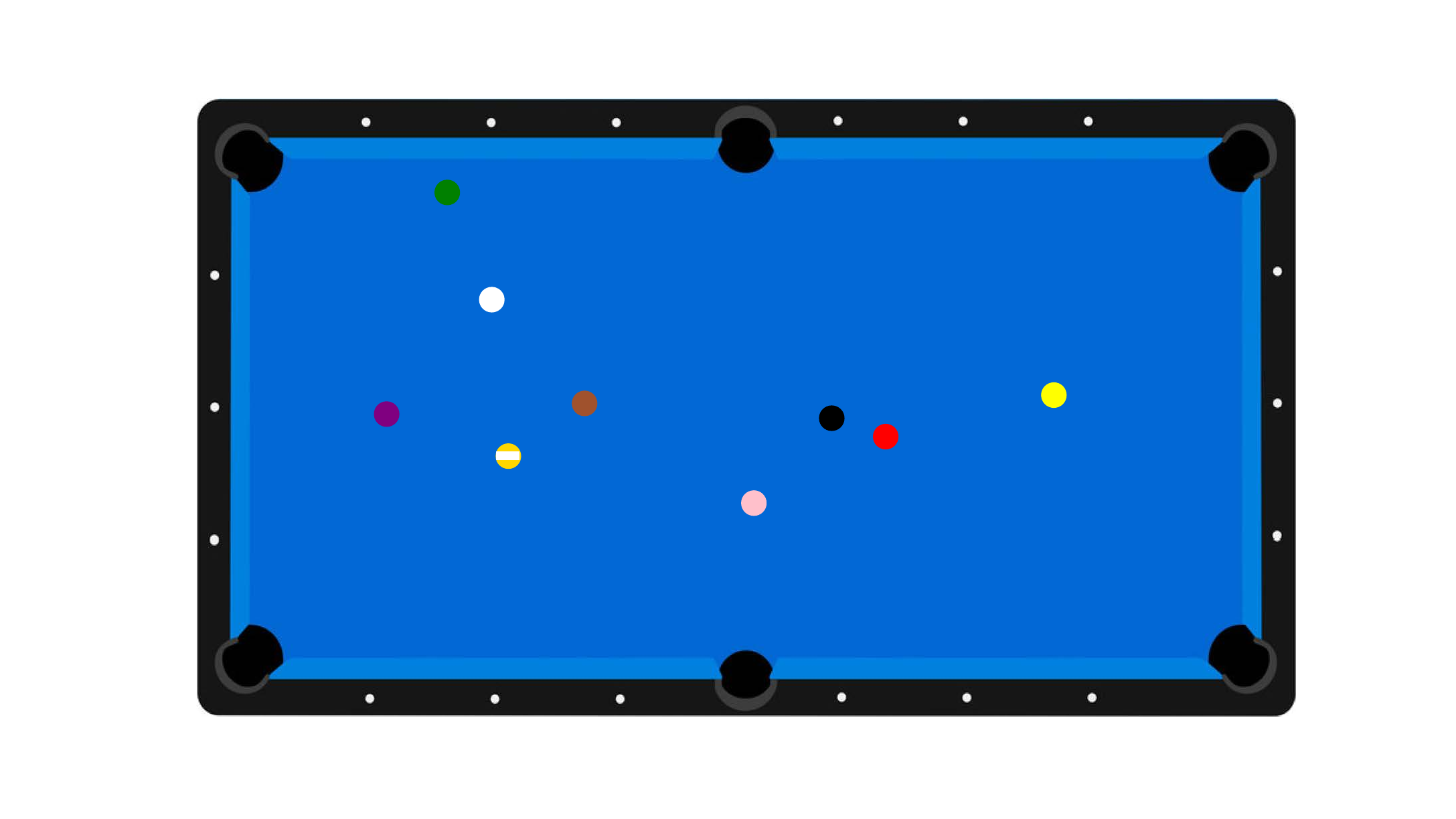}%-eps-converted-to
 \end{minipage}
 &
 \begin{minipage}{0.18\linewidth}
    \includegraphics[width=\linewidth]{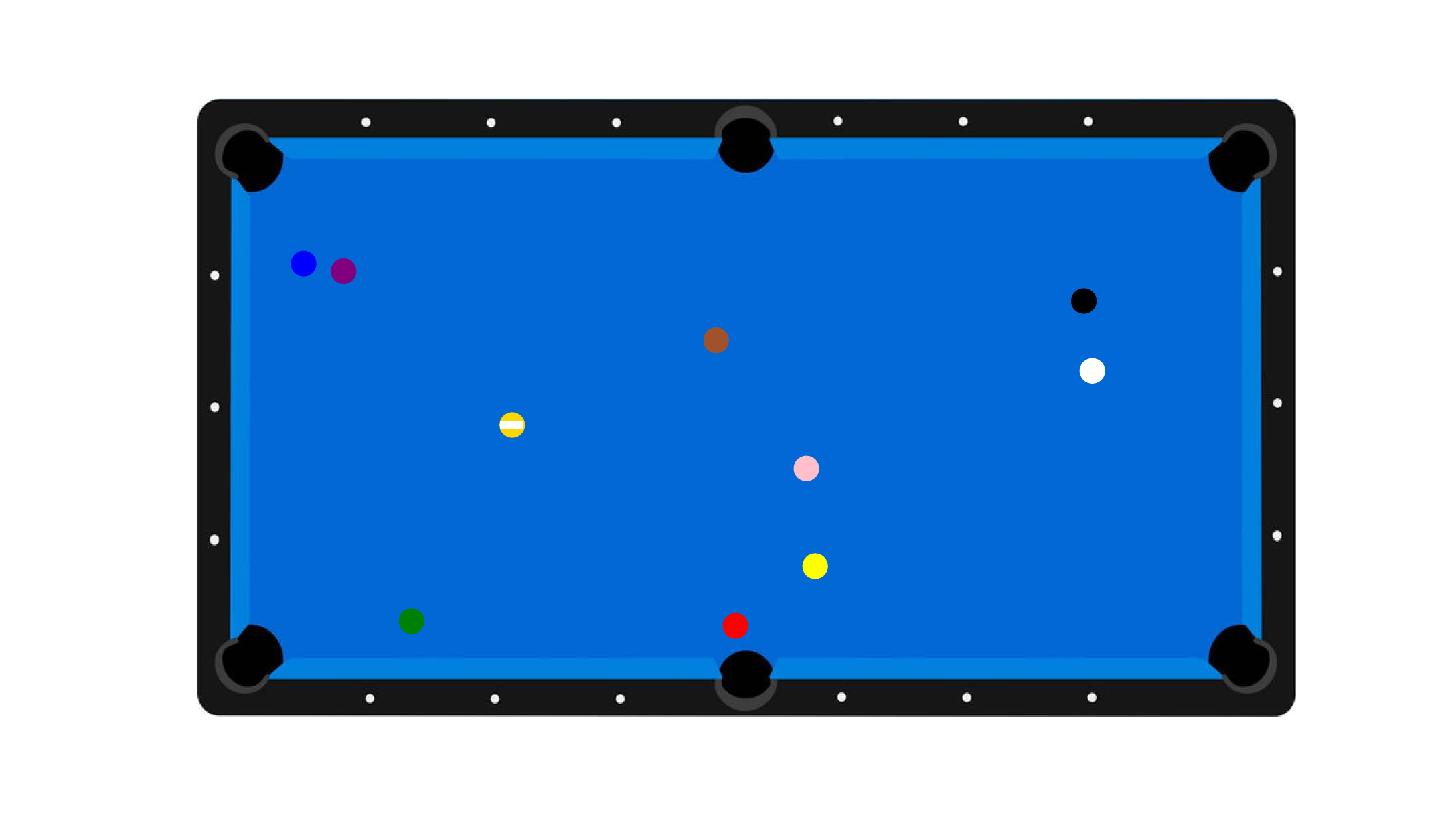}%-eps-converted-to
 \end{minipage}
 &
 \begin{minipage}{0.18\linewidth}
 \includegraphics[width=\linewidth]{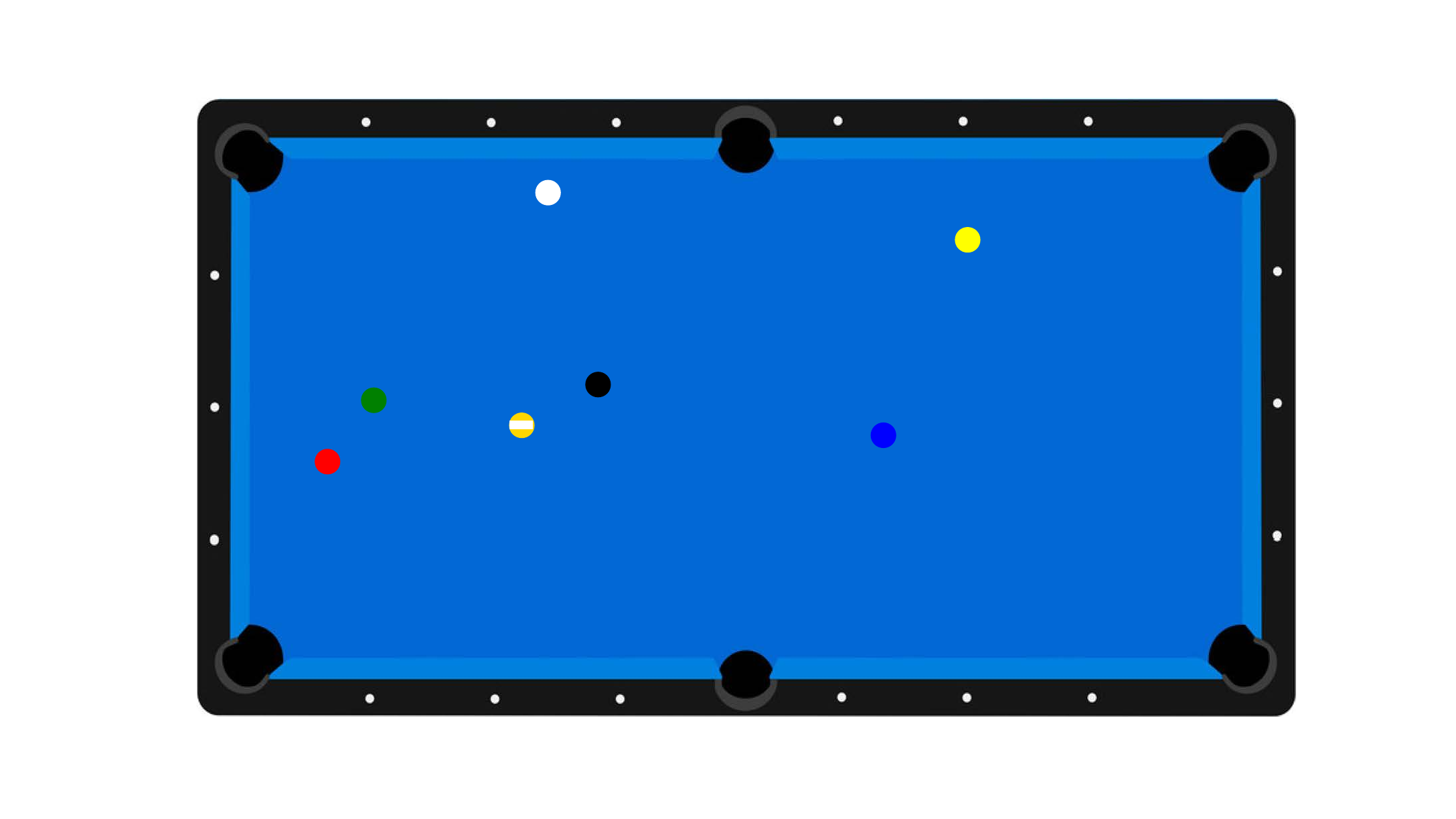}%-eps-converted-to
 \end{minipage}
 &
 \begin{minipage}{0.18\linewidth}
    \includegraphics[width=\linewidth]{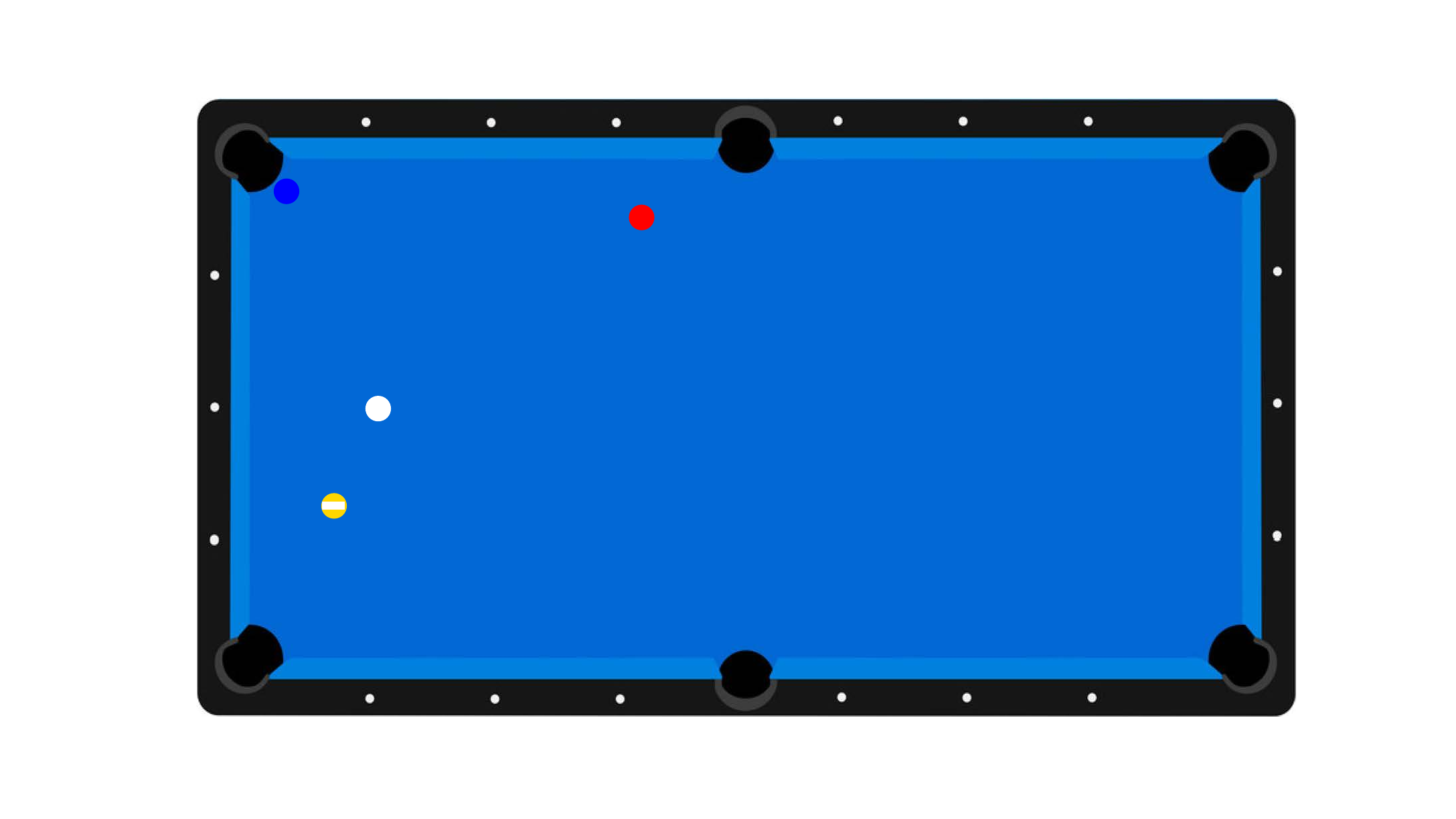}%-eps-converted-to
 \end{minipage}
 \\
 DTW 6 (0/10)
 &
 DTW 7 (1/10)
 &
 DTW 8 (1/10)
 &
 DTW 9 (1/10)
 &
 DTW 10 (0/10)
 \\
 \begin{minipage}{0.18\linewidth}
    \includegraphics[width=\linewidth]{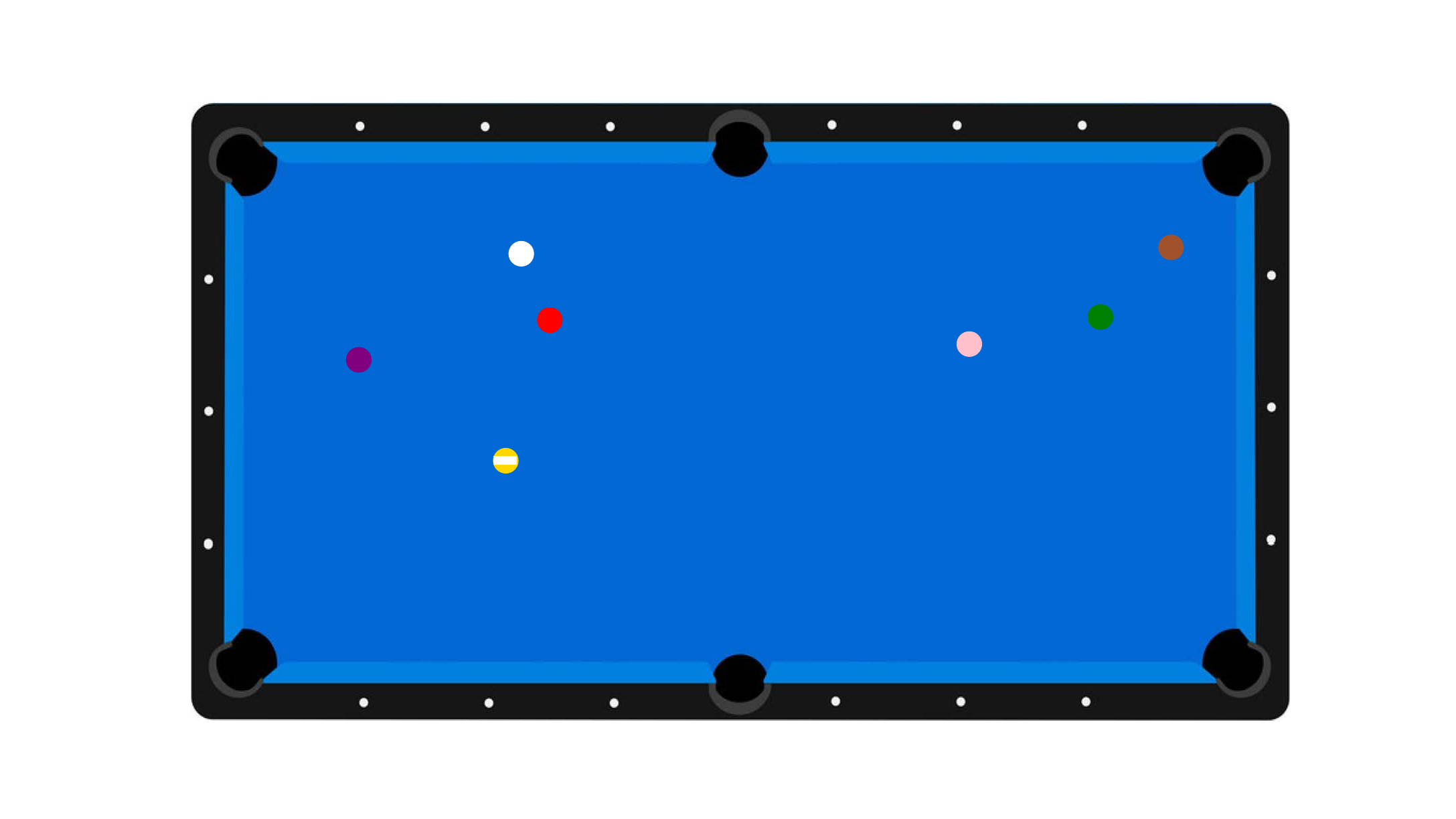}%-eps-converted-to
 \end{minipage}
 &
 \begin{minipage}{0.18\linewidth}
 \includegraphics[width=\linewidth]{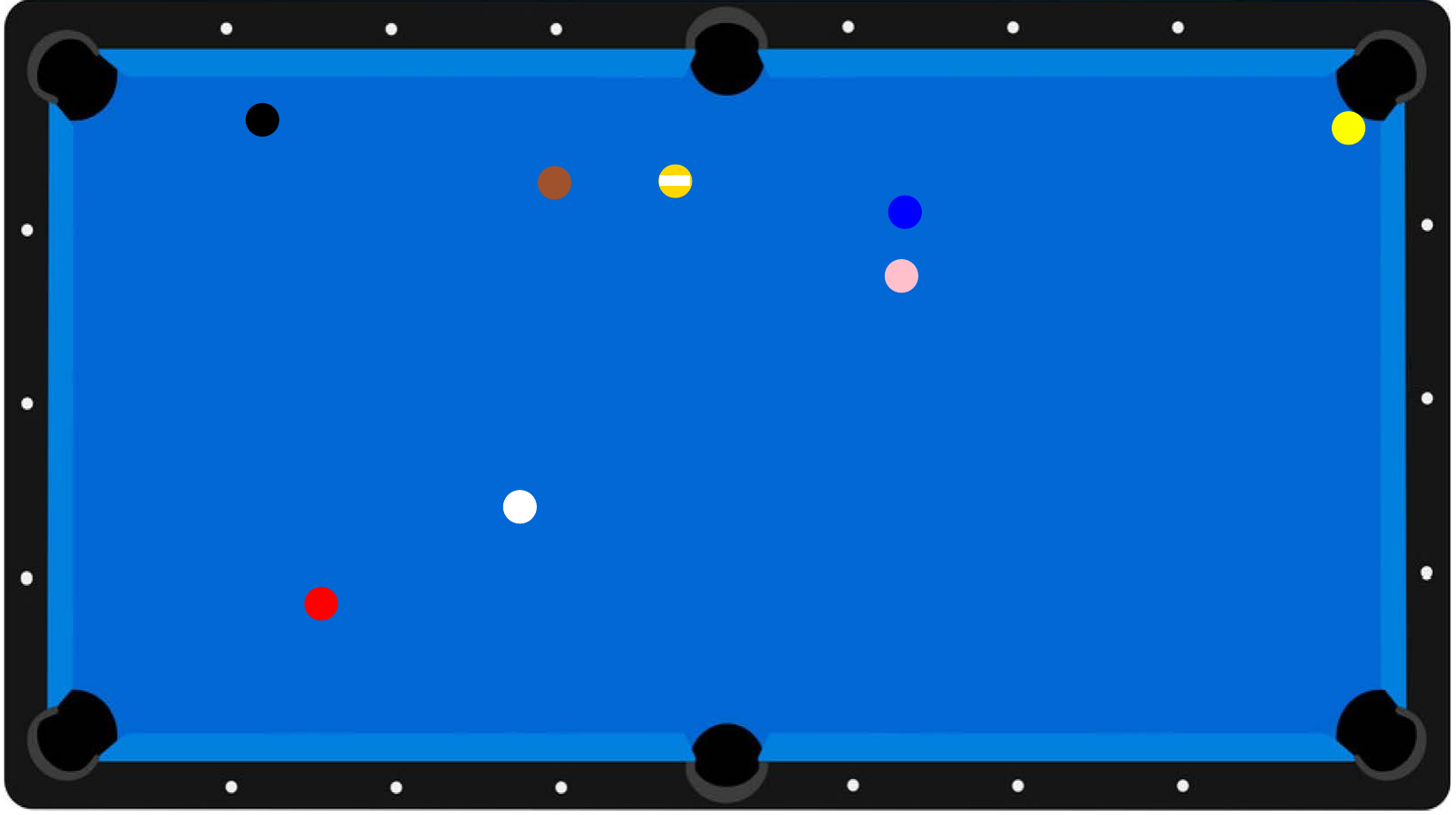}%-eps-converted-to
 \end{minipage}
 &
 \begin{minipage}{0.18\linewidth}
    \includegraphics[width=\linewidth]{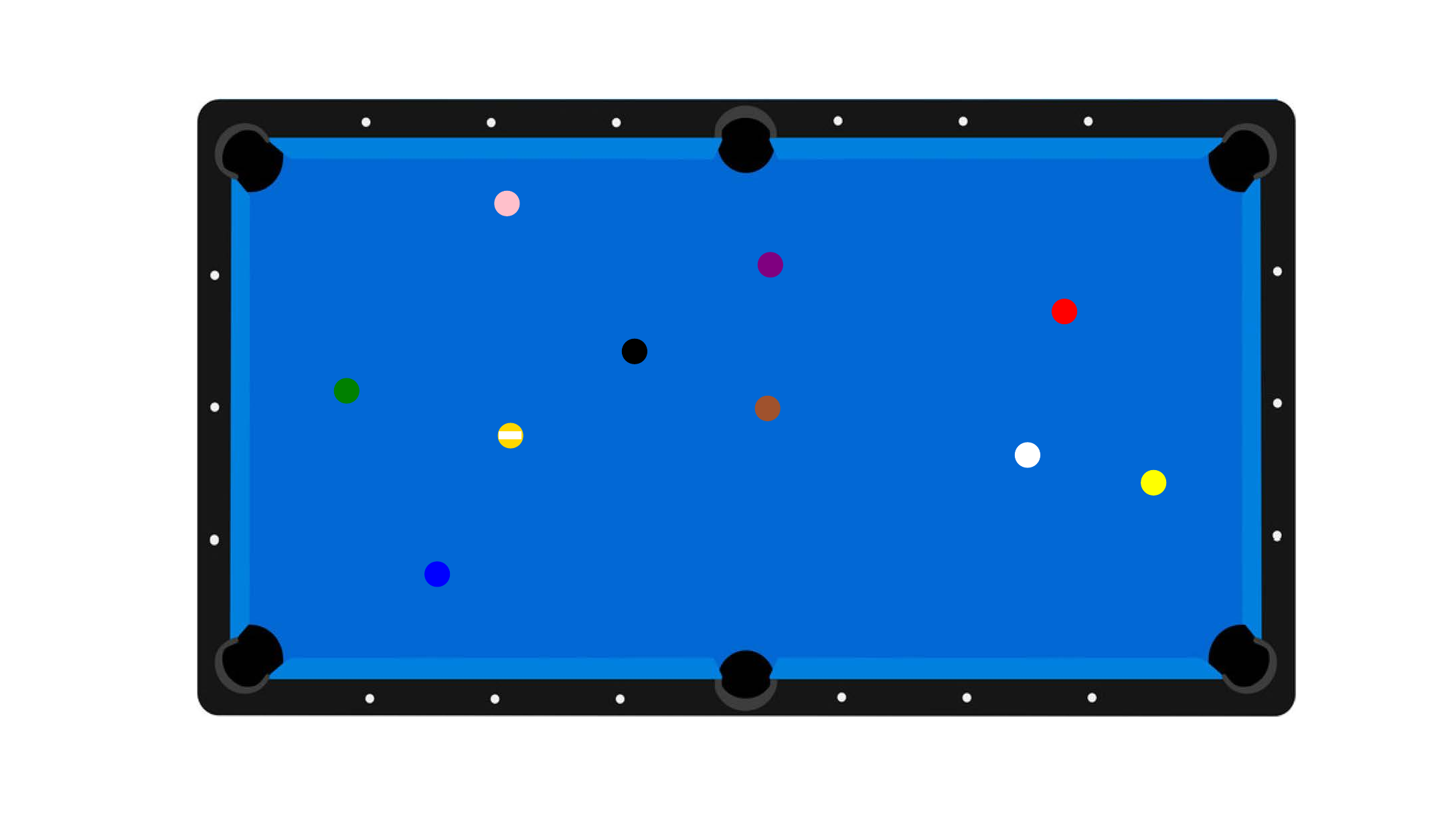}%-eps-converted-to
 \end{minipage}
 &
 \begin{minipage}{0.18\linewidth}
 \includegraphics[width=\linewidth]{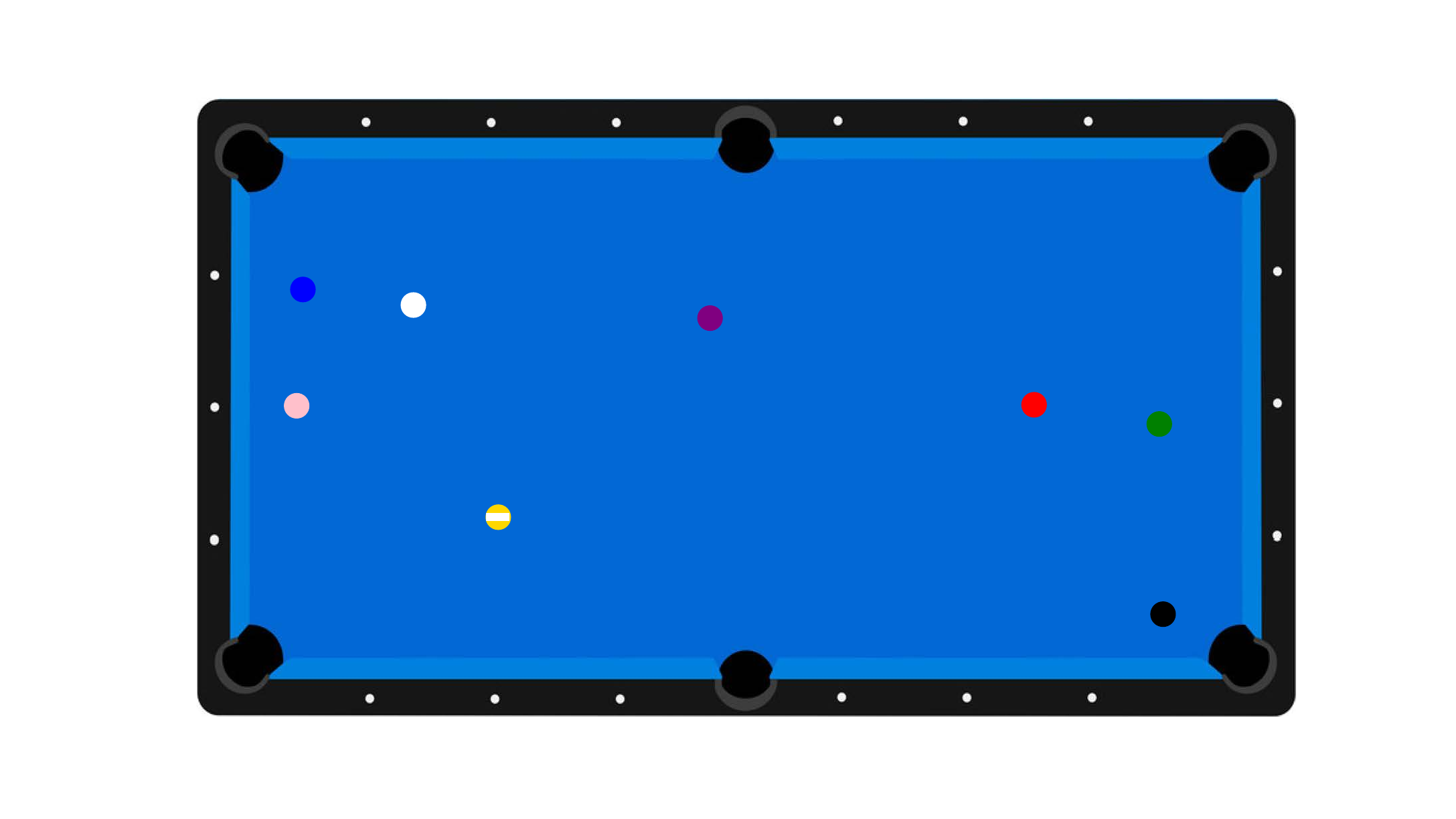}%-eps-converted-to
 \end{minipage}
 &
 \begin{minipage}{0.18\linewidth}
    \includegraphics[width=\linewidth]{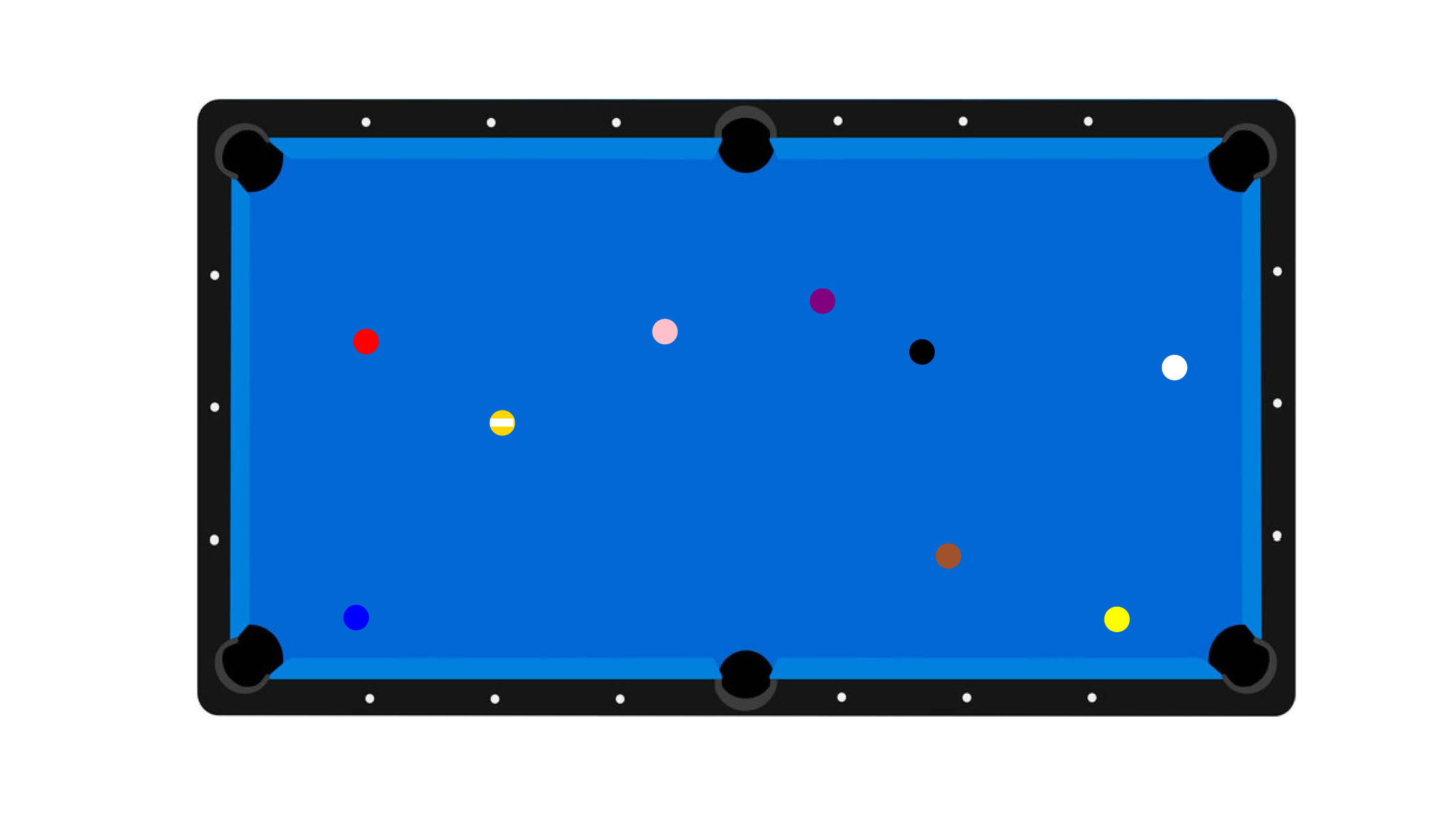}%-eps-converted-to
 \end{minipage}
 \\
 BL2Vec 6 (10/10)
 &
 BL2Vec 7 (9/10)
 &
 BL2Vec 8 (9/10)
 &
 BL2Vec 9 (9/10)
 &
 BL2Vec 10 (10/10)
\end{tabular}
%\vspace*{-3mm}
{\extension{\caption{Top-1 billiards layouts for Q6-Q10 returned by DTW and BL2Vec.}
\label{q6toq10}}}
%\vspace*{-4mm}
\end{figure*}

\begin{figure*}
\centering
\begin{tabular}{c c c c c}
  \begin{minipage}{0.18\linewidth}
  \includegraphics[width=\linewidth]{figures/user/dtw_63_0}%
  \end{minipage}
  &
  \begin{minipage}{0.18\linewidth}
    \includegraphics[width=\linewidth]{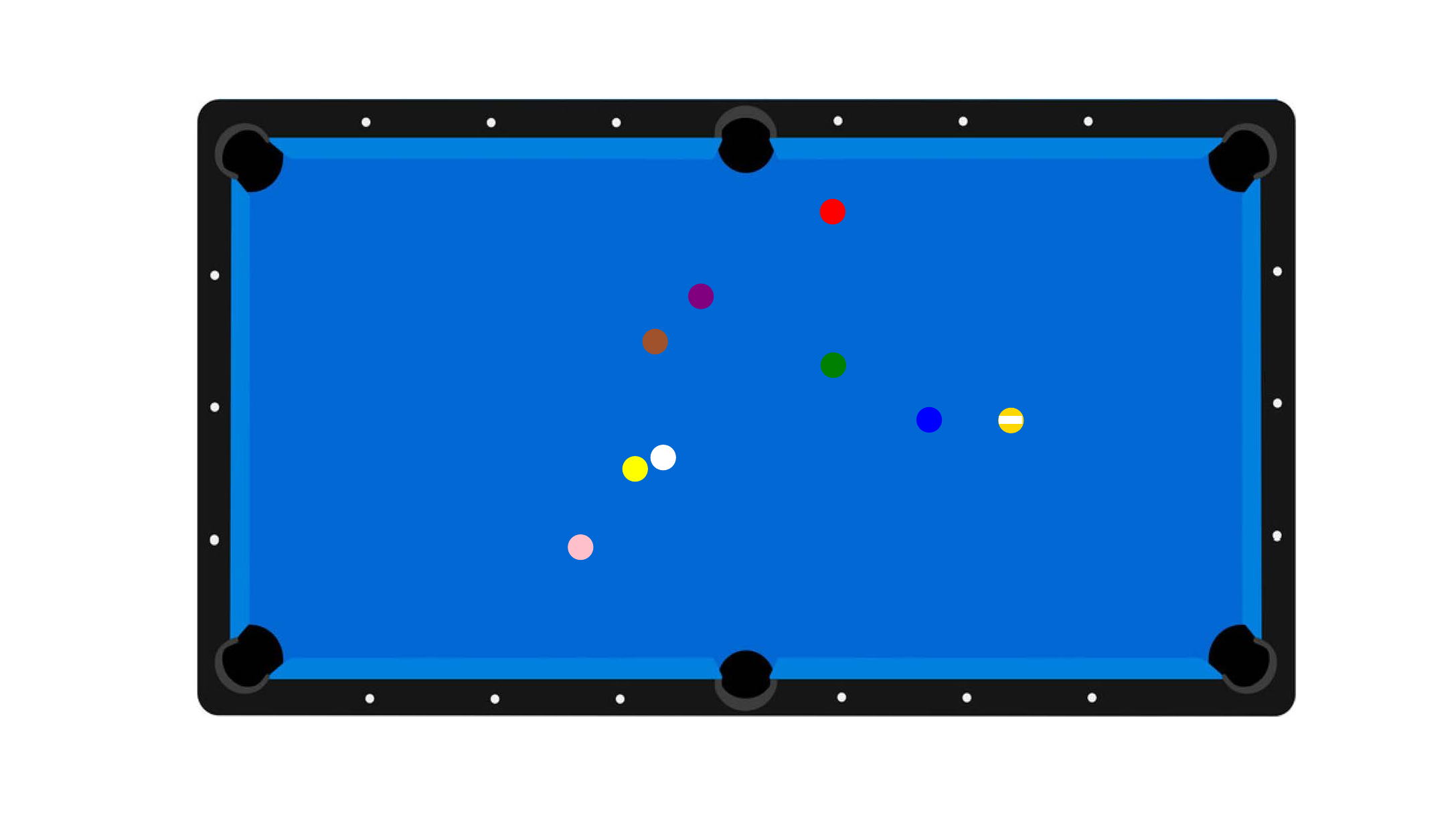}%
  \end{minipage}
  &
  \begin{minipage}{0.18\linewidth}
  \includegraphics[width=\linewidth]{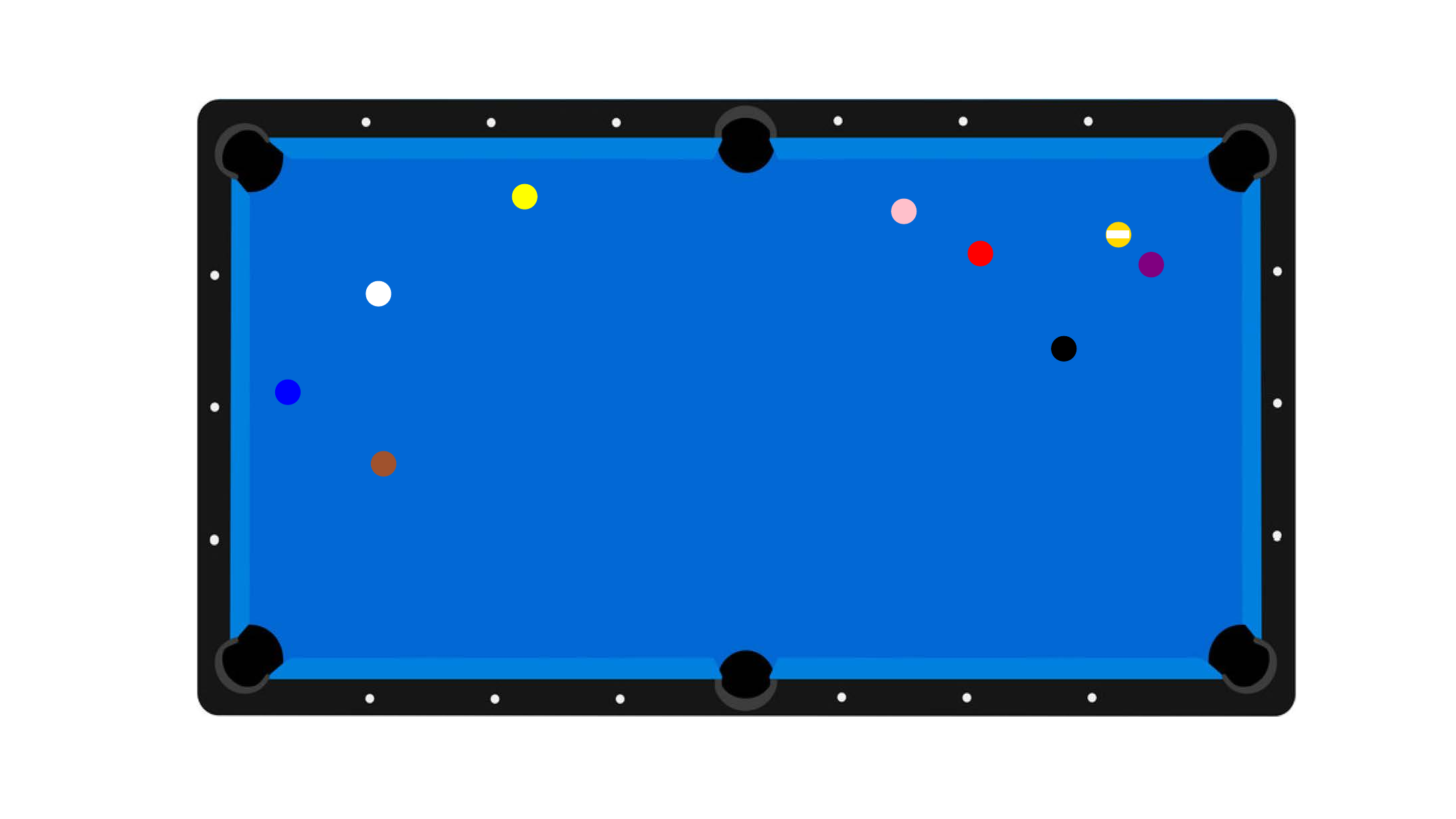}%
  \end{minipage}
  &
  \begin{minipage}{0.18\linewidth}
    \includegraphics[width=\linewidth]{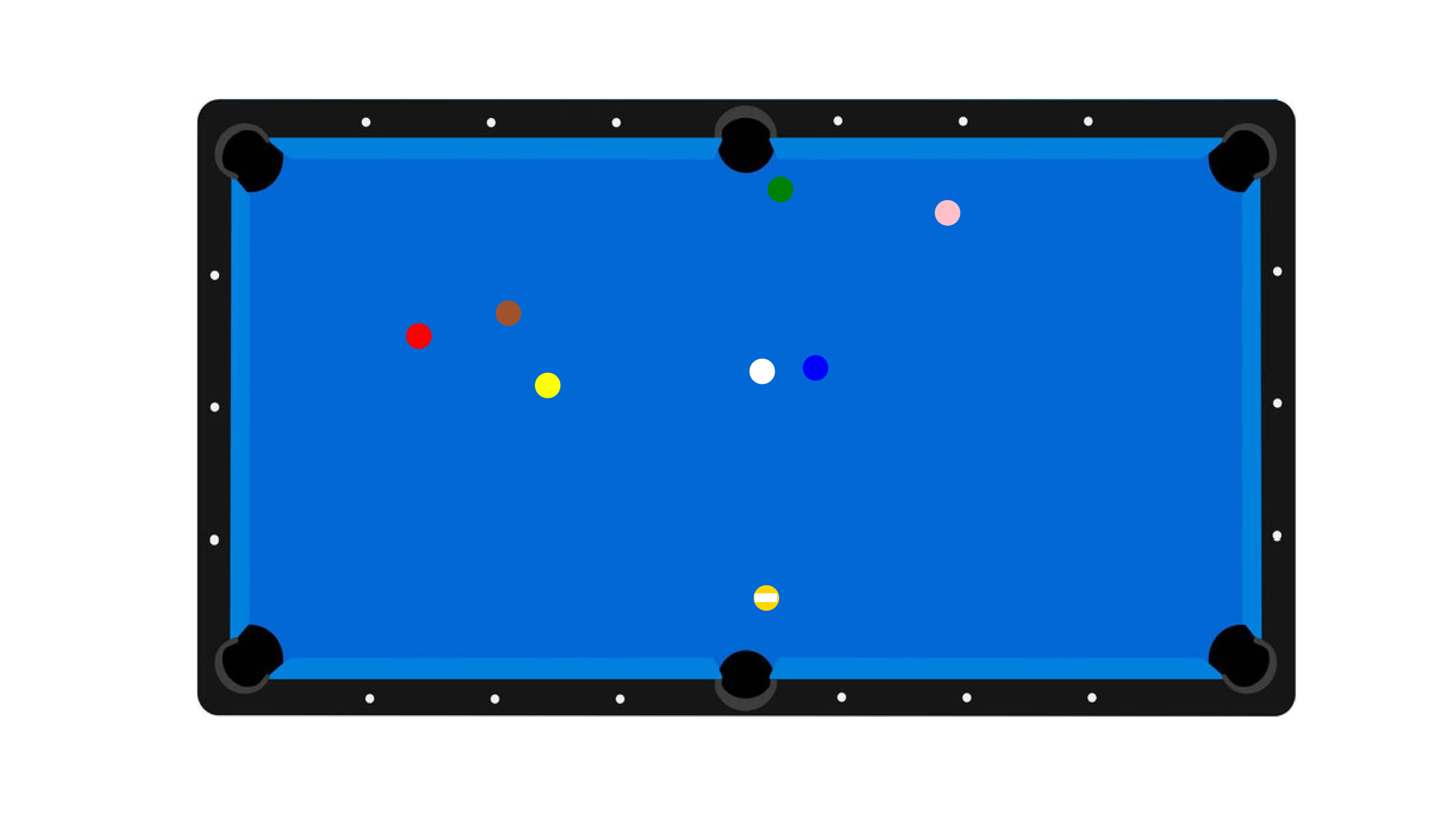}%
  \end{minipage}
  &
  \begin{minipage}{0.18\linewidth}
    \includegraphics[width=\linewidth]{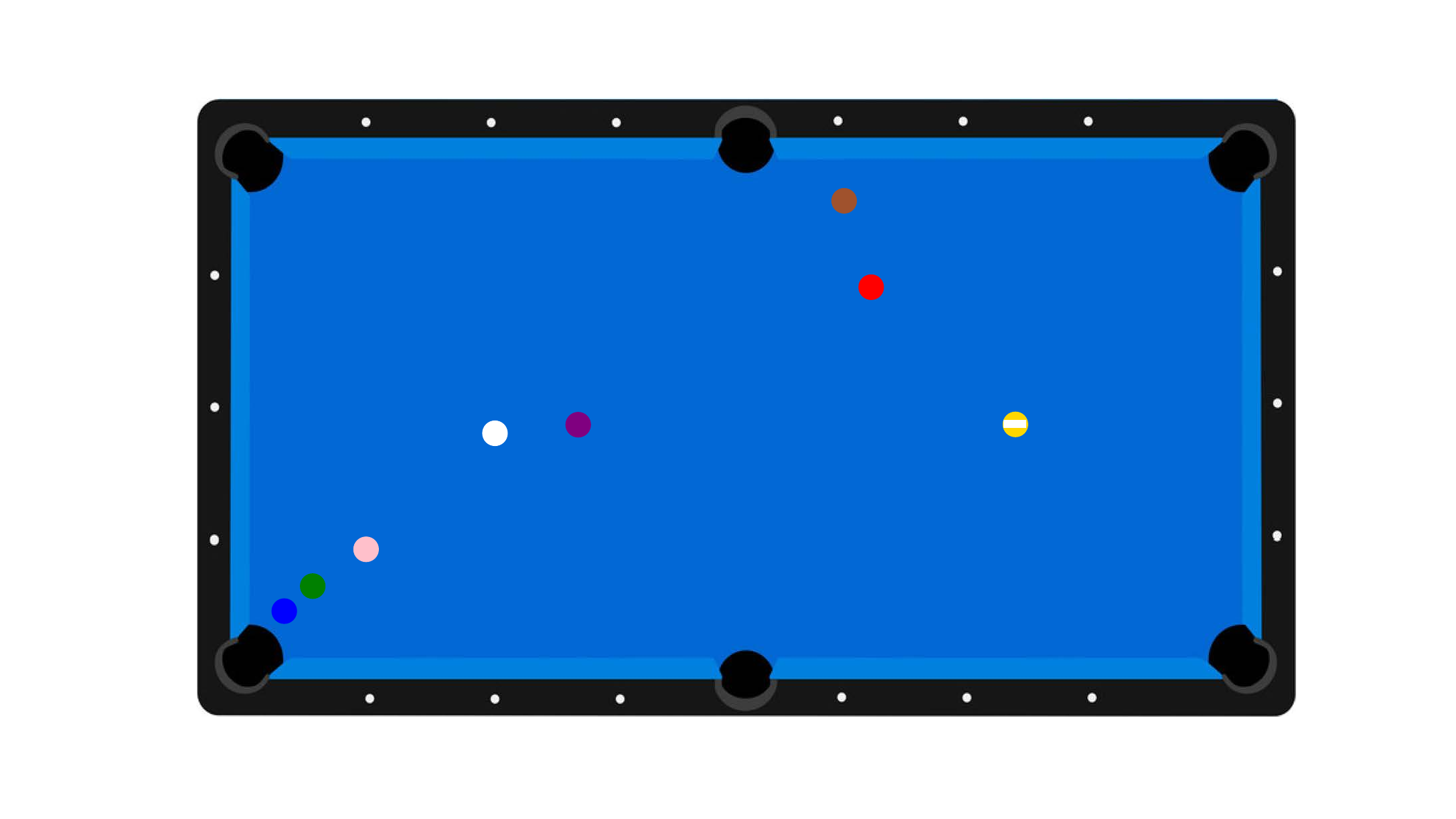}%
  \end{minipage}
  \\
  DTW 1 (1/10)
  &
  DTW 2
  &
  DTW 3
  &
  DTW 4
  &
  DTW 5
  \\
  \begin{minipage}{0.18\linewidth}
  \includegraphics[width=\linewidth]{figures/user/dml_63_0}%
  \end{minipage}
  &
  \begin{minipage}{0.18\linewidth}
    \includegraphics[width=\linewidth]{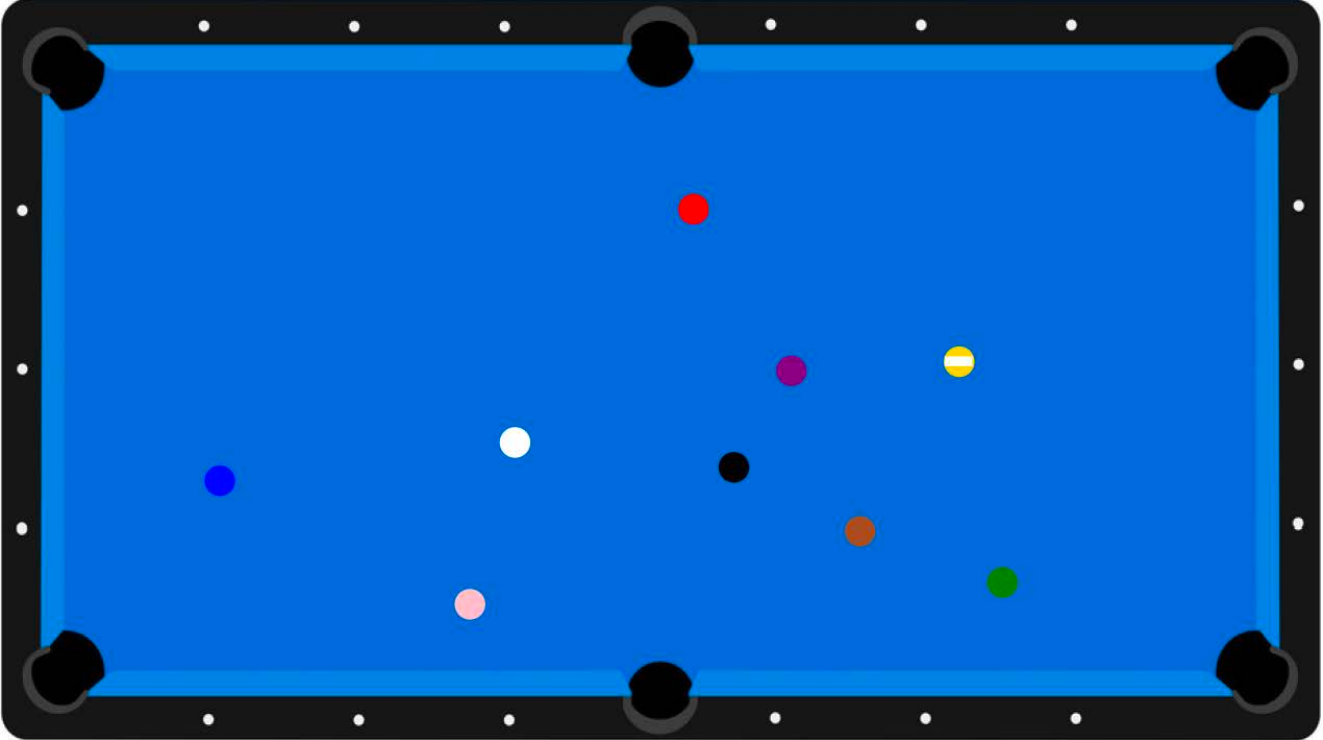}%
  \end{minipage}
  &
  \begin{minipage}{0.18\linewidth}
  \includegraphics[width=\linewidth]{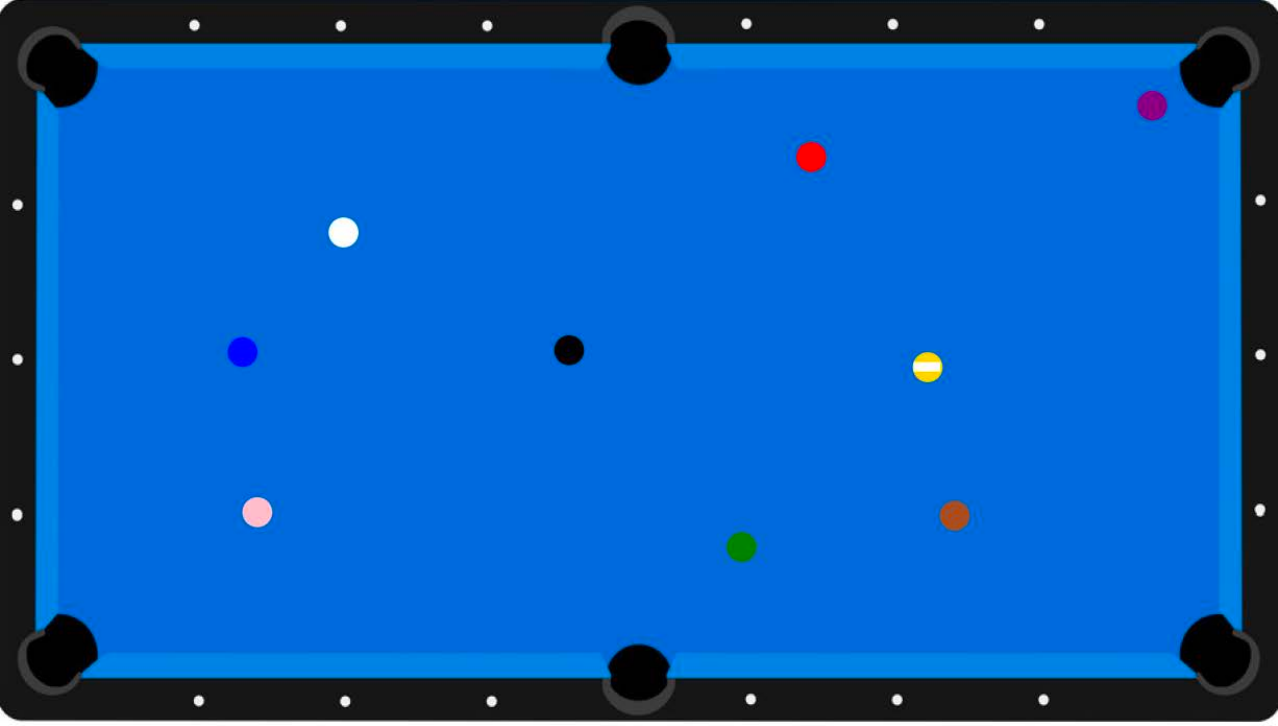}%
  \end{minipage}
  &
  \begin{minipage}{0.18\linewidth}
    \includegraphics[width=\linewidth]{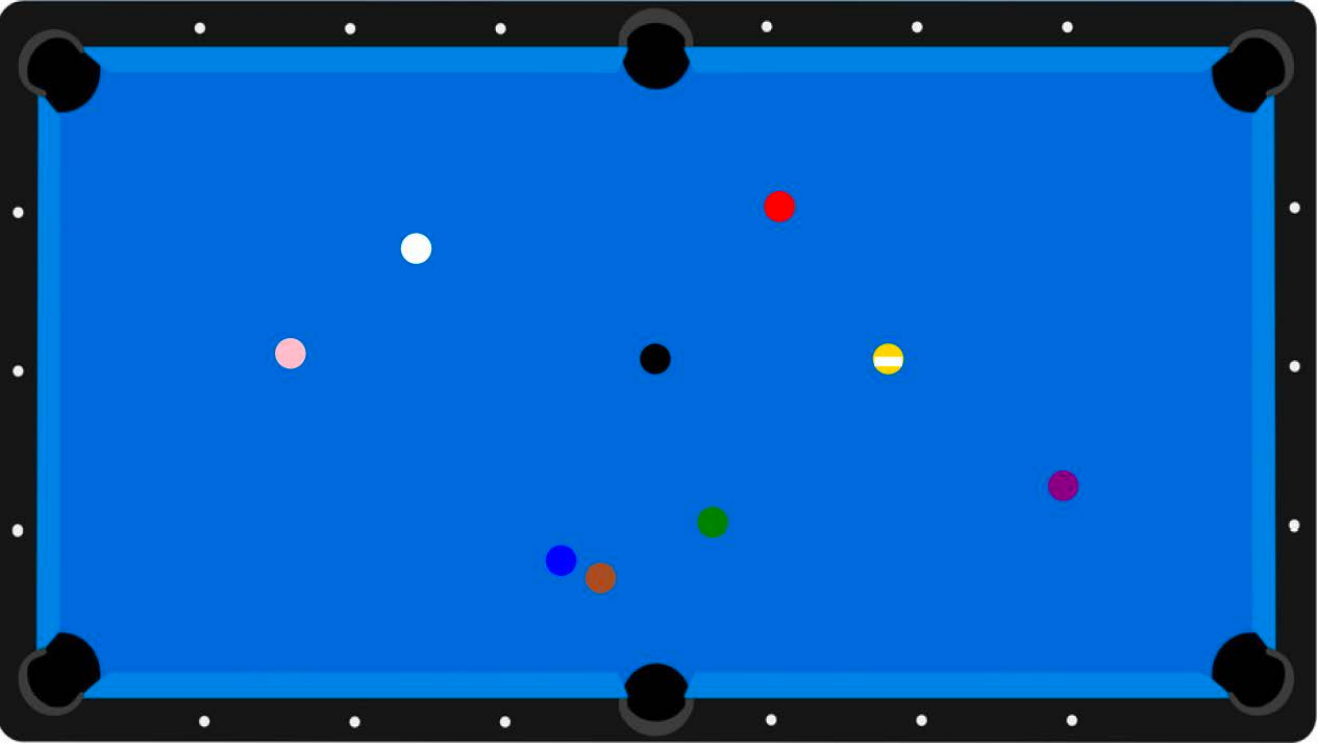}%
  \end{minipage}
  &
  \begin{minipage}{0.18\linewidth}
    \includegraphics[width=\linewidth]{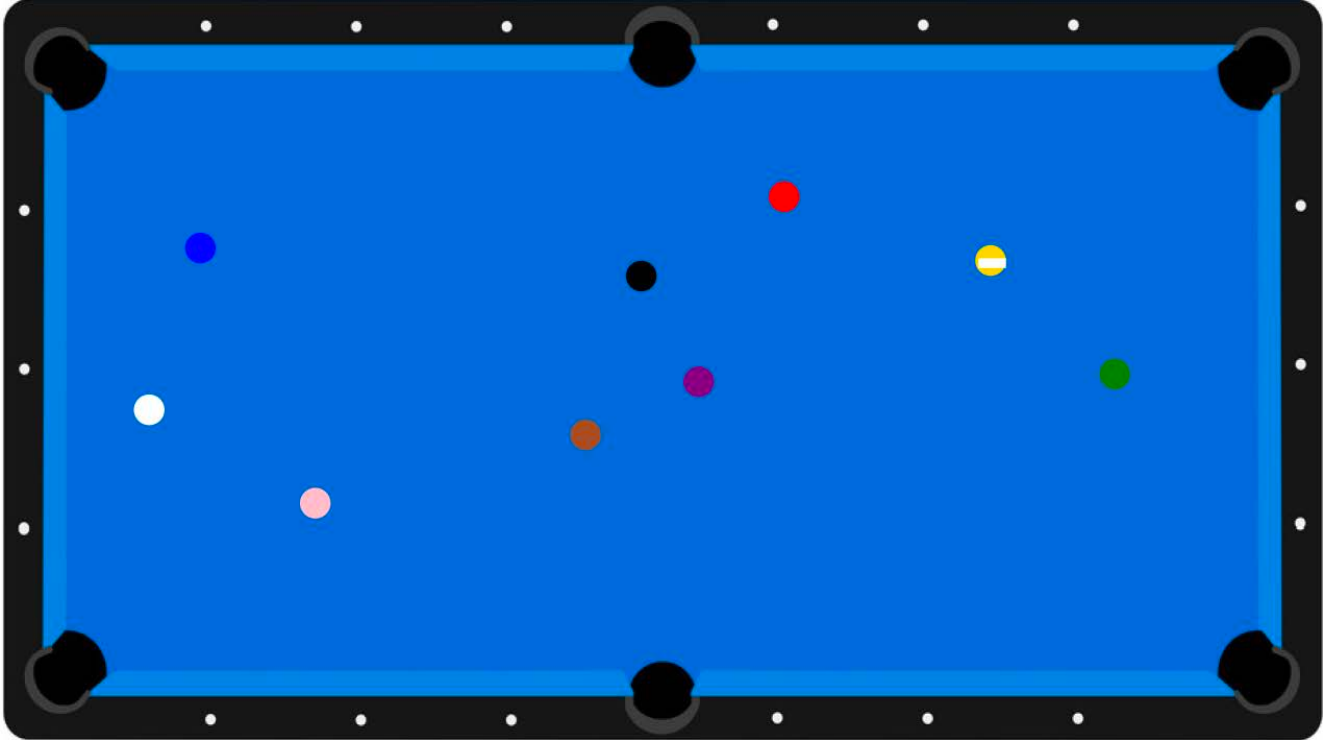}%
  \end{minipage}
  \\
  BL2Vec 1 (9/10)
  &
  BL2Vec 2
  &
  BL2Vec 3
  &
  BL2Vec 4
  &
  BL2Vec 5
  \\
\end{tabular}
%\vspace*{-2mm}
{\extension{\caption{Top-5 billiards layouts for Q1 returned by the DTW and BL2Vec (from left to right).}
%\vspace*{-4mm}
\label{top5}}}
\end{figure*}

\noindent \textbf{(8) User study.}
Since DTW performs the best among all baselines in the similar billiards layout retrieval task as shown in Figure~\ref{fig:effectiveness_similarity}, we conduct a user study to compare the BL2Vec with DTW. We randomly select 10 billiards layouts as the query set as shown in the first row in Figure~\ref{q1toq5} and Figure~\ref{q6toq10}, and for each query in the query set, we use BL2Vec and DTW to retrieve Top-1 billiards layouts from the target database. 
We invite ten volunteers with strong billiards knowledge to annotate the relevance of the retrieved results. More specifically, we firstly spent 15 minutes on introducing the background to the volunteers so that they understand these images as shown in Figure~\ref{q1toq5} and Figure~\ref{q6toq10}. Then for each of 10 queries, the volunteers specify the more similar result from the Top-1 result retrieved by BL2Vec and the Top-1 result retrieved by DTW. Note that the volunteers do not know which result is returned by which method in advance. We report the scores of the ten queries for both methods in Figure~\ref{fig:userstudy}. We observe that our BL2Vec outperforms DTW with expert knowledge. In particular, BL2Vec outperforms DTW for 8 queries out of the total 10 queries. In addition, BL2Vec gets 81\% votes while DTW gets only 19\% votes. We illustrate the Top-1 results of BL2Vec and DTW for Q1 to Q5 and Q6 to Q10 in Figure~\ref{q1toq5} and Figure~\ref{q6toq10}, respectively. From these figures, we observe the layouts retrieved by BL2Vec match the queries very well. For example, we find that the overall layout of the TOP-1 result by BL2Vec is very similar to Q1. %%and Top 2-5 results also maintain high consistency in the layouts. %{\color{blue}
DTW is slightly better than BL2Vec for Q4 and Q5 (i.e., 2\% votes). This is because the overall layouts returned by DTW are more similar to the queries; however, DTW falsely matches the balls, e.g., 9 balls in Q4, but the layout returned by DTW only has 7 balls and they are matched with different colors. It indicates that the pairwise point-matching solution is not suitable for the billiards layout retrieval task. In addition, we visualize the Top 5 results for Q1 returned by the DTW and BL2Vec in Figure~\ref{top5}, we observe BL2Vec maintains a high consistency between the retrieved Top 2-5 results and the query.
% , and outperforms the DTW based on the visual.
}}

%% file: conclusion.tex
\section{CONCLUSION}
\label{sec:conclusion}

%In this paper, we study the similar billiards layout retrieval problem. We develop the first deep learning-based solution called BL2Vec by learning the representations of billiards layouts for computing their similarity. BL2Vec captures the unique characteristics in a billiards layout and runs very fast, i.e., it runs in linear time with the size of the database when performing the similarity search. We collect a billiards layout dataset and conduct extensive experiments on the dataset, and the results verify that our method achieves the best effectiveness and competitive efficiency. 
%
%In the future, we plan to explore more methods for billiards related analytics tasks, e.g., realistic layouts generation and player performance analysis. 

{\extension{In this paper, we contribute a publicly available billiards layout dataset, which includes break shot layout data, trajectory data of strikes, details of strikes, etc. On the dataset, we investigate a few tasks including (1) prediction and (2) generation based on the break shot layout data,
% (3) similarity search based on the trajectory data,
and (3) similar billiards layout retrieval. Extensive experiments are conducted on the collected dataset, which verify the usefulness of the dataset and also the proposed methods for the tasks on the dataset. 
In the future, we plan to explore more methods for billiards related analytics tasks, e.g., player performance analysis, tactics discovery, etc.}}

\smallskip
\noindent\textbf{Acknowledgments:}
This research/project is supported by the National Research Foundation, Singapore under its AI Singapore Programme (AISG Award No: AISG-PhD/2021-08-024[T]). Any opinions, findings and conclusions or recommendations expressed in this material are those of the author(s) and do not reflect the views of National Research Foundation, Singapore. This project is partially supported by HKU-SCF FinTech Academy and Shenzhen-Hong Kong-Macao Science and Technology Plan Project (Category C Project: SGDX202108 23103537030) and Theme-based Research Scheme T35-710/20-R.

%% file: main.bbl

\begin{thebibliography}{61}

%%% the Web address.

\ifx \showCODEN    \undefined \def \showCODEN     #1{\unskip}     \fi
\ifx \showDOI      \undefined \def \showDOI       #1{#1}\fi
\ifx \showISBNx    \undefined \def \showISBNx     #1{\unskip}     \fi
\ifx \showISBNxiii \undefined \def \showISBNxiii  #1{\unskip}     \fi
\ifx \showISSN     \undefined \def \showISSN      #1{\unskip}     \fi
\ifx \showLCCN     \undefined \def \showLCCN      #1{\unskip}     \fi
\ifx \shownote     \undefined \def \shownote      #1{#1}          \fi
\ifx \showarticletitle \undefined \def \showarticletitle #1{#1}   \fi
\ifx \showURL      \undefined \def \showURL       {\relax}        \fi
% invisible to TeX
\providecommand\bibfield[2]{#2}
\providecommand\bibinfo[2]{#2}
\providecommand\natexlab[1]{#1}
\providecommand\showeprint[2][]{arXiv:#2}

\bibitem[Alahi et~al\mbox{.}(2016)]%
        {alahi2016social}
\bibfield{author}{\bibinfo{person}{Alexandre Alahi}, \bibinfo{person}{Kratarth
  Goel}, \bibinfo{person}{Vignesh Ramanathan}, \bibinfo{person}{Alexandre
  Robicquet}, \bibinfo{person}{Li Fei-Fei}, {and} \bibinfo{person}{Silvio
  Savarese}.} \bibinfo{year}{2016}\natexlab{}.
\newblock \showarticletitle{Social lstm: Human trajectory prediction in crowded
  spaces}. In \bibinfo{booktitle}{\emph{CVPR}}.
\newblock


\bibitem[Alt and Godau(1995)]%
        {alt1995computing}
\bibfield{author}{\bibinfo{person}{Helmut Alt} {and} \bibinfo{person}{Michael
  Godau}.} \bibinfo{year}{1995}\natexlab{}.
\newblock \showarticletitle{Computing the Fr{\'e}chet distance between two
  polygonal curves}.
\newblock \bibinfo{journal}{\emph{IJCGA}} \bibinfo{volume}{5},
  \bibinfo{number}{01n02} (\bibinfo{year}{1995}), \bibinfo{pages}{75--91}.
\newblock


\bibitem[Aoki et~al\mbox{.}(2017)]%
        {aoki2017luck}
\bibfield{author}{\bibinfo{person}{Raquel Aoki}, \bibinfo{person}{Renato~M
  Assuncao}, {and} \bibinfo{person}{Pedro~OS Vaz~de Melo}.}
  \bibinfo{year}{2017}\natexlab{}.
\newblock \showarticletitle{Luck is hard to beat: The difficulty of sports
  prediction}. In \bibinfo{booktitle}{\emph{SIGKDD}}. ACM,
  \bibinfo{pages}{1367--1376}.
\newblock


\bibitem[Archibald et~al\mbox{.}(2009)]%
        {archibald2009analysis}
\bibfield{author}{\bibinfo{person}{Christopher Archibald},
  \bibinfo{person}{Alon Altman}, {and} \bibinfo{person}{Yoav Shoham}.}
  \bibinfo{year}{2009}\natexlab{}.
\newblock \showarticletitle{Analysis of a Winning Computational Billiards
  Player.}. In \bibinfo{booktitle}{\emph{IJCAI}}, Vol.~\bibinfo{volume}{9}.
  Citeseer, \bibinfo{pages}{1377--1382}.
\newblock


\bibitem[{Arsomngern} et~al\mbox{.}(2021)]%
        {Arsomngern2021self}
\bibfield{author}{\bibinfo{person}{P. {Arsomngern}}, \bibinfo{person}{C.
  {Long}}, \bibinfo{person}{S. {Suwajanakorn}}, {and} \bibinfo{person}{S.
  {Nutanong}}.} \bibinfo{year}{2021}\natexlab{}.
\newblock \showarticletitle{Self-Supervised Deep Metric Learning for
  Pointsets}. In \bibinfo{booktitle}{\emph{ICDE}}.
\newblock


\bibitem[Bachman and Precup(2015)]%
        {bachman2015data}
\bibfield{author}{\bibinfo{person}{Philip Bachman} {and} \bibinfo{person}{Doina
  Precup}.} \bibinfo{year}{2015}\natexlab{}.
\newblock \showarticletitle{Data generation as sequential decision making}.
\newblock \bibinfo{journal}{\emph{arXiv}} (\bibinfo{year}{2015}).
\newblock


\bibitem[Bahdanau et~al\mbox{.}(2016)]%
        {bahdanau2016actor}
\bibfield{author}{\bibinfo{person}{Dzmitry Bahdanau}, \bibinfo{person}{Philemon
  Brakel}, \bibinfo{person}{Kelvin Xu}, \bibinfo{person}{Anirudh Goyal},
  \bibinfo{person}{Ryan Lowe}, \bibinfo{person}{Joelle Pineau},
  \bibinfo{person}{Aaron Courville}, {and} \bibinfo{person}{Yoshua Bengio}.}
  \bibinfo{year}{2016}\natexlab{}.
\newblock \showarticletitle{An actor-critic algorithm for sequence prediction}.
\newblock \bibinfo{journal}{\emph{arXiv}} (\bibinfo{year}{2016}).
\newblock


\bibitem[Bentley(1975)]%
        {bentley1975multidimensional}
\bibfield{author}{\bibinfo{person}{Jon~Louis Bentley}.}
  \bibinfo{year}{1975}\natexlab{}.
\newblock \showarticletitle{Multidimensional binary search trees used for
  associative searching}.
\newblock \bibinfo{journal}{\emph{Commun. ACM}} \bibinfo{volume}{18},
  \bibinfo{number}{9} (\bibinfo{year}{1975}), \bibinfo{pages}{509--517}.
\newblock


\bibitem[Bromley et~al\mbox{.}(1993)]%
        {bromley1993signature}
\bibfield{author}{\bibinfo{person}{Jane Bromley}, \bibinfo{person}{Isabelle
  Guyon}, \bibinfo{person}{Yann LeCun}, \bibinfo{person}{Eduard S{\"a}ckinger},
  {and} \bibinfo{person}{Roopak Shah}.} \bibinfo{year}{1993}\natexlab{}.
\newblock \showarticletitle{Signature verification using a" siamese" time delay
  neural network}.
\newblock \bibinfo{journal}{\emph{NIPS}}  \bibinfo{volume}{6}
  (\bibinfo{year}{1993}), \bibinfo{pages}{737--744}.
\newblock


\bibitem[Che et~al\mbox{.}(2017)]%
        {che2017maximum}
\bibfield{author}{\bibinfo{person}{Tong Che}, \bibinfo{person}{Yanran Li},
  \bibinfo{person}{Ruixiang Zhang}, \bibinfo{person}{R~Devon Hjelm},
  \bibinfo{person}{Wenjie Li}, \bibinfo{person}{Yangqiu Song}, {and}
  \bibinfo{person}{Yoshua Bengio}.} \bibinfo{year}{2017}\natexlab{}.
\newblock \showarticletitle{Maximum-likelihood augmented discrete generative
  adversarial networks}.
\newblock \bibinfo{journal}{\emph{arXiv}} (\bibinfo{year}{2017}).
\newblock


\bibitem[Chen and Ng(2004)]%
        {chen2004marriage}
\bibfield{author}{\bibinfo{person}{Lei Chen} {and} \bibinfo{person}{Raymond
  Ng}.} \bibinfo{year}{2004}\natexlab{}.
\newblock \showarticletitle{On the marriage of lp-norms and edit distance}. In
  \bibinfo{booktitle}{\emph{PVLDB}}. VLDB Endowment, \bibinfo{pages}{792--803}.
\newblock


\bibitem[Chen et~al\mbox{.}(2005)]%
        {chen2005robust}
\bibfield{author}{\bibinfo{person}{Lei Chen}, \bibinfo{person}{M~Tamer
  {\"O}zsu}, {and} \bibinfo{person}{Vincent Oria}.}
  \bibinfo{year}{2005}\natexlab{}.
\newblock \showarticletitle{Robust and fast similarity search for moving object
  trajectories}. In \bibinfo{booktitle}{\emph{SIGMOD}}. ACM,
  \bibinfo{pages}{491--502}.
\newblock


\bibitem[Cui et~al\mbox{.}(2023)]%
        {cui2023sportsmot}
\bibfield{author}{\bibinfo{person}{Yutao Cui}, \bibinfo{person}{Chenkai Zeng},
  \bibinfo{person}{Xiaoyu Zhao}, \bibinfo{person}{Yichun Yang},
  \bibinfo{person}{Gangshan Wu}, {and} \bibinfo{person}{Limin Wang}.}
  \bibinfo{year}{2023}\natexlab{}.
\newblock \showarticletitle{SportsMOT: A Large Multi-Object Tracking Dataset in
  Multiple Sports Scenes}.
\newblock \bibinfo{journal}{\emph{arXiv preprint arXiv:2304.05170}}
  (\bibinfo{year}{2023}).
\newblock


\bibitem[Decroos et~al\mbox{.}(2018)]%
        {decroos2018automatic}
\bibfield{author}{\bibinfo{person}{Tom Decroos}, \bibinfo{person}{Jan
  Van~Haaren}, {and} \bibinfo{person}{Jesse Davis}.}
  \bibinfo{year}{2018}\natexlab{}.
\newblock \showarticletitle{Automatic Discovery of Tactics in Spatio-Temporal
  Soccer Match Data}. In \bibinfo{booktitle}{\emph{SIGKDD}}. ACM,
  \bibinfo{pages}{223--232}.
\newblock


\bibitem[Di et~al\mbox{.}(2018)]%
        {di2018large}
\bibfield{author}{\bibinfo{person}{Mingyang Di}, \bibinfo{person}{Diego
  Klabjan}, \bibinfo{person}{Long Sha}, {and} \bibinfo{person}{Patrick Lucey}.}
  \bibinfo{year}{2018}\natexlab{}.
\newblock \showarticletitle{Large-Scale Adversarial Sports Play Retrieval with
  Learning to Rank}.
\newblock \bibinfo{journal}{\emph{TKDD}} \bibinfo{volume}{12},
  \bibinfo{number}{6} (\bibinfo{year}{2018}), \bibinfo{pages}{69}.
\newblock


\bibitem[Fedus et~al\mbox{.}(2018)]%
        {fedus2018maskgan}
\bibfield{author}{\bibinfo{person}{William Fedus}, \bibinfo{person}{Ian
  Goodfellow}, {and} \bibinfo{person}{Andrew~M Dai}.}
  \bibinfo{year}{2018}\natexlab{}.
\newblock \showarticletitle{Maskgan: better text generation via filling in
  the\_}.
\newblock \bibinfo{journal}{\emph{arXiv}} (\bibinfo{year}{2018}).
\newblock


\bibitem[Frank et~al\mbox{.}(2001)]%
        {frank2001time}
\bibfield{author}{\bibinfo{person}{Ray~J Frank}, \bibinfo{person}{Neil Davey},
  {and} \bibinfo{person}{Stephen~P Hunt}.} \bibinfo{year}{2001}\natexlab{}.
\newblock \showarticletitle{Time series prediction and neural networks}.
\newblock \bibinfo{journal}{\emph{JIRS}} (\bibinfo{year}{2001}).
\newblock


\bibitem[Goodfellow et~al\mbox{.}(2014)]%
        {goodfellow2014generative}
\bibfield{author}{\bibinfo{person}{Ian~J Goodfellow}, \bibinfo{person}{Jean
  Pouget-Abadie}, \bibinfo{person}{Mehdi Mirza}, \bibinfo{person}{Bing Xu},
  \bibinfo{person}{David Warde-Farley}, \bibinfo{person}{Sherjil Ozair},
  \bibinfo{person}{Aaron Courville}, {and} \bibinfo{person}{Yoshua Bengio}.}
  \bibinfo{year}{2014}\natexlab{}.
\newblock \showarticletitle{Generative adversarial networks}.
\newblock \bibinfo{journal}{\emph{arXiv}} (\bibinfo{year}{2014}).
\newblock


\bibitem[Guo et~al\mbox{.}(2018)]%
        {guo2018long}
\bibfield{author}{\bibinfo{person}{Jiaxian Guo}, \bibinfo{person}{Sidi Lu},
  \bibinfo{person}{Han Cai}, \bibinfo{person}{Weinan Zhang},
  \bibinfo{person}{Yong Yu}, {and} \bibinfo{person}{Jun Wang}.}
  \bibinfo{year}{2018}\natexlab{}.
\newblock \showarticletitle{Long text generation via adversarial training with
  leaked information}. In \bibinfo{booktitle}{\emph{AAAI}},
  Vol.~\bibinfo{volume}{32}.
\newblock


\bibitem[He and Sun(2015)]%
        {he2015convolutional}
\bibfield{author}{\bibinfo{person}{Kaiming He} {and} \bibinfo{person}{Jian
  Sun}.} \bibinfo{year}{2015}\natexlab{}.
\newblock \showarticletitle{Convolutional neural networks at constrained time
  cost}. In \bibinfo{booktitle}{\emph{CVPR}}. \bibinfo{pages}{5353--5360}.
\newblock


\bibitem[Henriques et~al\mbox{.}(2013)]%
        {henriques2013beyond}
\bibfield{author}{\bibinfo{person}{Joao~F Henriques}, \bibinfo{person}{Joao
  Carreira}, \bibinfo{person}{Rui Caseiro}, {and} \bibinfo{person}{Jorge
  Batista}.} \bibinfo{year}{2013}\natexlab{}.
\newblock \showarticletitle{Beyond hard negative mining: Efficient detector
  learning via block-circulant decomposition}. In
  \bibinfo{booktitle}{\emph{ICCV}}. \bibinfo{pages}{2760--2767}.
\newblock


\bibitem[Hoffer and Ailon(2015)]%
        {hoffer2015deep}
\bibfield{author}{\bibinfo{person}{Elad Hoffer} {and} \bibinfo{person}{Nir
  Ailon}.} \bibinfo{year}{2015}\natexlab{}.
\newblock \showarticletitle{Deep metric learning using triplet network}. In
  \bibinfo{booktitle}{\emph{SIMBAD}}. Springer, \bibinfo{pages}{84--92}.
\newblock


\bibitem[Huttenlocher et~al\mbox{.}(1993)]%
        {huttenlocher1993comparing}
\bibfield{author}{\bibinfo{person}{Daniel~P Huttenlocher},
  \bibinfo{person}{Gregory~A. Klanderman}, {and} \bibinfo{person}{William~J
  Rucklidge}.} \bibinfo{year}{1993}\natexlab{}.
\newblock \showarticletitle{Comparing images using the Hausdorff distance}.
\newblock \bibinfo{journal}{\emph{TPAMI}} \bibinfo{volume}{15},
  \bibinfo{number}{9} (\bibinfo{year}{1993}), \bibinfo{pages}{850--863}.
\newblock


\bibitem[Ju et~al\mbox{.}(2020)]%
        {ju2020interaction}
\bibfield{author}{\bibinfo{person}{Ce Ju}, \bibinfo{person}{Zheng Wang},
  \bibinfo{person}{Cheng Long}, \bibinfo{person}{Xiaoyu Zhang}, {and}
  \bibinfo{person}{Dong~Eui Chang}.} \bibinfo{year}{2020}\natexlab{}.
\newblock \showarticletitle{Interaction-aware kalman neural networks for
  trajectory prediction}. In \bibinfo{booktitle}{\emph{IEEE IV}}. IEEE,
  \bibinfo{pages}{1793--1800}.
\newblock


\bibitem[Kaya and Bilge(2019)]%
        {kaya2019deep}
\bibfield{author}{\bibinfo{person}{Mahmut Kaya} {and}
  \bibinfo{person}{Hasan~{\c{S}}akir Bilge}.} \bibinfo{year}{2019}\natexlab{}.
\newblock \showarticletitle{Deep metric learning: A survey}.
\newblock \bibinfo{journal}{\emph{Symmetry}} \bibinfo{volume}{11},
  \bibinfo{number}{9} (\bibinfo{year}{2019}), \bibinfo{pages}{1066}.
\newblock


\bibitem[Kusner and Hern{\'a}ndez-Lobato(2016)]%
        {kusner2016gans}
\bibfield{author}{\bibinfo{person}{Matt~J Kusner} {and}
  \bibinfo{person}{Jos{\'e}~Miguel Hern{\'a}ndez-Lobato}.}
  \bibinfo{year}{2016}\natexlab{}.
\newblock \showarticletitle{Gans for sequences of discrete elements with the
  gumbel-softmax distribution}.
\newblock \bibinfo{journal}{\emph{arXiv}} (\bibinfo{year}{2016}).
\newblock


\bibitem[LeCun et~al\mbox{.}(1998)]%
        {lecun1998gradient}
\bibfield{author}{\bibinfo{person}{Yann LeCun}, \bibinfo{person}{L{\'e}on
  Bottou}, \bibinfo{person}{Yoshua Bengio}, {and} \bibinfo{person}{Patrick
  Haffner}.} \bibinfo{year}{1998}\natexlab{}.
\newblock \showarticletitle{Gradient-based learning applied to document
  recognition}.
\newblock \bibinfo{journal}{\emph{IEEE}} \bibinfo{volume}{86},
  \bibinfo{number}{11} (\bibinfo{year}{1998}), \bibinfo{pages}{2278--2324}.
\newblock


\bibitem[Li et~al\mbox{.}(2018)]%
        {li2017diffusion}
\bibfield{author}{\bibinfo{person}{Yaguang Li}, \bibinfo{person}{Rose Yu},
  \bibinfo{person}{Cyrus Shahabi}, {and} \bibinfo{person}{Yan Liu}.}
  \bibinfo{year}{2018}\natexlab{}.
\newblock \showarticletitle{Diffusion convolutional recurrent neural network:
  Data-driven traffic forecasting}.
\newblock \bibinfo{journal}{\emph{ICLR}} (\bibinfo{year}{2018}).
\newblock


\bibitem[Lin et~al\mbox{.}(2017)]%
        {lin2017adversarial}
\bibfield{author}{\bibinfo{person}{Kevin Lin}, \bibinfo{person}{Dianqi Li},
  \bibinfo{person}{Xiaodong He}, \bibinfo{person}{Zhengyou Zhang}, {and}
  \bibinfo{person}{Ming-Ting Sun}.} \bibinfo{year}{2017}\natexlab{}.
\newblock \showarticletitle{Adversarial ranking for language generation}.
\newblock \bibinfo{journal}{\emph{NIPS}} (\bibinfo{year}{2017}).
\newblock


\bibitem[Lin et~al\mbox{.}(2004)]%
        {lin2004grey}
\bibfield{author}{\bibinfo{person}{ZM Lin}, \bibinfo{person}{Jr-Syu Yang},
  {and} \bibinfo{person}{CY Yang}.} \bibinfo{year}{2004}\natexlab{}.
\newblock \showarticletitle{Grey decision-making for a billiard robot}. In
  \bibinfo{booktitle}{\emph{ICSMC}}, Vol.~\bibinfo{volume}{6}. IEEE,
  \bibinfo{pages}{5350--5355}.
\newblock


\bibitem[Manandhar et~al\mbox{.}(2020)]%
        {manandhar2020learning}
\bibfield{author}{\bibinfo{person}{Dipu Manandhar}, \bibinfo{person}{Dan Ruta},
  {and} \bibinfo{person}{John Collomosse}.} \bibinfo{year}{2020}\natexlab{}.
\newblock \showarticletitle{Learning structural similarity of user interface
  layouts using graph networks}. In \bibinfo{booktitle}{\emph{ECCV}}.
  \bibinfo{pages}{730--746}.
\newblock


\bibitem[Mathavan et~al\mbox{.}(2010)]%
        {mathavan2010theoretical}
\bibfield{author}{\bibinfo{person}{Senthan Mathavan}, \bibinfo{person}{MR
  Jackson}, {and} \bibinfo{person}{Robert~M Parkin}.}
  \bibinfo{year}{2010}\natexlab{}.
\newblock \showarticletitle{A theoretical analysis of billiard ball dynamics
  under cushion impacts}.
\newblock \bibinfo{journal}{\emph{JMES}} \bibinfo{volume}{224},
  \bibinfo{number}{9} (\bibinfo{year}{2010}), \bibinfo{pages}{1863--1873}.
\newblock


\bibitem[Nie et~al\mbox{.}(2018)]%
        {nie2018relgan}
\bibfield{author}{\bibinfo{person}{Weili Nie}, \bibinfo{person}{Nina
  Narodytska}, {and} \bibinfo{person}{Ankit Patel}.}
  \bibinfo{year}{2018}\natexlab{}.
\newblock \showarticletitle{Relgan: Relational generative adversarial networks
  for text generation}. In \bibinfo{booktitle}{\emph{ICLR}}.
\newblock


\bibitem[Pan et~al\mbox{.}(2021)]%
        {pan2021can}
\bibfield{author}{\bibinfo{person}{Jing~Wen Pan}, \bibinfo{person}{John Komar},
  \bibinfo{person}{Shawn Bing~Kai Sng}, {and} \bibinfo{person}{Pui~Wah Kong}.}
  \bibinfo{year}{2021}\natexlab{}.
\newblock \showarticletitle{Can a Good Break Shot Determine the Game Outcome in
  9-Ball?}
\newblock \bibinfo{journal}{\emph{Frontiers in psychology}}
  \bibinfo{volume}{12} (\bibinfo{year}{2021}).
\newblock


\bibitem[Patil et~al\mbox{.}(2021)]%
        {patil2021layoutgmn}
\bibfield{author}{\bibinfo{person}{Akshay~Gadi Patil}, \bibinfo{person}{Manyi
  Li}, \bibinfo{person}{Matthew Fisher}, \bibinfo{person}{Manolis Savva}, {and}
  \bibinfo{person}{Hao Zhang}.} \bibinfo{year}{2021}\natexlab{}.
\newblock \showarticletitle{Layoutgmn: Neural graph matching for structural
  layout similarity}. In \bibinfo{booktitle}{\emph{CVPR}}.
  \bibinfo{pages}{11048--11057}.
\newblock


\bibitem[Pileggi et~al\mbox{.}(2012)]%
        {pileggi2012snapshot}
\bibfield{author}{\bibinfo{person}{Hannah Pileggi}, \bibinfo{person}{Charles~D
  Stolper}, \bibinfo{person}{J~Michael Boyle}, {and} \bibinfo{person}{John~T
  Stasko}.} \bibinfo{year}{2012}\natexlab{}.
\newblock \showarticletitle{Snapshot: Visualization to propel ice hockey
  analytics}.
\newblock \bibinfo{journal}{\emph{IEEE TVCG}} \bibinfo{volume}{18},
  \bibinfo{number}{12} (\bibinfo{year}{2012}), \bibinfo{pages}{2819--2828}.
\newblock


\bibitem[Qi et~al\mbox{.}(2017a)]%
        {qi2017pointnet}
\bibfield{author}{\bibinfo{person}{Charles~R Qi}, \bibinfo{person}{Hao Su},
  \bibinfo{person}{Kaichun Mo}, {and} \bibinfo{person}{Leonidas~J Guibas}.}
  \bibinfo{year}{2017}\natexlab{a}.
\newblock \showarticletitle{Pointnet: Deep learning on point sets for 3d
  classification and segmentation}. In \bibinfo{booktitle}{\emph{CVPR}}.
  \bibinfo{pages}{652--660}.
\newblock


\bibitem[Qi et~al\mbox{.}(2017b)]%
        {qi2017pointnet++}
\bibfield{author}{\bibinfo{person}{Charles~R Qi}, \bibinfo{person}{Li Yi},
  \bibinfo{person}{Hao Su}, {and} \bibinfo{person}{Leonidas~J Guibas}.}
  \bibinfo{year}{2017}\natexlab{b}.
\newblock \showarticletitle{Pointnet++: Deep hierarchical feature learning on
  point sets in a metric space}.
\newblock \bibinfo{journal}{\emph{arXiv}} (\bibinfo{year}{2017}).
\newblock


\bibitem[Ranu et~al\mbox{.}(2015)]%
        {ranu2015indexing}
\bibfield{author}{\bibinfo{person}{Sayan Ranu}, \bibinfo{person}{P Deepak},
  \bibinfo{person}{Aditya~D Telang}, \bibinfo{person}{Prasad Deshpande},
  \bibinfo{person}{Sriram Raghavan}, {et~al\mbox{.}}}
  \bibinfo{year}{2015}\natexlab{}.
\newblock \showarticletitle{Indexing and matching trajectories under
  inconsistent sampling rates}. In \bibinfo{booktitle}{\emph{ICDE}}. IEEE,
  \bibinfo{pages}{999--1010}.
\newblock


\bibitem[Rubner et~al\mbox{.}(2000)]%
        {rubner2000earth}
\bibfield{author}{\bibinfo{person}{Yossi Rubner}, \bibinfo{person}{Carlo
  Tomasi}, {and} \bibinfo{person}{Leonidas~J Guibas}.}
  \bibinfo{year}{2000}\natexlab{}.
\newblock \showarticletitle{The earth mover's distance as a metric for image
  retrieval}.
\newblock \bibinfo{journal}{\emph{IJCV}} \bibinfo{volume}{40},
  \bibinfo{number}{2} (\bibinfo{year}{2000}), \bibinfo{pages}{99--121}.
\newblock


\bibitem[Sapankevych and Sankar(2009)]%
        {sapankevych2009time}
\bibfield{author}{\bibinfo{person}{Nicholas~I Sapankevych} {and}
  \bibinfo{person}{Ravi Sankar}.} \bibinfo{year}{2009}\natexlab{}.
\newblock \showarticletitle{Time series prediction using support vector
  machines: a survey}.
\newblock \bibinfo{journal}{\emph{CIM}} (\bibinfo{year}{2009}).
\newblock


\bibitem[Sha et~al\mbox{.}(2016)]%
        {sha2016chalkboarding}
\bibfield{author}{\bibinfo{person}{Long Sha}, \bibinfo{person}{Patrick Lucey},
  \bibinfo{person}{Yisong Yue}, \bibinfo{person}{Peter Carr},
  \bibinfo{person}{Charlie Rohlf}, {and} \bibinfo{person}{Iain Matthews}.}
  \bibinfo{year}{2016}\natexlab{}.
\newblock \showarticletitle{Chalkboarding: A new spatiotemporal query paradigm
  for sports play retrieval}. In \bibinfo{booktitle}{\emph{IUI}}. ACM,
  \bibinfo{pages}{336--347}.
\newblock


\bibitem[Skianis et~al\mbox{.}(2020)]%
        {skianis2020rep}
\bibfield{author}{\bibinfo{person}{Konstantinos Skianis},
  \bibinfo{person}{Giannis Nikolentzos}, \bibinfo{person}{Stratis Limnios},
  {and} \bibinfo{person}{Michalis Vazirgiannis}.}
  \bibinfo{year}{2020}\natexlab{}.
\newblock \showarticletitle{Rep the set: Neural networks for learning set
  representations}. In \bibinfo{booktitle}{\emph{AISTATS}}. PMLR,
  \bibinfo{pages}{1410--1420}.
\newblock


\bibitem[Smith(2006)]%
        {smith2006running}
\bibfield{author}{\bibinfo{person}{Michael Smith}.}
  \bibinfo{year}{2006}\natexlab{}.
\newblock \showarticletitle{Running the table: An AI for computer billiards}.
  In \bibinfo{booktitle}{\emph{AAAI}}, Vol.~\bibinfo{volume}{21}.
  \bibinfo{pages}{994}.
\newblock


\bibitem[Smith(2007)]%
        {smith2007pickpocket}
\bibfield{author}{\bibinfo{person}{Michael Smith}.}
  \bibinfo{year}{2007}\natexlab{}.
\newblock \showarticletitle{PickPocket: A computer billiards shark}.
\newblock \bibinfo{journal}{\emph{Artificial Intelligence}}
  \bibinfo{volume}{171}, \bibinfo{number}{16-17} (\bibinfo{year}{2007}),
  \bibinfo{pages}{1069--1091}.
\newblock


\bibitem[Su et~al\mbox{.}(2020)]%
        {su2020survey}
\bibfield{author}{\bibinfo{person}{Han Su}, \bibinfo{person}{Shuncheng Liu},
  \bibinfo{person}{Bolong Zheng}, \bibinfo{person}{Xiaofang Zhou}, {and}
  \bibinfo{person}{Kai Zheng}.} \bibinfo{year}{2020}\natexlab{}.
\newblock \showarticletitle{A survey of trajectory distance measures and
  performance evaluation}.
\newblock \bibinfo{journal}{\emph{The VLDB Journal}} \bibinfo{volume}{29},
  \bibinfo{number}{1} (\bibinfo{year}{2020}), \bibinfo{pages}{3--32}.
\newblock


\bibitem[Vlachos et~al\mbox{.}(2002)]%
        {vlachos2002discovering}
\bibfield{author}{\bibinfo{person}{Michail Vlachos}, \bibinfo{person}{George
  Kollios}, {and} \bibinfo{person}{Dimitrios Gunopulos}.}
  \bibinfo{year}{2002}\natexlab{}.
\newblock \showarticletitle{Discovering similar multidimensional trajectories}.
  In \bibinfo{booktitle}{\emph{Proceedings 18th international conference on
  data engineering (ICDE)}}. IEEE, \bibinfo{pages}{673--684}.
\newblock


\bibitem[Wang et~al\mbox{.}(2021a)]%
        {wang2021survey}
\bibfield{author}{\bibinfo{person}{Sheng Wang}, \bibinfo{person}{Zhifeng Bao},
  \bibinfo{person}{J~Shane Culpepper}, {and} \bibinfo{person}{Gao Cong}.}
  \bibinfo{year}{2021}\natexlab{a}.
\newblock \showarticletitle{A survey on trajectory data management, analytics,
  and learning}.
\newblock \bibinfo{journal}{\emph{ACM Computing Surveys (CSUR)}}
  \bibinfo{volume}{54}, \bibinfo{number}{2} (\bibinfo{year}{2021}),
  \bibinfo{pages}{1--36}.
\newblock


\bibitem[Wang et~al\mbox{.}(2021b)]%
        {wang2021similar}
\bibfield{author}{\bibinfo{person}{Zheng Wang}, \bibinfo{person}{Cheng Long},
  {and} \bibinfo{person}{Gao Cong}.} \bibinfo{year}{2021}\natexlab{b}.
\newblock \showarticletitle{Similar Sports Play Retrieval with Deep
  Reinforcement Learning}.
\newblock \bibinfo{journal}{\emph{IEEE TKDE}} (\bibinfo{year}{2021}).
\newblock


\bibitem[Wang et~al\mbox{.}(2019)]%
        {wang2019effective}
\bibfield{author}{\bibinfo{person}{Zheng Wang}, \bibinfo{person}{Cheng Long},
  \bibinfo{person}{Gao Cong}, {and} \bibinfo{person}{Ce Ju}.}
  \bibinfo{year}{2019}\natexlab{}.
\newblock \showarticletitle{Effective and efficient sports play retrieval with
  deep representation learning}. In \bibinfo{booktitle}{\emph{SIGKDD}}.
  \bibinfo{pages}{499--509}.
\newblock


\bibitem[Weigend(2018)]%
        {weigend2018time}
\bibfield{author}{\bibinfo{person}{Andreas~S Weigend}.}
  \bibinfo{year}{2018}\natexlab{}.
\newblock \bibinfo{booktitle}{\emph{Time series prediction: forecasting the
  future and understanding the past}}.
\newblock \bibinfo{publisher}{Routledge}.
\newblock


\bibitem[Williams(1992)]%
        {williams1992simple}
\bibfield{author}{\bibinfo{person}{Ronald~J Williams}.}
  \bibinfo{year}{1992}\natexlab{}.
\newblock \showarticletitle{Simple statistical gradient-following algorithms
  for connectionist reinforcement learning}.
\newblock \bibinfo{journal}{\emph{Machine learning}} \bibinfo{volume}{8},
  \bibinfo{number}{3-4} (\bibinfo{year}{1992}), \bibinfo{pages}{229--256}.
\newblock


\bibitem[Yi et~al\mbox{.}(1998)]%
        {yi1998efficient}
\bibfield{author}{\bibinfo{person}{Byoung-Kee Yi}, \bibinfo{person}{HV
  Jagadish}, {and} \bibinfo{person}{Christos Faloutsos}.}
  \bibinfo{year}{1998}\natexlab{}.
\newblock \showarticletitle{Efficient retrieval of similar time sequences under
  time warping}. In \bibinfo{booktitle}{\emph{ICDE}}. IEEE,
  \bibinfo{pages}{201--208}.
\newblock


\bibitem[Yu et~al\mbox{.}(2017)]%
        {yu2017seqgan}
\bibfield{author}{\bibinfo{person}{Lantao Yu}, \bibinfo{person}{Weinan Zhang},
  \bibinfo{person}{Jun Wang}, {and} \bibinfo{person}{Yong Yu}.}
  \bibinfo{year}{2017}\natexlab{}.
\newblock \showarticletitle{Seqgan: Sequence generative adversarial nets with
  policy gradient}. In \bibinfo{booktitle}{\emph{AAAI}},
  Vol.~\bibinfo{volume}{31}.
\newblock


\bibitem[Yue et~al\mbox{.}(2014)]%
        {yue2014learning}
\bibfield{author}{\bibinfo{person}{Yisong Yue}, \bibinfo{person}{Patrick
  Lucey}, \bibinfo{person}{Peter Carr}, \bibinfo{person}{Alina Bialkowski},
  {and} \bibinfo{person}{Iain Matthews}.} \bibinfo{year}{2014}\natexlab{}.
\newblock \showarticletitle{Learning fine-grained spatial models for dynamic
  sports play prediction}. In \bibinfo{booktitle}{\emph{ICDM}}. IEEE,
  \bibinfo{pages}{670--679}.
\newblock


\bibitem[Zaheer et~al\mbox{.}(2017)]%
        {zaheer2017deep}
\bibfield{author}{\bibinfo{person}{Manzil Zaheer}, \bibinfo{person}{Satwik
  Kottur}, \bibinfo{person}{Siamak Ravanbakhsh}, \bibinfo{person}{Barnabas
  Poczos}, \bibinfo{person}{Ruslan Salakhutdinov}, {and}
  \bibinfo{person}{Alexander Smola}.} \bibinfo{year}{2017}\natexlab{}.
\newblock \showarticletitle{Deep sets}.
\newblock \bibinfo{journal}{\emph{arXiv}} (\bibinfo{year}{2017}).
\newblock


\bibitem[Zhang et~al\mbox{.}(2022a)]%
        {zhang2022predicting}
\bibfield{author}{\bibinfo{person}{Qianru Zhang}, \bibinfo{person}{Zheng Wang},
  \bibinfo{person}{Cheng Long}, {and} \bibinfo{person}{Siu-Ming Yiu}.}
  \bibinfo{year}{2022}\natexlab{a}.
\newblock \showarticletitle{On Predicting and Generating a Good Break Shot in
  Billiards Sports}. In \bibinfo{booktitle}{\emph{SDM}}. SIAM,
  \bibinfo{pages}{109--117}.
\newblock


\bibitem[Zhang et~al\mbox{.}(2022b)]%
        {TR}
\bibfield{author}{\bibinfo{person}{Qianru Zhang}, \bibinfo{person}{Zheng Wang},
  \bibinfo{person}{Cheng Long}, {and} \bibinfo{person}{Siu~Ming Yiu}.}
  \bibinfo{year}{2022}\natexlab{b}.
\newblock \bibinfo{title}{On Predicting and Generating a Good Break Shot in
  Billiards Sports (Supplementary Materials)}.
\newblock
  \bibinfo{howpublished}{\url{https://zhengwang125.github.io/paper/SDM\_Supplementary.pdf}}.
\newblock


\bibitem[Zhang and Wallace(2015)]%
        {zhang2015sensitivity}
\bibfield{author}{\bibinfo{person}{Ye Zhang} {and} \bibinfo{person}{Byron
  Wallace}.} \bibinfo{year}{2015}\natexlab{}.
\newblock \showarticletitle{A sensitivity analysis of (and practitioners' guide
  to) convolutional neural networks for sentence classification}.
\newblock \bibinfo{journal}{\emph{arXiv}} (\bibinfo{year}{2015}).
\newblock


\bibitem[Zheng et~al\mbox{.}(2010)]%
        {zheng2010geolife}
\bibfield{author}{\bibinfo{person}{Yu Zheng}, \bibinfo{person}{Xing Xie},
  \bibinfo{person}{Wei-Ying Ma}, {et~al\mbox{.}}}
  \bibinfo{year}{2010}\natexlab{}.
\newblock \showarticletitle{Geolife: A collaborative social networking service
  among user, location and trajectory.}
\newblock \bibinfo{journal}{\emph{IEEE DEB.}} \bibinfo{volume}{33},
  \bibinfo{number}{2} (\bibinfo{year}{2010}), \bibinfo{pages}{32--39}.
\newblock


\bibitem[Zhou et~al\mbox{.}(2020)]%
        {zhou2020informer}
\bibfield{author}{\bibinfo{person}{Haoyi Zhou}, \bibinfo{person}{Shanghang
  Zhang}, \bibinfo{person}{Jieqi Peng}, \bibinfo{person}{Shuai Zhang},
  \bibinfo{person}{Jianxin Li}, \bibinfo{person}{Hui Xiong}, {and}
  \bibinfo{person}{Wancai Zhang}.} \bibinfo{year}{2020}\natexlab{}.
\newblock \showarticletitle{Informer: Beyond Efficient Transformer for Long
  Sequence Time-Series Forecasting}.
\newblock \bibinfo{journal}{\emph{arXiv}} (\bibinfo{year}{2020}).
\newblock


\end{thebibliography}
